\documentclass{article}
\usepackage{comment}
\usepackage{todonotes}
\usepackage[export]{adjustbox}
\usepackage[english]{babel}
\usepackage[utf8]{inputenc}
\usepackage{capt-of}
\usepackage{amsmath}
\usepackage{tikz}
\usepackage{mathtools}
\usepackage{amsmath}
\usepackage{mathtools}
\usepackage{natbib}
\usepackage{bibentry}
\usepackage{adjustbox}
\usepackage{array}
\usepackage{booktabs}
\usepackage{dsfont}
\usepackage{tikz}
\usepackage{graphics}

\definecolor{tikbleu}{RGB}{74, 144, 226}
\definecolor{tikrouge}{RGB}{208, 2, 27}
\definecolor{colrwableu}{RGB}{74, 144, 226}
\definecolor{colrwarouge}{RGB}{208, 2, 27}

\usepackage{multirow}
\usepackage{hhline}
\usepackage[utf8]{inputenc} 
\usepackage[T1]{fontenc}    
\usepackage[hidelinks]{hyperref}
\usepackage{url}            
\usepackage{booktabs}       
\usepackage{amsfonts}       
\usepackage{nicefrac}       
\usepackage{microtype}      
\usepackage{amsthm}
\theoremstyle{definition}
\newtheorem{definition}{Definition}[section]

\usepackage{float}
\usepackage[caption = false]{subfig}
\usepackage{graphicx}

\usepackage[toc,page]{appendix}
\usepackage{algorithm}
\usepackage{algpseudocode}
\usepackage{xcolor}
\usepackage{amsthm,mathtools}
\theoremstyle{definition}

\usepackage{amsmath}
\usepackage{tikz}
\usetikzlibrary{patterns}
\usepackage{mathtools}
\usepackage{amsmath}
\usepackage{amsfonts}
\usepackage{amssymb}
\usepackage{bm}
\usepackage{mathtools}
\usepackage{natbib}
\usepackage{bibentry}
\usepackage{adjustbox}
\usepackage{array}
\usepackage{booktabs}
\usepackage{dsfont}
\usepackage{multirow}
\usepackage{hhline}
\usepackage{algorithm}
\usepackage{algpseudocode}
\usepackage{xcolor}
\newcommand{\indep}{\perp \!\!\! \perp}
\usepackage{fontawesome}
\usepackage[hidelinks]{hyperref}
\usepackage{graphics}

\definecolor{tikbleu}{RGB}{74, 144, 226}
\definecolor{tikrouge}{RGB}{208, 2, 27}
\definecolor{colrwableu}{RGB}{74, 144, 226}
\definecolor{colrwarouge}{RGB}{208, 2, 27}

\usepackage{fontawesome}

\title{Quantifying fairness and discrimination\\in predictive models}

\author{Arthur Charpentier$^{a*}$\\
        \small $^{a}$ Université du Québec à Montréal (UQAM), Montréal (Québec), Canada\\
        \small $^{*}$Corresponding author: \tt{charpentier.arthur@uqam.ca}}

\date{} 

\begin{document}

\maketitle

\abstract{The analysis of discrimination has long interested economists and lawyers. In recent years, the literature in computer science and machine learning has become interested in the subject, offering an interesting re-reading of the topic. These questions are the consequences of numerous criticisms of algorithms used to translate texts or to identify people in images. With the arrival of massive data, and the use of increasingly opaque algorithms, it is not surprising to have discriminatory algorithms, because it has become easy to have a proxy of a sensitive variable, by enriching the data indefinitely. According to \cite{kranzberg1986technology}, ``{\em technology is neither good nor bad, nor is it neutral}~", and therefore, ``{\em machine learning won't give you anything like gender neutrality `for free' that you didn't explicitely ask for}~", as claimed by \cite{kearns2019ethical}. In this article, we will come back to the general context, for predictive models in classification. We will present the main concepts of fairness, called group fairness, based on independence between the sensitive variable and the prediction, possibly conditioned on this or that information. We will finish by going further, by presenting the concepts of individual fairness. Finally, we will see how to correct a potential discrimination, in order to guarantee that a model is more ethical.}

\ 


\section{Introduction}

In a classification problem, given a collection of covariates $\boldsymbol{x}\in\mathcal{X}\subset\mathbb{R}^p$, we will estimate a model (such as a logistic regression) to compute a score $m(\boldsymbol{x})$ that will be used to predict a binary outcome $y\in\{0,1\}$. In standard econometrics, $m(\boldsymbol{x})$ is interpreted as a probability, $m:\mathcal{X}\to [0,1]$. The assignment to classes, $\widehat{y}$ will be done according to the interpretation of $y$ (here the occurrence -- or not -- of a specific risk) and $m(\boldsymbol{x})$, though a function  $m_{\tau}:\mathcal{X}\to \{0,1\}$, where $\tau$ is some chosen probability level.
\begin{itemize}
    \item In credit risk, $y$ will designate a credit default ($y=1$ in case of default), and often the credit score (in the FICO sense) will be all the higher that the risk is low. In other words, $y$ and the ``credit score", evolve in opposite directions: with this interpretation of the score, $\widehat{y}=m_{\tau}(\boldsymbol{x})=\boldsymbol{1}_{m(\boldsymbol{x})<\tau}$ (for some appropriate threshold $\tau$).
    \item In ``death insurance'', $y$ denotes a death ($y=1$ in case of death of a specific individual) and the ``score'' $m(\boldsymbol{x})$ will be the probability of dying within a specific time window. In other words, here, $m(\boldsymbol{x})$ and $y$ evolve in the same direction: with this interpretation of the score, $\widehat{y}=m_{\tau}(\boldsymbol{x})=\boldsymbol{1}_{m(\boldsymbol{x})>\tau}$.
\end{itemize}
For a regression problem, we will predict $y\in\mathbb{R}$ using a model $m(\boldsymbol{x})$, and we will consider $\widehat{y}=m(\boldsymbol{x})$. The subtle part in a classification problem is the intermediate passage through this score $m(\boldsymbol{x})$ (which will not have values in $\{0,1\}$ - like $y$ - but in $[0,1]$ or in $\mathbb{R}$). To avoid any confusion, the score $m(\boldsymbol{x})$ will be supposed to be evolve in the same direction as $y$~: large values of $m(\boldsymbol{x})$ are supposed to be associated with $y=1$. 
Finally, we suppose that there is a sensitive (or protected) attribute $s$, that is supposed to be binary, with $s\in\{0,1\}$, or $\{\text{\textcolor{tikbleu}{\scriptsize\faCircle},~\textcolor{tikrouge}{\scriptsize\faStop}}\}$, with bullets and squares) in illustrations. 

\subsection{Sensitive attribute and discrimination}

For economists, the study of discrimination is an old problem, as recalled in \cite{charles2011studying}, whose theoretical foundations were laid by \cite{becker2010economics}, and earlier considerations can be found in \cite{edgeworth1922equal} for instance. Other examples include \cite{phelps1972statistical}, who had tried to understand the origins of discrimination, and who had argued (in the context of racial discrimination) ``{\em what has been called racism -- similar remarks apply to sexism -- can be hypothesized to be the consequence of scientific management in the impersonal pursuit of maximum profit, not racial hostility or intolerance". This idea will be the basis of ``statistical discrimination}'', where the central question was to link discrimination with a rational behaviour, and therefore a notion of efficiency. \cite{bohren2019inaccurate} reminds us that statistical discrimination is sometimes qualified as ``{efficient discrimination}'', insofar as it is the optimal answer to a signal extraction problem. \cite{phelps1972statistical} and \cite{aigner1977statistical} have laid the groundwork for this statistical discrimination. For example, if the probability of being an ``offender" and of belonging to a group with a visible characteristic is higher, on average, than for other groups, then a policeman will be more likely to control an ``offender" if he controls a member of this group. There is therefore, at the group level, a ``statistical reason" which will be opposed to the individual principle of non-discrimination. Also, in the name of the efficiency of the procedure, Gary Becker defended ``racial profiling". One finds this argumentation in the fight against terrorism. As \cite{BeckerEthic} says ``{\em if young Moslem Middle Eastern males were in fact much more likely to commit terrorism against U.S. than were other groups, putting them through tighter security clearance would reduce current airport terrorism}'', in other words, ``racial profiling'' is ``effective'', even though ``{\em such profiling is `unfair' to the many young male Moslems who are not terrorists, and to the many minority shoppers who are honest}''. And he proposes a method for ``testing" efficiency ``{\em some profiling by governments and the private sector has been due to prejudice against various groups, not as a way of achieving efficiency. So it is crucial to be able to distinguish whether a profiling is efficient from whether it is evidence of discrimination. This distinction can be made in the terrorist field by keeping records on the fractions of young Moslem males and others who were searched and found with weapons or other evidence of intent to commit a terrorist act}''. Another typical example of ``statistical discrimination'', justified by economists on the grounds of economic efficiency, is discrimination in the hiring of young women (who might become pregnant, and (temporarily) interrupt her work). In the latter case, there is no need for statistics, since only a woman can become pregnant. Many countries now offer long parental leave, which can be taken by any parent, regardless of gender, which questions the economic efficiency of this profiling. For Gary Becker, this ``statistical reason" is, and must be, the only decision criterion used. That is, more or less, what actuaries have in mind when they mention ``a fair actuarial classification". As we can see, this search for efficiency raises many moral and ethical questions.

From the legal point of view, \cite{cornu2016vocabulaire}, presents discrimination as a ``{\em differentiation contrary to the principle of civil equality consisting in breaking this one to the detriment of certain physical persons because of their racial or confessional belonging, more generally of criteria on which the law forbids to found legal distinctions}'' (while admitting ``{\em more rarely, in a neutral sense, synonymous with distinction (not necessarily hateful)}'' which recalls the statistician's vision). Equality, which we find stated in the Universal Declaration of Human Rights, was initially perceived in a ``vertical" sense, imposing a constraint on the behavior of the State towards citizens. The law will impose ``horizontal" constraints in private law, whether in the labor market (during job interviews in particular) or in real estate (for renting housing). Article 23 of the Charter of Fundamental Rights of the European Union imposes the general principle that ``{\em equality between women and men must be ensured in all areas, including employment, work and pay}". This right cannot be invoked by ``women", but individually, by everyone (for example in a dispute with an employer). This ``right to equal treatment" belongs to a person, as an individual, and not in his capacity as a member of a sexual group (for example). The law says that an individual cannot be treated differently because of his or her membership to such a group, especially a group to which he or she has not chosen to belong to. This individualistic view of the law is strongly opposed to the mutual and collective concept of insurance, as insurers aim for a kind of equality within the group, based on averages, and not at the individual level (see \cite{charpentier2022assuranceGB} for a specific discussion in the context of insurance).

From a more moral point of view, according to the principle of choice mentioned by \cite{lippert2007nothing}, people should not be subjected to disadvantageous treatment because of something that does not reflect their own choices. This may explain why discrimination on the basis of gender, race, ethnicity, or genetics is widely seen as morally problematic, as argued by \cite{daniels2004functions}, \cite{palmer2007insurance}, or \cite{avraham2014towards}. However, the principle of choice is not violated if individuals have imposed additional costs as a result of their choice. How about discrimination on the basis of religious beliefs?
As we can see, the term ``discrimination" seems to encompass all sorts of realities. In particular, using the typology of \cite{Thomsen2017} and \cite{Khaitan2017}, we can distinguish {direct discrimination} and {indirect discrimination}. In a nutshell, indirect discrimination (proxy discrimination or statistical discrimination) means that instead of discriminating according to a sensitive attribute $s$ (which would not be allowed by law), a variable $x_j$, which is highly correlated with $s$, is used.

Big data and machine learning have provided an opportunity to revisit this topic that has been explored by lawyers, economists, philosophers and statisticians for the past fifty years, or longer. The aim here is to revisit these ideas, to shed new light on them, with a focus on risk management, and explore possible solutions. Lawyers, in particular, have discussed these predictive models in the context of justice, also called ``actuarial justice", as \cite{thomas2007some},  \cite{harcourt2011} or \cite{RothschildElyassi}.

The idea of bias and algorithmic discrimination is not a new one, as shown for instance by \cite{pedreshi2008discrimination}. However, since then the number of examples has continued to increase. ``{\em AI biases caused 80\% of black mortgage applicants to be rejected}'' claimed \cite{KoriHale}, or ``{\em How the use of AI risks recreating the inequity of the insurance industry of the previous century}" from \cite{wired}. Pursuing \citeauthor{david2015there}'s \citeyear{david2015there} analysis, \cite{McKinsey2017technology} announced that artificial intelligence would disrupt the workplace (including the insurance and banking sectors). These replacements raise questions, and compel the market and the regulator to be cautious. \cite{bergstrom2021calling} note, with a touch of irony, that there are people writing a bill of rights for robots, or devising ways to protect humanity from super-intelligent, Terminator-like machines, but that getting into the details of algorithmic auditing is often seen as boring, but necessary. To solve the problems that AI is creating now, we need to understand the data and algorithms we already use for more mundane purposes.

\subsection{Examples of discrimination}

In many countries, there are legal texts stipulating that all individuals must have equal opportunities with other individuals, without being hindered by discriminatory practices based on race, national or ethnic origin, color, religion, age, sex, sexual orientation, gender identity or expression, marital status, family status, genetic characteristics, disability, among others. A challenge for econometricians and statisticians is that those protected or sensitive attributes are personal, with strong concern with privacy. \cite{Kelly2021WP} reminds that ``{\em often the location data is used to determine what stores people visit. Things like sexual orientation are used to determine what demographics to target}".
Each type of data can reveal something about our interests and preferences, our opinions, our hobbies and our social interactions. For example, a MIT study\footnote{Project \href{https://immersion.media.mit.edu/}{\scriptsize{\sffamily https://immersion.media.mit.edu/}}.} demonstrated how email metadata can be used to map our lives, showing the changing dynamics of our professional and personal networks. This data can be used to infer personal information, including a person's background, religion or beliefs, political views, sexual orientation and gender identity, social relationships or health. For example, it is possible to infer our specific health conditions simply by connecting the dots between a series of phone calls. For \cite{Mayer2016privacy}, the law currently treats call content and metadata separately and makes it easier for government agencies to obtain metadata, in part because it assumes that it should not be possible to infer specific sensitive details about individuals from metadata alone.\cite{Chakraborty2013Framework} reminds us that current approaches to privacy protection, typically defined in multi-user contexts, rely on anonymization to prevent such sensitive behavior from being traced back to the user - a strategy that does not apply if the user's identity is already known. 
In 2015, as told in \cite{miracle2016anonymization}, Noah Deneau wondered if it would be possible to identify devout Muslim drivers in New York City by looking at anonymized data and inactive drivers during the five times of the day they are supposed to pray. He quickly searched for drivers who were not very active during the 30-45 minute Muslim prayer period and was able to find four examples of drivers who might fit this pattern. This brings to mind \cite{Gambs2010showme}, who conducted an investigation on a dataset containing mobility data of taxi drivers in the San Francisco Bay Area. By finding places where the taxi's GPS sensor was turned off for a long period of time (e.g. two hours), they were able to infer the interests of the drivers. For 20 of the 90 analyzed users, they were able to locate a plausible home in a small neighborhood. They even confirmed these results for 10 users by using a satellite view of the area: It showed the presence of a yellow taxi parked in front of the driver's supposed home. \cite{dalenius1977towards} introduced an interesting concept of privacy. Nothing about an individual should be learned from a dataset if it cannot be learned without having access to the dataset. We will return to this idea when we define the fairness criteria, and when we require that the protected variable $s$ cannot be predicted from the data, and from the predictions.
This idea will be used later on to insure that a model is non-discriminatory. Recently, \cite{fairLB} provided several examples of possible sensitive attribute in the context of insurance, where predictive models are used constantly (for pricing, fraud detection, etc).

In the discrimination literature, we note either $p$ (for {p}rotected variable), $s$ (for {s}ensitive variable), or even $a$ (for {protected attribute). In the causal inference literature, we use $t$ (for {t}reatment). We will use here the notation $s$ for the sensitive variable, while $m$ will be used for the score. We will refer to $s=0$ as the (assumed) advantaged population, and $s=1$ as the disadvantaged population, as in the literature in causal inference, where $t=0$ is the control group, and $t=1$ the treated group. \cite{scott2009dictionary} point out in their dictionary, that in common language, discrimination is about ``treating unfairly". But in social sciences and humanities, ``{\em most sociological analyses of discrimination concentrate on patterns of dominance and oppression, viewed as expressions of a struggle for power and privilege}". In other words, for discrimination to occur, there must be a ``favored" and a ``disadvantaged" group, a ``dominant" and a ``dominated" group, according to a (often implicit) power criterion, or even a ``majority" group and a ``minority" group (this terminology, introduced by \cite{wirth1941morale}, can lead to confusion because these terms are often defined according to a criterion of group size). Here, we assume that the favored and disadvantaged group are not related by some domination power, only to risk occurrence, $y$. More specifically, since here $y=1$ denotes the occurrence of a ``bad'' event (not a moral sense, but related to a big economic loss), $\mathbb{P}[Y=1|S=1]>\mathbb{P}[Y=1|S=0]$ means that $0$ is the favored group (the ``less risky"), while $1 $ is the disadvantaged one (the ``more risky").

\subsection{Notations on classifiers}

As we have already mentioned, the binary classification (when $y\in\{0,1\}$) is a bit particular because of the intermediate construction of a score $m(\boldsymbol{x})$, before constructing the classifier $\widehat{y}$ {\em stricto sensu} (and $\widehat{y}\in\{0,1\})$. By analogy with logistic or probit regressions, as discussed earlier, the score $m$ will be a function $\mathcal{X}\to[0,1]$, where $\mathcal{X}\subset\mathbb{R}^p$, and with, for example\footnote{Some articles define the score as $\boldsymbol{x}^\top\boldsymbol{\beta}$, which has values in $\mathbb{R}$. The score we define is an increasing function of this linear combination. Note that here, $\Phi$ denotes the cumulative distribution function of some $\mathcal{N}(0,1)$ variable.}
$$
m(\boldsymbol{x})=\frac{\exp[\boldsymbol{x}^\top\boldsymbol{\beta}]}{1+\exp[\boldsymbol{x}^\top\boldsymbol{\beta}]}\text{ or }
m(\boldsymbol{x})=\Phi(\boldsymbol{x}^\top\boldsymbol{\beta}),
$$
respectively for the logistic model and the probit model.

A standard tool to describe the performance of $m$ is to use the ROC curve, as discussed in \cite{charpentier2018econometrics}.
Figure \ref{fig:classifier:confusion} schematically resumes the analysis of a (linear) classifier, with a confusion matrix on the right, which will be used as a basis to build the ROC curve,
$$
(\mathbb{P}[m(\boldsymbol{X})> t|Y=0],\mathbb{P}[m(\boldsymbol{X})> t|Y=1])_{t\in[0,1]},
$$
 or, noting $\widehat{y}=m_{t}(\boldsymbol{x})=\boldsymbol{1}_{m(\boldsymbol{x})> t}$ for threshold $t$,
$$
(\mathbb{P}[\widehat{Y}=1|Y=0],\mathbb{P}[\widehat{Y}=1|Y=1])=(\text{FPR},\text{TPR}),
$$
meaning that the ROC curve is the true positive rate (TPR) plotted against the false positive rate (FPR), as the threshold $t$ varies from $0$ to $1$. In the example of Figure \ref{fig:classifier:confusion}, the false positive rate (FPR) is 2/7 (out of the 7 blue bullets \text{\textcolor{tikbleu}{\scriptsize\faCircle}}, 2 points are misclassified (since they are in the red region)  and announced as positive), that is 28.57\%; the true positive rate (TPR) is 5/6 (out of 6 red squares \text{\textcolor{tikrouge}{\scriptsize\faStop}}, 1 point is misclassified (being in the blue region), and announced negative), that is 83.33\%.

\begin{figure}[!ht]
    \centering
      \include{tikz/tik-16.tex} 
    \vspace{-1.3cm}\caption{Construction of the confusion matrix for a classifier, $\widehat{y}=\boldsymbol{1}(x_1+x_2>t)$, where the bullets $\text{\textcolor{tikbleu}{\scriptsize\faCircle}}$ represent points $y=0$ and the squares $\text{\textcolor{tikrouge}{\scriptsize\faStop}}$ represent points $y=1$. The zone below, in the south-west part, corresponds to the predictions $\widehat{y}=0$ and the upper zone, in the north-east part, to the predictions $\widehat{y}=1$. The square points in the lower area, and the bullet points in the upper area are bad classifications, corresponding to errors, respectively false negatives and false positives.}
    \label{fig:classifier:confusion}
\end{figure}

In Figure \ref{fig:ROC:1}, we can visualize the distribution of a credit score (in accordance with the approach in econometrics, statistics and machine learning, a low $m(\boldsymbol{x})$ score indicates a good risk, and therefore less chance of occurrence of that risk).
We can visualize on the left the distributions of the score, conditionally on $y$, with respectively in dotted line the density of $m(\boldsymbol{X})$ when $y=1$ and in solid line, that of $m(\boldsymbol{X})$ when $y=0$. Let us suppose that the threshold allowing to change class is $60\%$, so that $\widehat{y}=1$ if $m(\boldsymbol{x})>60\%$. We can observe on the left that
$$
\begin{cases}
\mathbb{P}[m(\boldsymbol{X})> 60\%|Y=1]=
\mathbb{P}[\widehat{Y}=1|Y=1]\sim 66.3\%~\text{true positive rate}\\
\mathbb{P}[m(\boldsymbol{X})> 60\%|Y=0]=
\mathbb{P}[\widehat{Y}=1|Y=0]\sim 9.6\%~\text{false positive rate}
\end{cases}
$$
(because $\widehat{y}=1$ corresponds to a ``positive''). 

\begin{figure}[!ht]
    \centering
    \includegraphics[width=.99\textwidth]{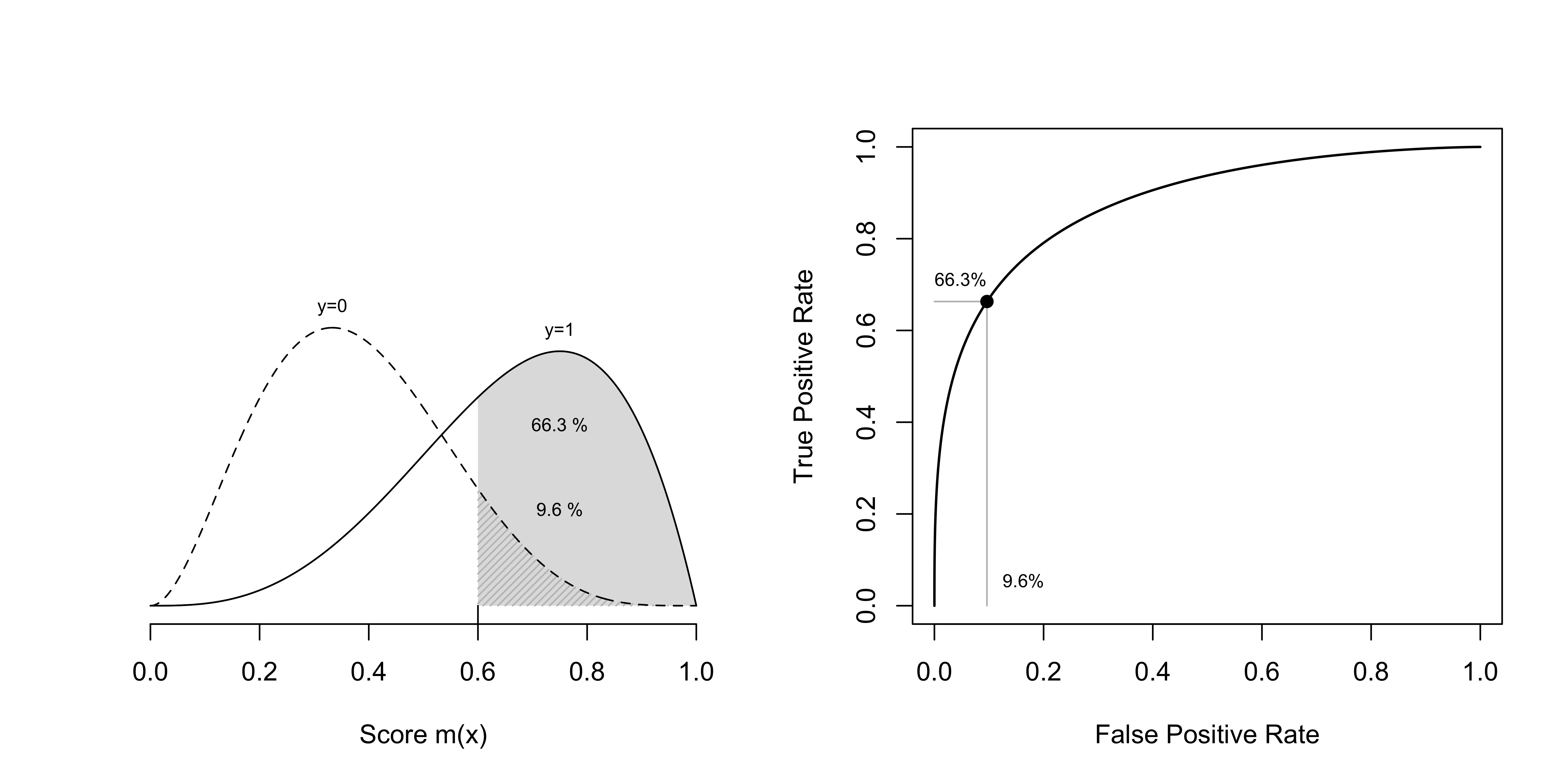}
    \caption{Distributions of $S$ score conditional on $y=1$ and $y=0$, left, and ROC curve, right, for a score $n$, with $y=1$ denoting the occurrence of a risk. The areas on the left and the point on the right of the ROC curve correspond to a threshold $\tau$ of 60\%. For that specific threshold, the true positive rate (the gray on the right of $0.6$, associated with $y=1$) of 66.3\% while the false positive rate (the gray dashed area on the right of $0.6$, associated with $y=0$) of 9.6\% .}
    \label{fig:ROC:1}
\end{figure}

To illustrate more precisely the construction, let us consider a toy dataset, like the one in \cite{kearns2019ethical}, as in Figure \ref{fig:circle:square:1}. In our case, we consider bullet individuals $\text{\textcolor{tikbleu}{\scriptsize\faCircle}}$ ($s=0$) and square individuals $\text{\textcolor{tikrouge}{\scriptsize\faStop}}$ ($s=1$), which will be our sensitive variable $s$.
Bullet individuals $\text{\textcolor{tikbleu}{\scriptsize\faCircle}}$ ($s=0$) are said to be favored since the distribution of $m(\boldsymbol{X})$ is less skewed towards that right than the distribution of $m(\boldsymbol{X})$  of square individuals $\text{\textcolor{tikrouge}{\scriptsize\faStop}}$, therefore considered disadvantaged. Favored individuals are less facing the risk we model here than disadvantaged ones. 
In Figure \ref{fig:circle:square:1}, at the $m(\boldsymbol{x}_i)\in[0,1]$ level, we observe a bullet $\text{\textcolor{tikbleu}{\scriptsize\faCircle}}$ or square $\text{\textcolor{tikrouge}{\scriptsize\faStop}}$, locating on the $[0,1]$ scale according to the value of $m(\boldsymbol{x}_i)$. The value given below is $y_i\in\{0,1\}$.

\begin{figure}[!ht]
    \centering
  \tikzset{every picture/.style={line width=0.75pt}} 

\begin{tikzpicture}[x=0.75pt,y=0.75pt,yscale=-1,xscale=1]

\draw    (27.37,40.1) -- (501.37,40.7) ;
\draw    (28.37,32.1) -- (28.37,49.1) ;
\draw    (501.37,33.1) -- (501.37,50.1) ;
\draw  [fill={rgb, 255:red, 74; green, 144; blue, 226 }  ,fill opacity=1 ] (61,40) .. controls (61,36.69) and (63.69,34) .. (67,34) .. controls (70.31,34) and (73,36.69) .. (73,40) .. controls (73,43.31) and (70.31,46) .. (67,46) .. controls (63.69,46) and (61,43.31) .. (61,40) -- cycle ;
\draw  [fill={rgb, 255:red, 74; green, 144; blue, 226 }  ,fill opacity=1 ] (88,40) .. controls (88,36.69) and (90.69,34) .. (94,34) .. controls (97.31,34) and (100,36.69) .. (100,40) .. controls (100,43.31) and (97.31,46) .. (94,46) .. controls (90.69,46) and (88,43.31) .. (88,40) -- cycle ;
\draw  [fill={rgb, 255:red, 74; green, 144; blue, 226 }  ,fill opacity=1 ] (106,40) .. controls (106,36.69) and (108.69,34) .. (112,34) .. controls (115.31,34) and (118,36.69) .. (118,40) .. controls (118,43.31) and (115.31,46) .. (112,46) .. controls (108.69,46) and (106,43.31) .. (106,40) -- cycle ;
\draw  [fill={rgb, 255:red, 74; green, 144; blue, 226 }  ,fill opacity=1 ] (125,40) .. controls (125,36.69) and (127.69,34) .. (131,34) .. controls (134.31,34) and (137,36.69) .. (137,40) .. controls (137,43.31) and (134.31,46) .. (131,46) .. controls (127.69,46) and (125,43.31) .. (125,40) -- cycle ;
\draw  [fill={rgb, 255:red, 74; green, 144; blue, 226 }  ,fill opacity=1 ] (143,40) .. controls (143,36.69) and (145.69,34) .. (149,34) .. controls (152.31,34) and (155,36.69) .. (155,40) .. controls (155,43.31) and (152.31,46) .. (149,46) .. controls (145.69,46) and (143,43.31) .. (143,40) -- cycle ;
\draw  [fill={rgb, 255:red, 74; green, 144; blue, 226 }  ,fill opacity=1 ] (162,40) .. controls (162,36.69) and (164.69,34) .. (168,34) .. controls (171.31,34) and (174,36.69) .. (174,40) .. controls (174,43.31) and (171.31,46) .. (168,46) .. controls (164.69,46) and (162,43.31) .. (162,40) -- cycle ;
\draw  [fill={rgb, 255:red, 74; green, 144; blue, 226 }  ,fill opacity=1 ] (251,40) .. controls (251,36.69) and (253.69,34) .. (257,34) .. controls (260.31,34) and (263,36.69) .. (263,40) .. controls (263,43.31) and (260.31,46) .. (257,46) .. controls (253.69,46) and (251,43.31) .. (251,40) -- cycle ;
\draw  [fill={rgb, 255:red, 74; green, 144; blue, 226 }  ,fill opacity=1 ] (269,40) .. controls (269,36.69) and (271.69,34) .. (275,34) .. controls (278.31,34) and (281,36.69) .. (281,40) .. controls (281,43.31) and (278.31,46) .. (275,46) .. controls (271.69,46) and (269,43.31) .. (269,40) -- cycle ;
\draw [color={rgb, 255:red, 65; green, 117; blue, 5 }  ,draw opacity=1 ]   (247.5,12) -- (248.5,82) ;
\draw  [fill={rgb, 255:red, 208; green, 2; blue, 27 }  ,fill opacity=1 ] (183,34) -- (194.5,34) -- (194.5,46) -- (183,46) -- cycle ;
\draw  [fill={rgb, 255:red, 208; green, 2; blue, 27 }  ,fill opacity=1 ] (200,34) -- (211.5,34) -- (211.5,46) -- (200,46) -- cycle ;
\draw  [fill={rgb, 255:red, 208; green, 2; blue, 27 }  ,fill opacity=1 ] (218,34) -- (229.5,34) -- (229.5,46) -- (218,46) -- cycle ;
\draw  [fill={rgb, 255:red, 208; green, 2; blue, 27 }  ,fill opacity=1 ] (234,34) -- (245.5,34) -- (245.5,46) -- (234,46) -- cycle ;
\draw  [fill={rgb, 255:red, 208; green, 2; blue, 27 }  ,fill opacity=1 ] (288,34) -- (299.5,34) -- (299.5,46) -- (288,46) -- cycle ;
\draw  [fill={rgb, 255:red, 208; green, 2; blue, 27 }  ,fill opacity=1 ] (305,34) -- (316.5,34) -- (316.5,46) -- (305,46) -- cycle ;
\draw  [fill={rgb, 255:red, 208; green, 2; blue, 27 }  ,fill opacity=1 ] (323,34) -- (334.5,34) -- (334.5,46) -- (323,46) -- cycle ;
\draw  [fill={rgb, 255:red, 208; green, 2; blue, 27 }  ,fill opacity=1 ] (339,34) -- (350.5,34) -- (350.5,46) -- (339,46) -- cycle ;
\draw  [fill={rgb, 255:red, 208; green, 2; blue, 27 }  ,fill opacity=1 ] (357,34) -- (368.5,34) -- (368.5,46) -- (357,46) -- cycle ;
\draw  [fill={rgb, 255:red, 208; green, 2; blue, 27 }  ,fill opacity=1 ] (372,34) -- (383.5,34) -- (383.5,46) -- (372,46) -- cycle ;
\draw  [fill={rgb, 255:red, 208; green, 2; blue, 27 }  ,fill opacity=1 ] (390.5,34) -- (402,34) -- (402,46) -- (390.5,46) -- cycle ;
\draw  [fill={rgb, 255:red, 208; green, 2; blue, 27 }  ,fill opacity=1 ] (405,34) -- (416.5,34) -- (416.5,46) -- (405,46) -- cycle ;
\draw  [fill={rgb, 255:red, 208; green, 2; blue, 27 }  ,fill opacity=1 ] (423,34) -- (434.5,34) -- (434.5,46) -- (423,46) -- cycle ;
\draw  [fill={rgb, 255:red, 208; green, 2; blue, 27 }  ,fill opacity=1 ] (439,34) -- (450.5,34) -- (450.5,46) -- (439,46) -- cycle ;
\draw  [fill={rgb, 255:red, 208; green, 2; blue, 27 }  ,fill opacity=1 ] (456,34) -- (467.5,34) -- (467.5,46) -- (456,46) -- cycle ;
\draw  [fill={rgb, 255:red, 208; green, 2; blue, 27 }  ,fill opacity=1 ] (474,34) -- (485.5,34) -- (485.5,46) -- (474,46) -- cycle ;

\draw (107,55) node [anchor=north west][inner sep=0.75pt]  [xscale=0.9,yscale=0.9] [align=left] {1};
\draw (145,55) node [anchor=north west][inner sep=0.75pt]  [xscale=0.9,yscale=0.9] [align=left] {1};
\draw (163,55) node [anchor=north west][inner sep=0.75pt]  [xscale=0.9,yscale=0.9] [align=left] {1};
\draw (218,55) node [anchor=north west][inner sep=0.75pt]  [xscale=0.9,yscale=0.9] [align=left] {1};
\draw (253,55) node [anchor=north west][inner sep=0.75pt]  [xscale=0.9,yscale=0.9] [align=left] {1};
\draw (271,55) node [anchor=north west][inner sep=0.75pt]  [xscale=0.9,yscale=0.9] [align=left] {1};
\draw (339,56) node [anchor=north west][inner sep=0.75pt]  [xscale=0.9,yscale=0.9] [align=left] {1};
\draw (356,56) node [anchor=north west][inner sep=0.75pt]  [xscale=0.9,yscale=0.9] [align=left] {1};
\draw (372,56) node [anchor=north west][inner sep=0.75pt]  [xscale=0.9,yscale=0.9] [align=left] {1};
\draw (391,56) node [anchor=north west][inner sep=0.75pt]  [xscale=0.9,yscale=0.9] [align=left] {1};
\draw (423,56) node [anchor=north west][inner sep=0.75pt]  [xscale=0.9,yscale=0.9] [align=left] {1};
\draw (440,56) node [anchor=north west][inner sep=0.75pt]  [xscale=0.9,yscale=0.9] [align=left] {1};
\draw (458,56) node [anchor=north west][inner sep=0.75pt]  [xscale=0.9,yscale=0.9] [align=left] {1};
\draw (475,56) node [anchor=north west][inner sep=0.75pt]  [xscale=0.9,yscale=0.9] [align=left] {1};
\draw (63,55) node [anchor=north west][inner sep=0.75pt]  [xscale=0.9,yscale=0.9] [align=left] {0};
\draw (88,55) node [anchor=north west][inner sep=0.75pt]  [xscale=0.9,yscale=0.9] [align=left] {0};
\draw (126,55) node [anchor=north west][inner sep=0.75pt]  [xscale=0.9,yscale=0.9] [align=left] {0};
\draw (183,56) node [anchor=north west][inner sep=0.75pt]  [xscale=0.9,yscale=0.9] [align=left] {0};
\draw (200,56) node [anchor=north west][inner sep=0.75pt]  [xscale=0.9,yscale=0.9] [align=left] {0};
\draw (235,55) node [anchor=north west][inner sep=0.75pt]  [xscale=0.9,yscale=0.9] [align=left] {0};
\draw (406,56) node [anchor=north west][inner sep=0.75pt]  [xscale=0.9,yscale=0.9] [align=left] {0};
\draw (289,55) node [anchor=north west][inner sep=0.75pt]  [xscale=0.9,yscale=0.9] [align=left] {0};
\draw (305,55) node [anchor=north west][inner sep=0.75pt]  [xscale=0.9,yscale=0.9] [align=left] {0};
\draw (323,55) node [anchor=north west][inner sep=0.75pt]  [xscale=0.9,yscale=0.9] [align=left] {0};
\draw (24,11) node [anchor=north west][inner sep=0.75pt]  [xscale=0.9,yscale=0.9] [align=left] {0};
\draw (497,12) node [anchor=north west][inner sep=0.75pt]  [xscale=0.9,yscale=0.9] [align=left] {1};
\draw (220,85) node [anchor=north west][inner sep=0.75pt]  [color={rgb, 255:red, 65; green, 117; blue, 5 }  ,opacity=1 ,xscale=0.9,yscale=0.9] [align=left] {thresold};

\end{tikzpicture}
  \vspace{-.6cm}
    \caption{Scale of the score $m(\boldsymbol{x})\in[0,1]$, with a sensitive attribute $s\in\{\text{\textcolor{tikbleu}{\scriptsize\faCircle}},\text{\textcolor{tikrouge}{\scriptsize\faStop}}\}$, and a variable of interest $y\in\{0,1\}$, inspired by \cite{kearns2019ethical}. The very first point on the left corresponds to $(y_i,s_i,\boldsymbol{x}_i)$ such that $y_i=0$, $s_i=\text{\textcolor{tikbleu}{\scriptsize\faCircle}}$ (or $0$), $\boldsymbol{x}_i$ are unobserved components in some space $\mathcal{X}$ such that $m(\boldsymbol{x}_i)\approx 10\%$, and since it is lower than $\tau\approx 45\%$, $\widehat{y}_i=0$. }
    \label{fig:circle:square:1}
\end{figure}
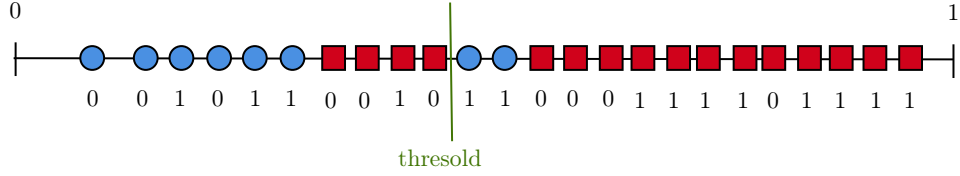

The ROC curve is the curve obtained by representing the true positive rates according to the false positive rates, by changing the threshold. It is therefore the parametric curve
$$
C(t)=\lbrace\mathbb{P}[m(\boldsymbol{X})> t|Y=0],\mathbb{P}[m(\boldsymbol{X})> t|Y=1]\rbrace,~\text{for }t\in[0,1],
$$
when the score $m(\boldsymbol{X})$ and $Y$ evolve in the same direction (a high score indicates a high risk). We define the convex envelope $\mathcal{C}$, as in \cite{hardt2016equality}. The convex envelope is interesting because it allows us to describe the set of classifiers that can be constructed from the $s$ score.
On the left of Figure \ref{fig:ROC:11}, we can see the convex envelope $\mathcal{C}$ of the ROC curve of the example in Figure \ref{fig:circle:square:1}. $\mathcal{C}$ is here a quadrilateral, the edges consisting of four segments. For example the segment [{\small\sffamily AB}] is obtained by using the classifier $s$, but by drawing the threshold at random: either the threshold associated to the point {\small\sffamily A}, or the threshold associated to the point {\small\sffamily B}.

For the right-hand side of Figure \ref{fig:ROC:11}, recall that the {\em accuracy} (noted $a$) associated with a confusion matrix is the proportion of good prediction, $\mathbb{P}[\widehat{Y}=Y]$, that is
$$
a = \frac{\text{TP}+\text{TN}}{\text{P}+\text{N}}=
 \frac{\text{TPR}\cdot \text{TPR}+(1-\text{FPR})\cdot \text{N}}{\text{P}+\text{N}}
$$
The iso-accuracy curves have the equation
$$
\text{TPR} = \frac{\text{N}}{\text{P}}\cdot \text{FPR} + 
\frac{a\cdot[\text{P}+\text{N}]-\text{N}}{\text{P}}
$$
which is linear in FPR: they are (parallel) lines of slope $\text{N}/\text{P}$, corresponding to the ratio $\mathbb{P}[Y=0]/\mathbb{P}[Y=1]$. The curve with the highest accuracy will be the highest, and it is ``{\em tangent}" to the ROC curve in {\small\sffamily B}.

\begin{figure}[!ht]
    \centering
     \includegraphics[width=.99\textwidth]{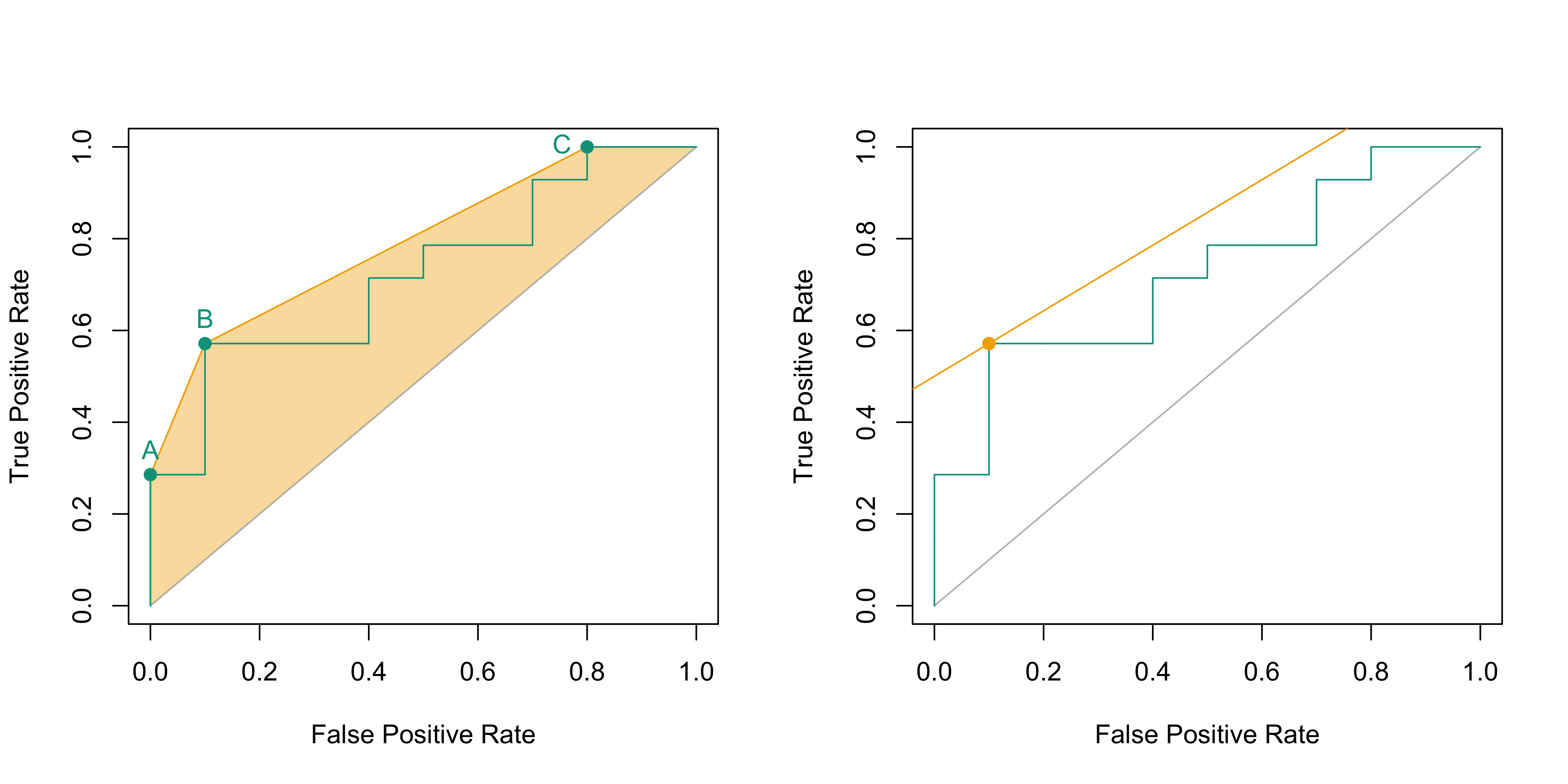}
    \caption{On the left, ROC curve $C$ and its convex envelope $\mathcal{C}$. On the right, optimal precision from the N/P ratio}.
    \label{fig:ROC:11}
\end{figure}



One can also construct $s$-conditional ROC curves, for subgroups characterized by the value of the sensitive attribute,
$$
C_s(t) = \lbrace\mathbb{P}[m(\boldsymbol{X})> t|Y=0,S=s],\mathbb{P}[m(\boldsymbol{X})> t|Y=1,S=s]\rbrace,~\text{for }t\in[0,1],
$$
for the two classes $s=0$ (or bullet points $\text{\textcolor{tikbleu}{\scriptsize\faCircle}}$ in the application in the next section) and $s=1$ (the square points $\text{\textcolor{tikrouge}{\scriptsize\faStop}}$), as well as their convex envelope $\mathcal{C}_s$ (as on Figure \ref{fig:circle:square:p:ROC}).
Note that we can also write $C(t)=\text{TPR}\circ\text{FPR}^{-1}(t)$, where $\text{FRP}(t)=\mathbb{P}[m(\boldsymbol{X})> t|Y=0]$ and $\text{TPR}(t)=\mathbb{P}[m(\boldsymbol{X})> t|Y=1]$. In other words, the ROC curve is obtained from the two survival functions of $m(\boldsymbol{X})$ FPR and TPR (respectively conditional on $T=0$ and $T=1)$. For the $s$-conditional ROC curve, $C_s(t)=\text{TPR}_s\circ\text{TPR}_s^{-1}(t)$, where $\text{FRP}_s(t)=\mathbb{P}[m(\boldsymbol{X})> t|Y=0,S=s]$ and $\text{TPR}_s(t)=\mathbb{P}[m(\boldsymbol{X})> t|Y=1,S=s]$. An finally, the AUC, the area under the curve, is then written
$$
\text{AUC} = \int_0^1 C(t)dt = \int_0^1 \text{TPR}\circ\text{FPR}^{-1}(t)dt.
$$

\begin{figure}[!ht]
    \centering
     \includegraphics[width=.98\textwidth]{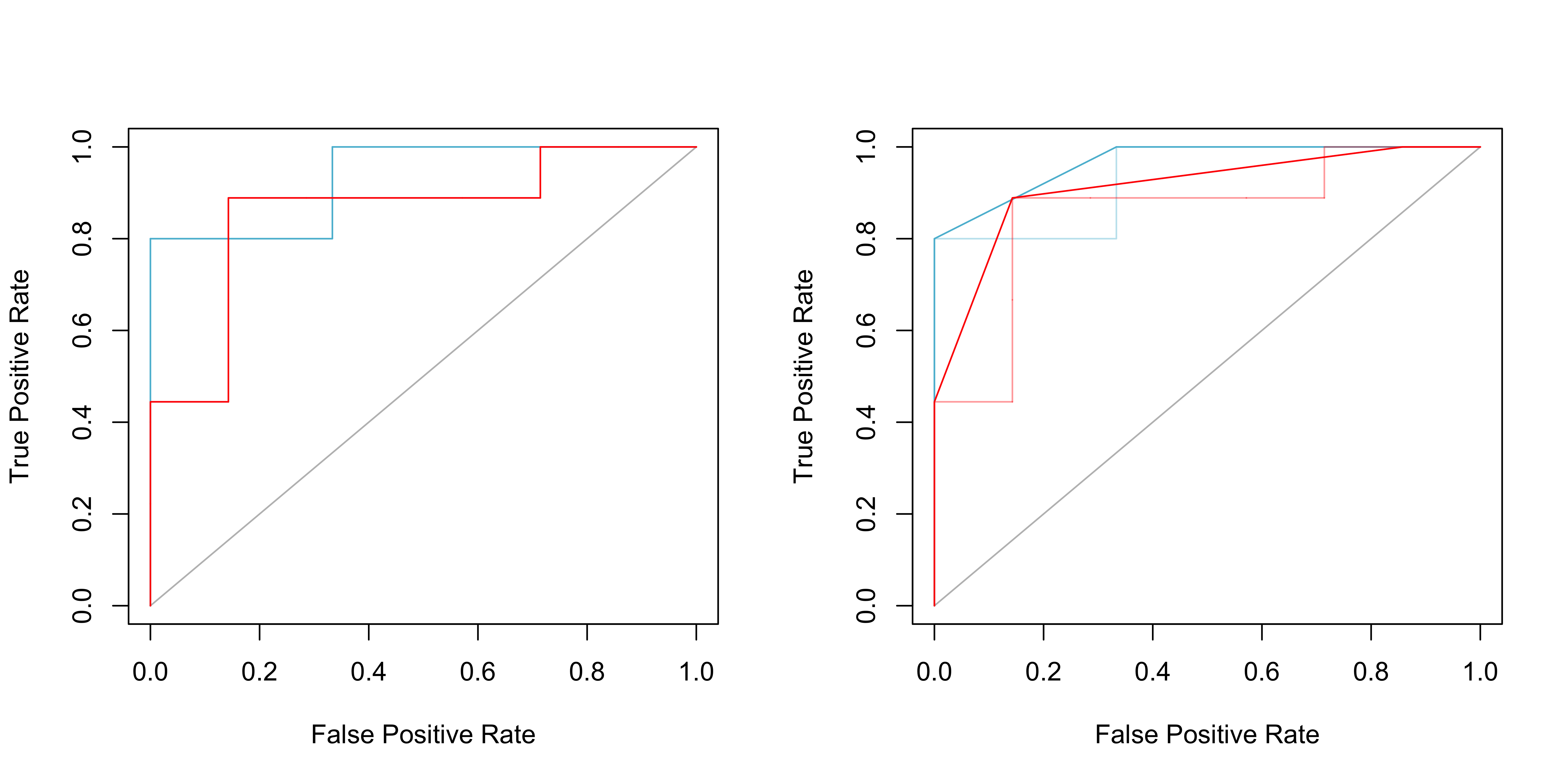}
    \caption{ROC curves $C_s$ on the left, and the convex envelope on the right $\mathcal{C}_s$, for \textcolor{colrwableu}{$s=0$} (bullet individuals $\text{\textcolor{tikbleu}{\scriptsize\faCircle}}$) and \textcolor{colrwarouge}{$s=1$} (square individuals $\text{\textcolor{tikrouge}{\scriptsize\faStop}}$).  }
    \label{fig:circle:square:p:ROC}
\end{figure}

\begin{figure}[!ht]
    \centering
     \includegraphics[width=\textwidth]{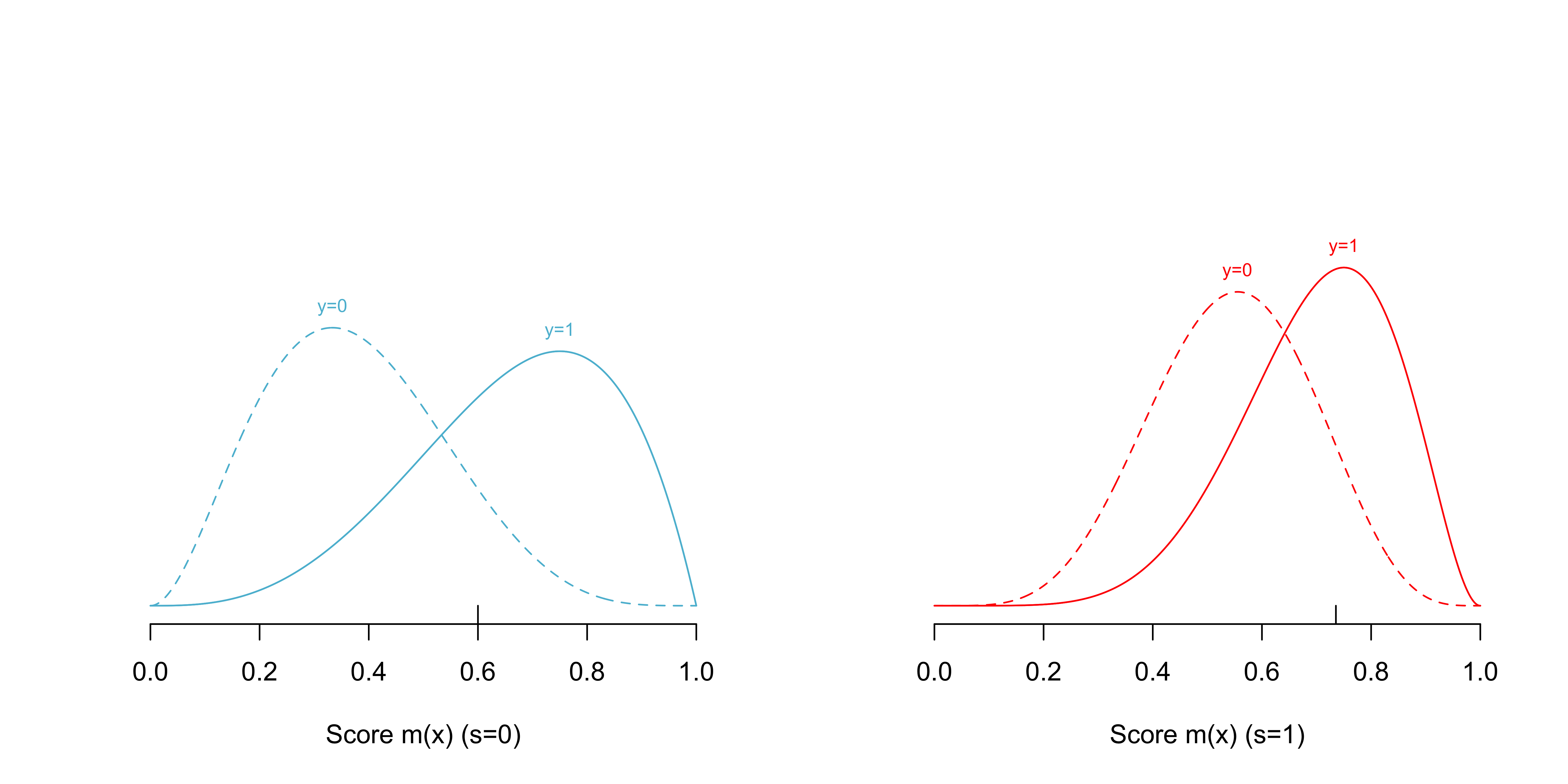}
    \caption{Conditional distributions of $m(\boldsymbol{X})| Y=1,S=0$ and $m(\boldsymbol{X})| Y=0,S=0$ ($S=0$, or $\text{\textcolor{tikbleu}{\scriptsize\faCircle}}$), on the left (population assumed \textcolor{colrwableu}{favored}), and conditional distributions of $m(\boldsymbol{X})| Y=1,S=1$ and $m(\boldsymbol{X})| Y=0,S=1$ ($S=1$, or $\text{\textcolor{tikrouge}{\scriptsize\faStop}}$), on the right (population assumed \textcolor{colrwarouge}{defavored}).}
    \label{fig:ROC:2}
\end{figure}

\subsection{Agenda}

In section \ref{sec:unawareness}, we will present the first concept of fairness, called ``fairness by unawareness". The idea will be to remove the sensitive attribute from the data, and to not use it. Even if that approach is still very popular, it is a very bad idea: not only it does not make discrimination disappear (with massive data, one can expect to get easily a proxy of the sensitive attributed), but it becomes impossible then both to quantify properly ``statistical discrimination", and to correct it. Thus, we will discuss two classes of fairness principles. The first one, in section \ref{sec:group} is called ``group fairness". In part \ref{sec:indep}, we will recall classical results on independence, conditional independence, as well as measures of dependence, with the correlation, as well as the maximal correlation. Then we will present Demographic Parity in part \ref{subsec:demog:parity} and Equalized Odds in part \ref{subsec:equal:opp}, those two concepts being probably the most popular ones. In part \ref{subsec:equal:opp}, we will present other measures than the one obtained by asking for the equality of ``true positive rates" (for instance the ``false positive rates" equality or equality of AUC). In part \ref{sub:con:demo:parity}, we will discuss extensions of Demographic Parity by adding some exogenous variables. In part \ref{sub:class:balance:2} we will discuss the idea of class balance parity, as well as calibration (or accuracy) parity. The last concept we will discuss in the concept on non-reconstruction, in part \ref{sub:non:reconstruct}. In part \ref{sug:comparison:group}, we will give a final overview, as well as a discussion about confidence intervals and uncertainty in part \ref{sub:conf:int}, and some practical example in part \ref{sub:implementation}. In section \ref{sec:eq:indiv}, we will introduce ``individual fairness" principles. In part \ref{sub:lipschitz} we will present the Lipschitz property while in part \ref{sub:counterfactual}, we discuss ``counterfactual fairness", and its connections with causal inference (``{\em what would have been $y$ is $s$ had been 1, instead of 0?}"). Finally, in section \ref{sec:correction}, we present standard techniques used to correct a possible discrimination. More precisely, we will focus on three approaches, namely pre-processing in part \ref{sub:pre:processing}, in-processin in part \ref{sub:in:processing}, and post-processing in part \ref{sub:post:processing}.

\section{Fairness by unawareness}\label{sec:unawareness}

The most popular approach to having a fair classifier is to prohibit the use of a protected variable in a predictive model. This approach is called ``{\em fairness by ignorance}'' or ``{\em fairness through unawareness}''.

\begin{definition}[Fairness Through Unawareness, \cite{Kusner17}]
We will speak of fairness through unawareness if the sensitive attribute $s$ is not explicitly used in the decision function $\widehat{y}$, i.e. neither in the construction of the score $m$, nor in the choice of the threshold level $\tau$, allowing to pass from $m$ to $\widehat{y}$.
\end{definition}

It is assumed here that the threshold choice does not depend on the sensitive criterion $s$, as in Figure \ref{fig:ROC:3}. 
In this case, for the favored population (curve \textcolor{colrwableu}{blue} on the left, $s=0$), more people in this sub-population have a lower score (and therefore a lower risk) than for the disadvantaged population (curve \textcolor{colrwarouge}{red} on the right, $s=1$).

\begin{figure}[!ht]
    \centering
     \includegraphics[width=\textwidth]{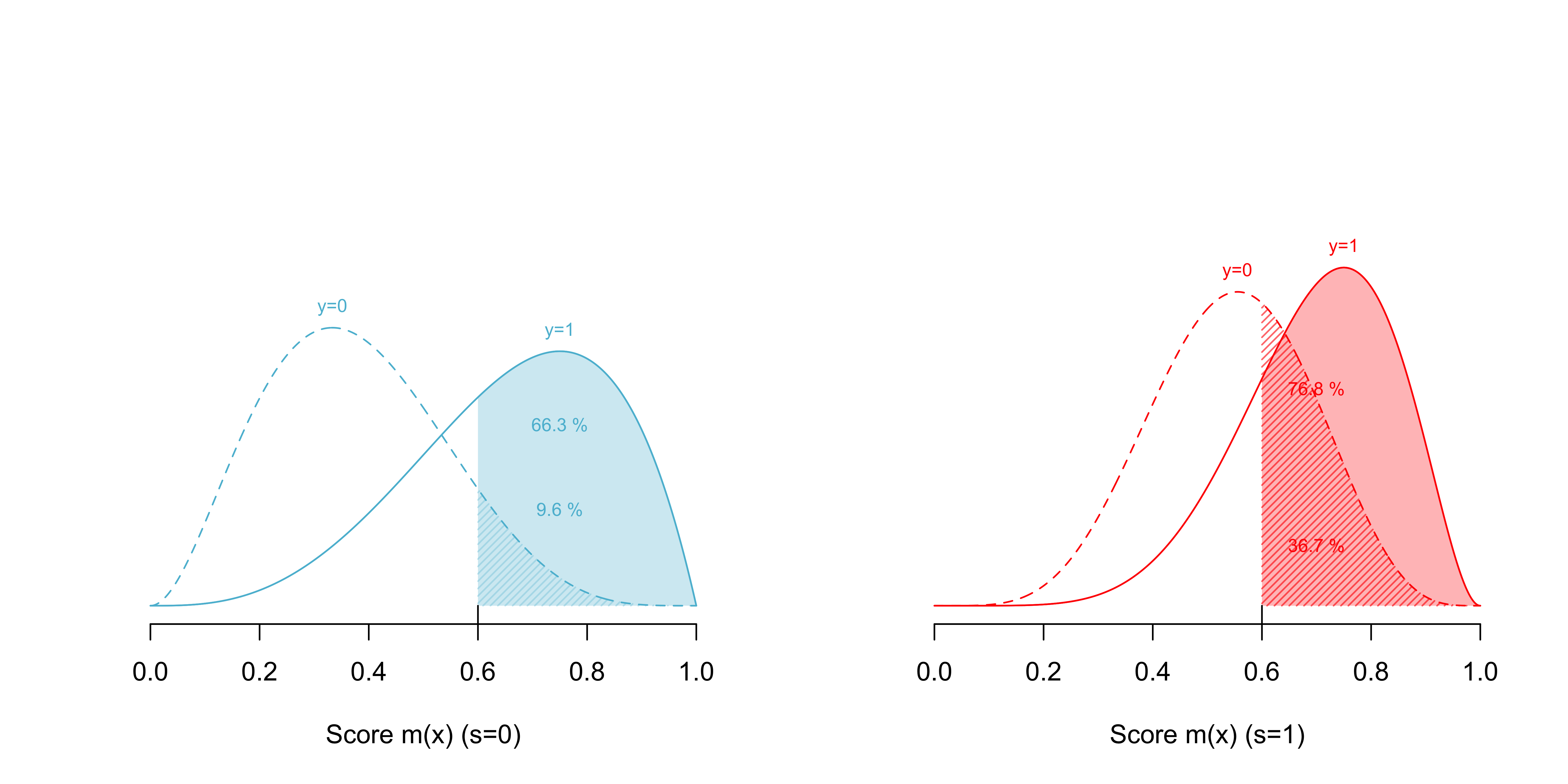}
    \caption{Distributions of $m(\boldsymbol{X})$ conditional on $y=1,s=1$ and $m(\boldsymbol{X})$ conditional on $y=0,s=1$ on the right (population assumed \textcolor{colrwarouge}{disadvantaged}), and distributions of $m(\boldsymbol{X})$ conditional on $y=1,s=0$ and $s$ conditional on $y=0,s=1$, on the left (population assumed \textcolor{colrwableu}{favored}). The threshold $\tau$ to go from $\widehat{y}=0$ to $\widehat{y}=1$ is set to 60\%.}
    \label{fig:ROC:3}
\end{figure}

Removing a sensitive variable from the training database might seem like a step forward, to insure more fairness of models. However, some of the unprotected predictor variables may in fact be (highly) correlated with the sensitive variable, allowing discrimination to go on and even eliminate all highly correlated variables as well. But this comes at a price, because each deletion of a variable also deletes valuable information for the prediction task.
\cite{gajane2017formalizing}, \cite{vzliobaite2017measuring}, \cite{verma2018fairness}, or \cite{friedler2019comparative} have identified several concepts of algorithmic fairness. 
Most definitions of fairness are based on group fairness, which addresses statistical fairness in the population as a whole. In addition, individual fairness states that similar individuals should be treated the same, regardless of their group membership. In this blog, we will focus primarily on group fairness, which has the following three definitions: (i) demographic parity, (ii) equality of opportunity, and (iii) equality of opportunity. We will now examine these in turn, before proposing some extensions, inspired by \cite{vzliobaite2017measuring} (which mentions about twenty measures) and \cite{gajane2017formalizing} (which considers seven major approaches).

\section{Group-level fairness}\label{sec:group}


\subsection{Independence and conditional independence}\label{sec:indep}

As we will see in this section, all group-level fairness concepts are related to independence, between the prediction $\widehat{Y}$ and the sensitive attribute $S$, possibly conditional on other variables, e.g. between $\widehat{Y}$ and $S$ conditional on $Y$ (``separation condition''), or $Y$ and $S$ conditional on $\widehat{Y}$ (``sufficiency condition''). Recall that independence is characterized by a separability condition, for (generally) two discrete random variables $X$ and $Y$,
$$
X\indep Y ~\Longleftrightarrow ~ \exists g,h: ~\mathbb{P}[X=x,Y=y]=g(x)\cdot h(y).
$$ 
A weaker notion is the second-order version, $X\perp Y$, meaning simply that the partial correlation is zero,
$$X\indep Y ~\Longrightarrow ~X\perp Y~ \Longleftrightarrow~
\mathbb{E}[(X-\mathbb{E}(X))(Y-\mathbb{E}(Y))] = 0,
$$
i.e. $\text{cov}(X,Y)=0$ or $\text{cor}(X,Y)=0$, Pearson's linear correlation is null, for continuous random variables. A characterization of independence can be obtained using the maximal correlation, introduced in \cite{hirschfeld_1935}, \cite{Gebelein} and \cite{renyi1959measures}, defined as the supremum of $\text{cor}(f(X),g(Y))$, for all functions $f$ and $g$ such that the correlation exists,
$$
\text{cor}^\star(X,Y)=\underset{f,g}{\max}\left\lbrace \text{cor}(f(X),g(Y))\right\rbrace.
$$
For computational reasons, it is necessary to normalize functions $f$ and $g$~: 
let $\mathcal{S}_x=\{f:\mathcal{X}\rightarrow\mathbb{R}:\mathbb{E}[f(X)]=0\text{ and }\mathbb{E}[f(X)^2]=1\}$ and similarly $\mathcal{S}_y$, and then
$$
\text{cor}^\star(X,Y)=\underset{f\in\mathcal{S}_x,g\in\mathcal{S}_y}{\max}\left\lbrace \mathbb{E}[f(X)g(Y)]\right\rbrace.
$$
As proved in \cite{renyi1959measures}, 
$$X\indep Y ~\Longleftrightarrow
\text{cor}^\star(X,Y)=0.
$$

For the conditional Independence, recall that
$$
X\indep Y~|~Z ~\Longleftrightarrow ~ \exists g,h: ~\mathbb{P}[X=x,Y=y,Z=z]=g(x,z)\cdot h(y,z).
$$ 
Note that if $X\indep Y~|~Z$ and $X\indep Z$ then we have unconditional independence between our two variables, $X\indep Y$. And if we have both $X\indep Y~|~Z$ and $X\indep Y$ then either $X\indep Z$ or $Y\indep Z$.
Note that
$$
X\indep Y~|~Z ~\Longleftrightarrow ~ Y\indep X~|~Z.
$$ Moreover
$$
X\indep Y~|~Z ~\text{and}~U=h(Y)~\Longleftrightarrow ~ X\indep U~|~Z.
$$
A weaker notion is the second-order version, $X\perp Y~|~Z$, meaning simply that the partial correlation is zero,
$$X\indep Y~|~Z ~\Longrightarrow
\mathbb{E}[(X-\mathbb{E}(X|Z))( Y-\mathbb{E}(Y|Z))] = 0,
$$
meaning that the conditional correlation is null. And as earlier, a characterization can be obtained using a conditional maximal correlation~: let $\mathcal{S}_{x|z}=\{f:\mathcal{X}\rightarrow\mathbb{R}:\mathbb{E}[f(X)|Z]=0\text{ and }\mathbb{E}[f(X)^2|Z]=1\}$ and similarly for  $\mathcal{S}_{y|z}$,
$$
\text{cor}^\star(X,Y|Z)=\underset{f\in\mathcal{S}_{x|z},g\in\mathcal{S}_{y|z}}{\max}\Big\lbrace \mathbb{E}[f(X)g(Y)|Z]\Big\rbrace.
$$

\subsection{Demographic Parity}\label{subsec:demog:parity}

As pointed out by \cite{caton2020fairness}, there are several ways to define (formally) the fairness of a classifier, or of a model. For example, one can wish for independence between the score and the group membership, $m(\boldsymbol{X})\indep S$, or between the prediction (as a class) and the sensitive variable $\widehat{Y}\indep S$.

\begin{definition}[Demographic Parity, \cite{corbettdavies2017algorithmic}, \cite{agarwal2019trade}]
A decision function $\widehat{y}$ satisfies demographic parity if $\widehat{Y}\indep S$, i.e.
$$
\mathbb{P}[\widehat{Y}=y|S=0] = 
\mathbb{P}[\widehat{Y}=y|S=1],~\forall y\in\{0,1\}~ -\text{ classification},
$$
or
$$
\mathbb{P}[\widehat{Y}\leq y|S=0] = 
\mathbb{P}[\widehat{Y}\leq y|S=1],~\forall y\in\mathbb{R}~ -\text{ regression}.
$$
For the later, an implication will be 
$
\mathbb{E}[\widehat{Y}|S=0] = 
\mathbb{E}[\widehat{Y}|S=1] 
$.
\end{definition}

The last characterization is equivalent to the two others if $y$ and $\widehat{y}$ are binary variables. In the case where $y$ is continuous, the second property corresponds to a notion of ``strong" demographic fairness while the last one corresponds to a notion of ``weak" demographic fairness (the second one implying the third one, but not the reverse).

This demographic fairness, also called ``statistical parity", simply requires that the fraction of blue applicants who are granted credit be approximately the same as the fraction of red applicants who are granted credit. By symmetry, the rejection proportions must be identical. Using the same threshold $\tau$ on the score, to grant credit, as in Figure \ref{fig:circle:square:1}, we see that statistical parity is not achieved:
$$
\mathbb{P}[\widehat{Y}=1|S=\text{\textcolor{tikbleu}{\scriptsize\faCircle}}]=\mathbb{P}[m(\boldsymbol{X})>\tau|S=\text{\textcolor{tikbleu}{\scriptsize\faCircle}}]=\frac{2}{8}=25\%\text{ et }\mathbb{P}[m(\boldsymbol{X})>\tau|S=\text{\textcolor{tikrouge}{\scriptsize\faStop}}]=\frac{12}{16}=75\%,
$$
so that
$$
\mathbb{P}[\widehat{Y}=1|S=\text{\textcolor{tikbleu}{\scriptsize\faCircle}}]\neq\mathbb{P}[\widehat{Y}=1|S=\text{\textcolor{tikrouge}{\scriptsize\faStop}}].
$$
Statistical parity is certainly a form of fairness, but it is generally weak and imperfect. And as the left side of Figure \ref{fig:circle:square:1:ROC} shows, it has nothing to do with the quality of the predictive model. 

\begin{figure}[!ht]
    \centering
     \includegraphics[width=.8\textwidth]{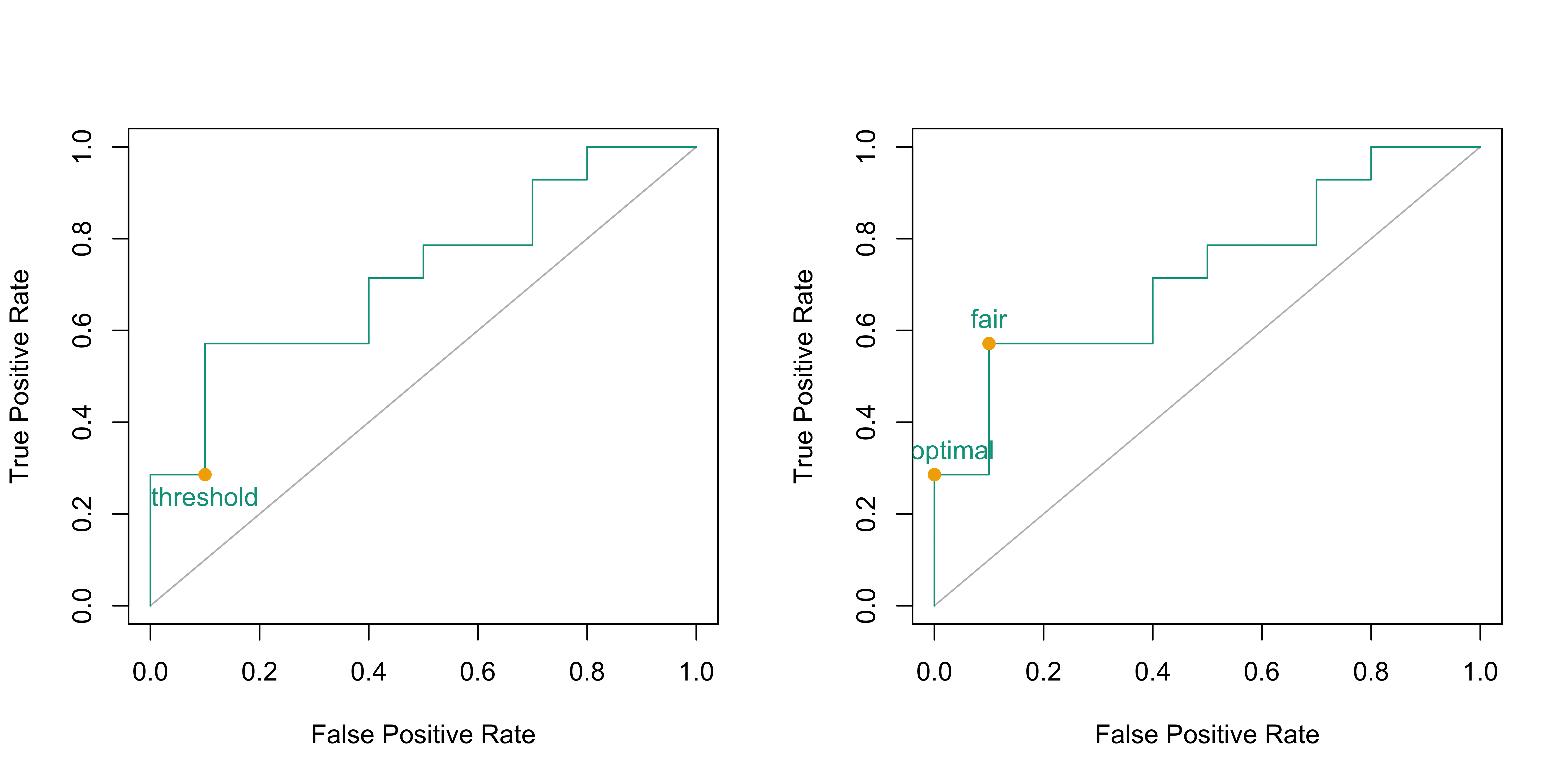}
    \caption{ROC curve on the example data in Figure \ref{fig:circle:square:1} with points in $\{\text{\textcolor{tikbleu}{\scriptsize\faCircle}},\text{\textcolor{tikrouge}{\scriptsize\faStop}}\}$, with two distinct levels for the threshold on the right, respectively the ``fair'' (Figure \ref{fig:circle:square:3}) and ``optimal'' (Figure \ref{fig:circle:square:5}).}
    \label{fig:circle:square:1:ROC}
\end{figure}

Suppose that
$$
\mathbb{P}[Y=1|S=\text{\textcolor{tikbleu}{\scriptsize\faCircle}}]=\frac{1}{4}=25\%\text{ et }\mathbb{P}[Y=1|S=\text{\textcolor{tikrouge}{\scriptsize\faStop}}]=\frac{3}{4}=75\%,
$$
and that the law of $Y$ depends {\em only} on $S$. In this case, imposing statistical parity means choosing the wrong model, because the perfect model would give
$$
\mathbb{P}[\widehat{Y}=1|S=\text{\textcolor{tikbleu}{\scriptsize\faCircle}}]=\frac{1}{4}=25\%\text{ et }\mathbb{P}[\widehat{Y}=1|S=\text{\textcolor{tikrouge}{\scriptsize\faStop}}]=\frac{3}{4}=75\%.
$$
On Figure \ref{fig:circle:square:3}, we can see the ``optimal'' threshold, in the sense of maximum predictive power, minimizing the rate of error committed, visible on Figure \ref{fig:circle:square:4}, on the left.

\begin{figure}[!ht]
    \centering
  \tikzset{every picture/.style={line width=0.75pt}} 

\begin{tikzpicture}[x=0.75pt,y=0.75pt,yscale=-1,xscale=1]

\draw    (27.37,40.1) -- (501.37,40.7) ;
\draw    (28.37,32.1) -- (28.37,49.1) ;
\draw    (501.37,33.1) -- (501.37,50.1) ;
\draw  [fill={rgb, 255:red, 74; green, 144; blue, 226 }  ,fill opacity=1 ] (61,40) .. controls (61,36.69) and (63.69,34) .. (67,34) .. controls (70.31,34) and (73,36.69) .. (73,40) .. controls (73,43.31) and (70.31,46) .. (67,46) .. controls (63.69,46) and (61,43.31) .. (61,40) -- cycle ;
\draw  [fill={rgb, 255:red, 74; green, 144; blue, 226 }  ,fill opacity=1 ] (88,40) .. controls (88,36.69) and (90.69,34) .. (94,34) .. controls (97.31,34) and (100,36.69) .. (100,40) .. controls (100,43.31) and (97.31,46) .. (94,46) .. controls (90.69,46) and (88,43.31) .. (88,40) -- cycle ;
\draw  [fill={rgb, 255:red, 74; green, 144; blue, 226 }  ,fill opacity=1 ] (106,40) .. controls (106,36.69) and (108.69,34) .. (112,34) .. controls (115.31,34) and (118,36.69) .. (118,40) .. controls (118,43.31) and (115.31,46) .. (112,46) .. controls (108.69,46) and (106,43.31) .. (106,40) -- cycle ;
\draw  [fill={rgb, 255:red, 74; green, 144; blue, 226 }  ,fill opacity=1 ] (125,40) .. controls (125,36.69) and (127.69,34) .. (131,34) .. controls (134.31,34) and (137,36.69) .. (137,40) .. controls (137,43.31) and (134.31,46) .. (131,46) .. controls (127.69,46) and (125,43.31) .. (125,40) -- cycle ;
\draw  [fill={rgb, 255:red, 74; green, 144; blue, 226 }  ,fill opacity=1 ] (143,40) .. controls (143,36.69) and (145.69,34) .. (149,34) .. controls (152.31,34) and (155,36.69) .. (155,40) .. controls (155,43.31) and (152.31,46) .. (149,46) .. controls (145.69,46) and (143,43.31) .. (143,40) -- cycle ;
\draw  [fill={rgb, 255:red, 74; green, 144; blue, 226 }  ,fill opacity=1 ] (162,40) .. controls (162,36.69) and (164.69,34) .. (168,34) .. controls (171.31,34) and (174,36.69) .. (174,40) .. controls (174,43.31) and (171.31,46) .. (168,46) .. controls (164.69,46) and (162,43.31) .. (162,40) -- cycle ;
\draw  [fill={rgb, 255:red, 74; green, 144; blue, 226 }  ,fill opacity=1 ] (251,40) .. controls (251,36.69) and (253.69,34) .. (257,34) .. controls (260.31,34) and (263,36.69) .. (263,40) .. controls (263,43.31) and (260.31,46) .. (257,46) .. controls (253.69,46) and (251,43.31) .. (251,40) -- cycle ;
\draw  [fill={rgb, 255:red, 74; green, 144; blue, 226 }  ,fill opacity=1 ] (269,40) .. controls (269,36.69) and (271.69,34) .. (275,34) .. controls (278.31,34) and (281,36.69) .. (281,40) .. controls (281,43.31) and (278.31,46) .. (275,46) .. controls (271.69,46) and (269,43.31) .. (269,40) -- cycle ;
\draw [color={rgb, 255:red, 65; green, 117; blue, 5 }  ,draw opacity=1 ]   (336.5,17) -- (337.5,87) ;
\draw  [fill={rgb, 255:red, 208; green, 2; blue, 27 }  ,fill opacity=1 ] (183,34) -- (194.5,34) -- (194.5,46) -- (183,46) -- cycle ;
\draw  [fill={rgb, 255:red, 208; green, 2; blue, 27 }  ,fill opacity=1 ] (200,34) -- (211.5,34) -- (211.5,46) -- (200,46) -- cycle ;
\draw  [fill={rgb, 255:red, 208; green, 2; blue, 27 }  ,fill opacity=1 ] (218,34) -- (229.5,34) -- (229.5,46) -- (218,46) -- cycle ;
\draw  [fill={rgb, 255:red, 208; green, 2; blue, 27 }  ,fill opacity=1 ] (234,34) -- (245.5,34) -- (245.5,46) -- (234,46) -- cycle ;
\draw  [fill={rgb, 255:red, 208; green, 2; blue, 27 }  ,fill opacity=1 ] (288,34) -- (299.5,34) -- (299.5,46) -- (288,46) -- cycle ;
\draw  [fill={rgb, 255:red, 208; green, 2; blue, 27 }  ,fill opacity=1 ] (305,34) -- (316.5,34) -- (316.5,46) -- (305,46) -- cycle ;
\draw  [fill={rgb, 255:red, 208; green, 2; blue, 27 }  ,fill opacity=1 ] (323,34) -- (334.5,34) -- (334.5,46) -- (323,46) -- cycle ;
\draw  [fill={rgb, 255:red, 208; green, 2; blue, 27 }  ,fill opacity=1 ] (339,34) -- (350.5,34) -- (350.5,46) -- (339,46) -- cycle ;
\draw  [fill={rgb, 255:red, 208; green, 2; blue, 27 }  ,fill opacity=1 ] (357,34) -- (368.5,34) -- (368.5,46) -- (357,46) -- cycle ;
\draw  [fill={rgb, 255:red, 208; green, 2; blue, 27 }  ,fill opacity=1 ] (372,34) -- (383.5,34) -- (383.5,46) -- (372,46) -- cycle ;
\draw  [fill={rgb, 255:red, 208; green, 2; blue, 27 }  ,fill opacity=1 ] (390.5,34) -- (402,34) -- (402,46) -- (390.5,46) -- cycle ;
\draw  [fill={rgb, 255:red, 208; green, 2; blue, 27 }  ,fill opacity=1 ] (405,34) -- (416.5,34) -- (416.5,46) -- (405,46) -- cycle ;
\draw  [fill={rgb, 255:red, 208; green, 2; blue, 27 }  ,fill opacity=1 ] (423,34) -- (434.5,34) -- (434.5,46) -- (423,46) -- cycle ;
\draw  [fill={rgb, 255:red, 208; green, 2; blue, 27 }  ,fill opacity=1 ] (439,34) -- (450.5,34) -- (450.5,46) -- (439,46) -- cycle ;
\draw  [fill={rgb, 255:red, 208; green, 2; blue, 27 }  ,fill opacity=1 ] (456,34) -- (467.5,34) -- (467.5,46) -- (456,46) -- cycle ;
\draw  [fill={rgb, 255:red, 208; green, 2; blue, 27 }  ,fill opacity=1 ] (474,34) -- (485.5,34) -- (485.5,46) -- (474,46) -- cycle ;

\draw (107,55) node [anchor=north west][inner sep=0.75pt]  [xscale=0.9,yscale=0.9] [align=left] {1};
\draw (145,55) node [anchor=north west][inner sep=0.75pt]  [xscale=0.9,yscale=0.9] [align=left] {1};
\draw (163,55) node [anchor=north west][inner sep=0.75pt]  [xscale=0.9,yscale=0.9] [align=left] {1};
\draw (218,55) node [anchor=north west][inner sep=0.75pt]  [xscale=0.9,yscale=0.9] [align=left] {1};
\draw (253,55) node [anchor=north west][inner sep=0.75pt]  [xscale=0.9,yscale=0.9] [align=left] {1};
\draw (271,55) node [anchor=north west][inner sep=0.75pt]  [xscale=0.9,yscale=0.9] [align=left] {1};
\draw (339,56) node [anchor=north west][inner sep=0.75pt]  [xscale=0.9,yscale=0.9] [align=left] {1};
\draw (356,56) node [anchor=north west][inner sep=0.75pt]  [xscale=0.9,yscale=0.9] [align=left] {1};
\draw (372,56) node [anchor=north west][inner sep=0.75pt]  [xscale=0.9,yscale=0.9] [align=left] {1};
\draw (391,56) node [anchor=north west][inner sep=0.75pt]  [xscale=0.9,yscale=0.9] [align=left] {1};
\draw (423,56) node [anchor=north west][inner sep=0.75pt]  [xscale=0.9,yscale=0.9] [align=left] {1};
\draw (440,56) node [anchor=north west][inner sep=0.75pt]  [xscale=0.9,yscale=0.9] [align=left] {1};
\draw (458,56) node [anchor=north west][inner sep=0.75pt]  [xscale=0.9,yscale=0.9] [align=left] {1};
\draw (475,56) node [anchor=north west][inner sep=0.75pt]  [xscale=0.9,yscale=0.9] [align=left] {1};
\draw (63,55) node [anchor=north west][inner sep=0.75pt]  [xscale=0.9,yscale=0.9] [align=left] {0};
\draw (88,55) node [anchor=north west][inner sep=0.75pt]  [xscale=0.9,yscale=0.9] [align=left] {0};
\draw (126,55) node [anchor=north west][inner sep=0.75pt]  [xscale=0.9,yscale=0.9] [align=left] {0};
\draw (183,56) node [anchor=north west][inner sep=0.75pt]  [xscale=0.9,yscale=0.9] [align=left] {0};
\draw (200,56) node [anchor=north west][inner sep=0.75pt]  [xscale=0.9,yscale=0.9] [align=left] {0};
\draw (235,55) node [anchor=north west][inner sep=0.75pt]  [xscale=0.9,yscale=0.9] [align=left] {0};
\draw (406,56) node [anchor=north west][inner sep=0.75pt]  [xscale=0.9,yscale=0.9] [align=left] {0};
\draw (289,55) node [anchor=north west][inner sep=0.75pt]  [xscale=0.9,yscale=0.9] [align=left] {0};
\draw (305,55) node [anchor=north west][inner sep=0.75pt]  [xscale=0.9,yscale=0.9] [align=left] {0};
\draw (323,55) node [anchor=north west][inner sep=0.75pt]  [xscale=0.9,yscale=0.9] [align=left] {0};
\draw (24,11) node [anchor=north west][inner sep=0.75pt]  [xscale=0.9,yscale=0.9] [align=left] {0};
\draw (497,12) node [anchor=north west][inner sep=0.75pt]  [xscale=0.9,yscale=0.9] [align=left] {1};
\draw (309,89) node [anchor=north west][inner sep=0.75pt]  [color={rgb, 255:red, 65; green, 117; blue, 5 }  ,opacity=1 ,xscale=0.9,yscale=0.9] [align=left] {\begin{minipage}[lt]{39.03pt}\setlength\topsep{0pt}
\begin{center}
optimal\\thresold
\end{center}

\end{minipage}};

\end{tikzpicture}
    \vspace{-.8cm}
    \caption{$s\in\{\text{\textcolor{tikbleu}{\scriptsize\faCircle}},\text{\textcolor{tikrouge}{\scriptsize\faStop}}\}$, $y\in\{0,1\}$, via \cite{kearns2019ethical}.}
    \label{fig:circle:square:3}
\end{figure}

\begin{figure}[!ht]
    \centering
     \includegraphics[width=\textwidth]{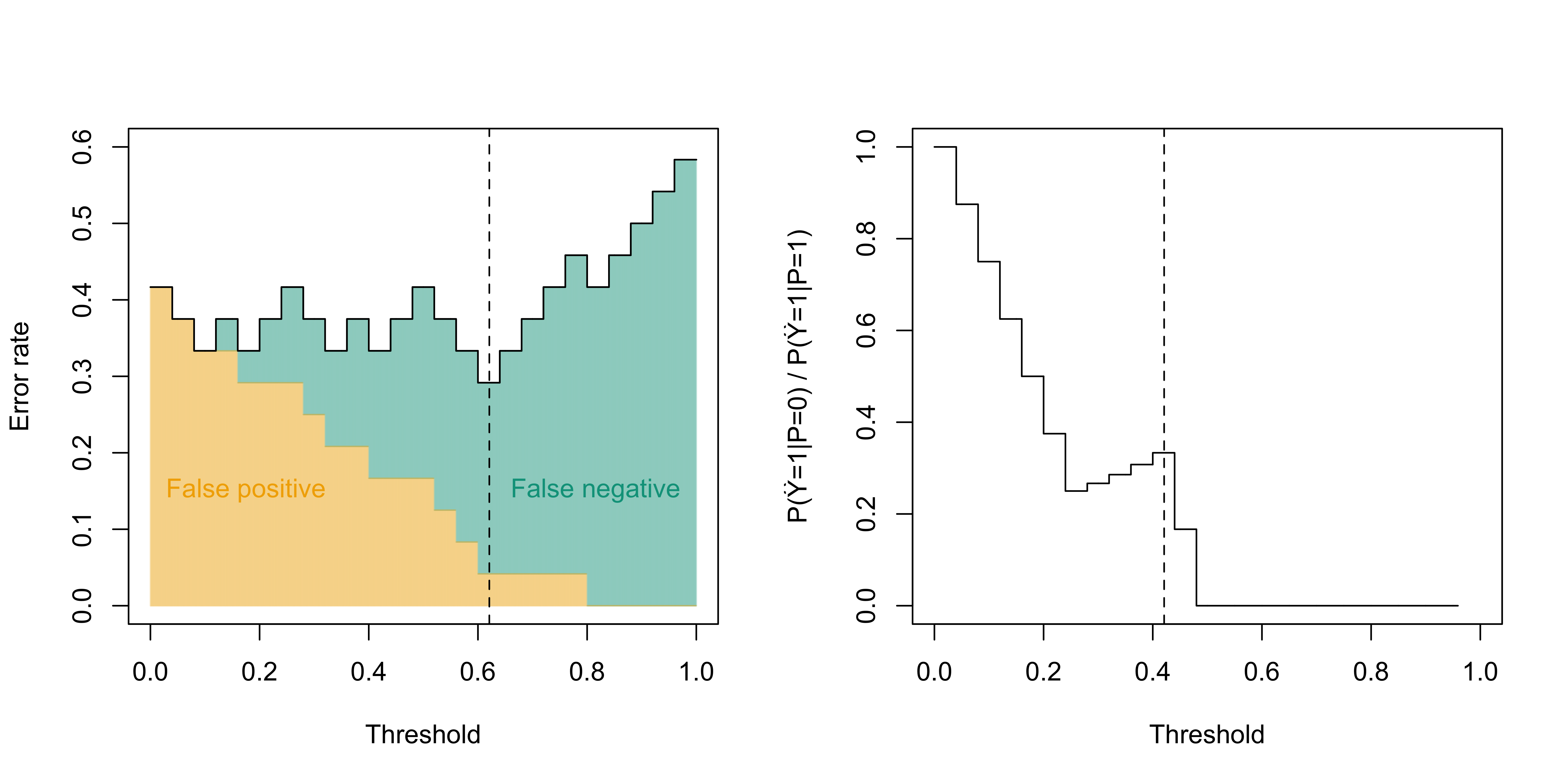}
    \caption{False-positive and false-negative rates, left, and evolution of the $\displaystyle{{{\mathbb{P}}[\widehat{Y}=1|S=0]}/{
\mathbb{P}[\widehat{Y}=1|S=1]}}$, as a function of the threshold level used.}
    \label{fig:circle:square:4}
\end{figure}

In Figure \ref{fig:circle:square:5}, we can see the ``fair" (or at least ``fairer") threshold, in the sense that $t\mapsto
\mathbb{P}[m(\boldsymbol{X})>t|S=\text{\textcolor{tikbleu}{\scriptsize\faCircle}}]/\mathbb{P}[m(\boldsymbol{X})>t|S=\text{\textcolor{tikrouge}{\scriptsize\faStop}}]$ is as large as possible (i.e. $
\mathbb{P}[\widehat{Y}=1|S=\text{\textcolor{tikbleu}{\scriptsize\faCircle}}]/\mathbb{P}[\widehat{Y}=1|S=\text{\textcolor{tikrouge}{\scriptsize\faStop}}]$ is as large as possible).

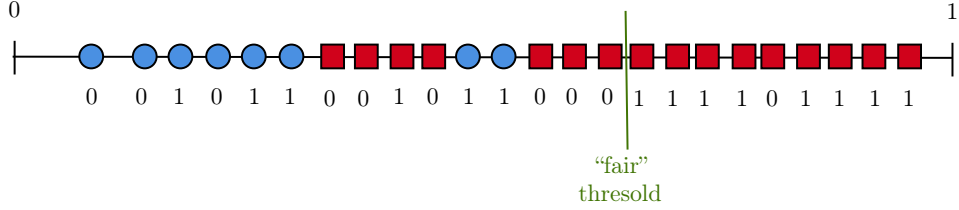
\begin{figure}[!ht]
    \centering
  \tikzset{every picture/.style={line width=0.75pt}} 

\begin{tikzpicture}[x=0.75pt,y=0.75pt,yscale=-1,xscale=1]

\draw    (27.37,40.1) -- (501.37,40.7) ;
\draw    (28.37,32.1) -- (28.37,49.1) ;
\draw    (501.37,33.1) -- (501.37,50.1) ;
\draw  [fill={rgb, 255:red, 74; green, 144; blue, 226 }  ,fill opacity=1 ] (61,40) .. controls (61,36.69) and (63.69,34) .. (67,34) .. controls (70.31,34) and (73,36.69) .. (73,40) .. controls (73,43.31) and (70.31,46) .. (67,46) .. controls (63.69,46) and (61,43.31) .. (61,40) -- cycle ;
\draw  [fill={rgb, 255:red, 74; green, 144; blue, 226 }  ,fill opacity=1 ] (88,40) .. controls (88,36.69) and (90.69,34) .. (94,34) .. controls (97.31,34) and (100,36.69) .. (100,40) .. controls (100,43.31) and (97.31,46) .. (94,46) .. controls (90.69,46) and (88,43.31) .. (88,40) -- cycle ;
\draw  [fill={rgb, 255:red, 74; green, 144; blue, 226 }  ,fill opacity=1 ] (106,40) .. controls (106,36.69) and (108.69,34) .. (112,34) .. controls (115.31,34) and (118,36.69) .. (118,40) .. controls (118,43.31) and (115.31,46) .. (112,46) .. controls (108.69,46) and (106,43.31) .. (106,40) -- cycle ;
\draw  [fill={rgb, 255:red, 74; green, 144; blue, 226 }  ,fill opacity=1 ] (125,40) .. controls (125,36.69) and (127.69,34) .. (131,34) .. controls (134.31,34) and (137,36.69) .. (137,40) .. controls (137,43.31) and (134.31,46) .. (131,46) .. controls (127.69,46) and (125,43.31) .. (125,40) -- cycle ;
\draw  [fill={rgb, 255:red, 74; green, 144; blue, 226 }  ,fill opacity=1 ] (143,40) .. controls (143,36.69) and (145.69,34) .. (149,34) .. controls (152.31,34) and (155,36.69) .. (155,40) .. controls (155,43.31) and (152.31,46) .. (149,46) .. controls (145.69,46) and (143,43.31) .. (143,40) -- cycle ;
\draw  [fill={rgb, 255:red, 74; green, 144; blue, 226 }  ,fill opacity=1 ] (162,40) .. controls (162,36.69) and (164.69,34) .. (168,34) .. controls (171.31,34) and (174,36.69) .. (174,40) .. controls (174,43.31) and (171.31,46) .. (168,46) .. controls (164.69,46) and (162,43.31) .. (162,40) -- cycle ;
\draw  [fill={rgb, 255:red, 74; green, 144; blue, 226 }  ,fill opacity=1 ] (251,40) .. controls (251,36.69) and (253.69,34) .. (257,34) .. controls (260.31,34) and (263,36.69) .. (263,40) .. controls (263,43.31) and (260.31,46) .. (257,46) .. controls (253.69,46) and (251,43.31) .. (251,40) -- cycle ;
\draw  [fill={rgb, 255:red, 74; green, 144; blue, 226 }  ,fill opacity=1 ] (269,40) .. controls (269,36.69) and (271.69,34) .. (275,34) .. controls (278.31,34) and (281,36.69) .. (281,40) .. controls (281,43.31) and (278.31,46) .. (275,46) .. controls (271.69,46) and (269,43.31) .. (269,40) -- cycle ;
\draw [color={rgb, 255:red, 65; green, 117; blue, 5 }  ,draw opacity=1 ]   (336.5,17) -- (337.5,87) ;
\draw  [fill={rgb, 255:red, 208; green, 2; blue, 27 }  ,fill opacity=1 ] (183,34) -- (194.5,34) -- (194.5,46) -- (183,46) -- cycle ;
\draw  [fill={rgb, 255:red, 208; green, 2; blue, 27 }  ,fill opacity=1 ] (200,34) -- (211.5,34) -- (211.5,46) -- (200,46) -- cycle ;
\draw  [fill={rgb, 255:red, 208; green, 2; blue, 27 }  ,fill opacity=1 ] (218,34) -- (229.5,34) -- (229.5,46) -- (218,46) -- cycle ;
\draw  [fill={rgb, 255:red, 208; green, 2; blue, 27 }  ,fill opacity=1 ] (234,34) -- (245.5,34) -- (245.5,46) -- (234,46) -- cycle ;
\draw  [fill={rgb, 255:red, 208; green, 2; blue, 27 }  ,fill opacity=1 ] (288,34) -- (299.5,34) -- (299.5,46) -- (288,46) -- cycle ;
\draw  [fill={rgb, 255:red, 208; green, 2; blue, 27 }  ,fill opacity=1 ] (305,34) -- (316.5,34) -- (316.5,46) -- (305,46) -- cycle ;
\draw  [fill={rgb, 255:red, 208; green, 2; blue, 27 }  ,fill opacity=1 ] (323,34) -- (334.5,34) -- (334.5,46) -- (323,46) -- cycle ;
\draw  [fill={rgb, 255:red, 208; green, 2; blue, 27 }  ,fill opacity=1 ] (339,34) -- (350.5,34) -- (350.5,46) -- (339,46) -- cycle ;
\draw  [fill={rgb, 255:red, 208; green, 2; blue, 27 }  ,fill opacity=1 ] (357,34) -- (368.5,34) -- (368.5,46) -- (357,46) -- cycle ;
\draw  [fill={rgb, 255:red, 208; green, 2; blue, 27 }  ,fill opacity=1 ] (372,34) -- (383.5,34) -- (383.5,46) -- (372,46) -- cycle ;
\draw  [fill={rgb, 255:red, 208; green, 2; blue, 27 }  ,fill opacity=1 ] (390.5,34) -- (402,34) -- (402,46) -- (390.5,46) -- cycle ;
\draw  [fill={rgb, 255:red, 208; green, 2; blue, 27 }  ,fill opacity=1 ] (405,34) -- (416.5,34) -- (416.5,46) -- (405,46) -- cycle ;
\draw  [fill={rgb, 255:red, 208; green, 2; blue, 27 }  ,fill opacity=1 ] (423,34) -- (434.5,34) -- (434.5,46) -- (423,46) -- cycle ;
\draw  [fill={rgb, 255:red, 208; green, 2; blue, 27 }  ,fill opacity=1 ] (439,34) -- (450.5,34) -- (450.5,46) -- (439,46) -- cycle ;
\draw  [fill={rgb, 255:red, 208; green, 2; blue, 27 }  ,fill opacity=1 ] (456,34) -- (467.5,34) -- (467.5,46) -- (456,46) -- cycle ;
\draw  [fill={rgb, 255:red, 208; green, 2; blue, 27 }  ,fill opacity=1 ] (474,34) -- (485.5,34) -- (485.5,46) -- (474,46) -- cycle ;

\draw (107,55) node [anchor=north west][inner sep=0.75pt]  [xscale=0.9,yscale=0.9] [align=left] {1};
\draw (145,55) node [anchor=north west][inner sep=0.75pt]  [xscale=0.9,yscale=0.9] [align=left] {1};
\draw (163,55) node [anchor=north west][inner sep=0.75pt]  [xscale=0.9,yscale=0.9] [align=left] {1};
\draw (218,55) node [anchor=north west][inner sep=0.75pt]  [xscale=0.9,yscale=0.9] [align=left] {1};
\draw (253,55) node [anchor=north west][inner sep=0.75pt]  [xscale=0.9,yscale=0.9] [align=left] {1};
\draw (271,55) node [anchor=north west][inner sep=0.75pt]  [xscale=0.9,yscale=0.9] [align=left] {1};
\draw (339,56) node [anchor=north west][inner sep=0.75pt]  [xscale=0.9,yscale=0.9] [align=left] {1};
\draw (356,56) node [anchor=north west][inner sep=0.75pt]  [xscale=0.9,yscale=0.9] [align=left] {1};
\draw (372,56) node [anchor=north west][inner sep=0.75pt]  [xscale=0.9,yscale=0.9] [align=left] {1};
\draw (391,56) node [anchor=north west][inner sep=0.75pt]  [xscale=0.9,yscale=0.9] [align=left] {1};
\draw (423,56) node [anchor=north west][inner sep=0.75pt]  [xscale=0.9,yscale=0.9] [align=left] {1};
\draw (440,56) node [anchor=north west][inner sep=0.75pt]  [xscale=0.9,yscale=0.9] [align=left] {1};
\draw (458,56) node [anchor=north west][inner sep=0.75pt]  [xscale=0.9,yscale=0.9] [align=left] {1};
\draw (475,56) node [anchor=north west][inner sep=0.75pt]  [xscale=0.9,yscale=0.9] [align=left] {1};
\draw (63,55) node [anchor=north west][inner sep=0.75pt]  [xscale=0.9,yscale=0.9] [align=left] {0};
\draw (88,55) node [anchor=north west][inner sep=0.75pt]  [xscale=0.9,yscale=0.9] [align=left] {0};
\draw (126,55) node [anchor=north west][inner sep=0.75pt]  [xscale=0.9,yscale=0.9] [align=left] {0};
\draw (183,56) node [anchor=north west][inner sep=0.75pt]  [xscale=0.9,yscale=0.9] [align=left] {0};
\draw (200,56) node [anchor=north west][inner sep=0.75pt]  [xscale=0.9,yscale=0.9] [align=left] {0};
\draw (235,55) node [anchor=north west][inner sep=0.75pt]  [xscale=0.9,yscale=0.9] [align=left] {0};
\draw (406,56) node [anchor=north west][inner sep=0.75pt]  [xscale=0.9,yscale=0.9] [align=left] {0};
\draw (289,55) node [anchor=north west][inner sep=0.75pt]  [xscale=0.9,yscale=0.9] [align=left] {0};
\draw (305,55) node [anchor=north west][inner sep=0.75pt]  [xscale=0.9,yscale=0.9] [align=left] {0};
\draw (323,55) node [anchor=north west][inner sep=0.75pt]  [xscale=0.9,yscale=0.9] [align=left] {0};
\draw (24,11) node [anchor=north west][inner sep=0.75pt]  [xscale=0.9,yscale=0.9] [align=left] {0};
\draw (497,12) node [anchor=north west][inner sep=0.75pt]  [xscale=0.9,yscale=0.9] [align=left] {1};
\draw (309,89) node [anchor=north west][inner sep=0.75pt]  [color={rgb, 255:red, 65; green, 117; blue, 5 }  ,opacity=1 ,xscale=0.9,yscale=0.9] [align=left] {\begin{minipage}[lt]{39.03pt}\setlength\topsep{0pt}
\begin{center}
``fair''\\thresold
\end{center}

\end{minipage}};

\end{tikzpicture}
    \vspace{-.8cm}\caption{Choosing a ``fair" fairness threshold, via \cite{kearns2019ethical}.}
    \label{fig:circle:square:5}
\end{figure}

To summarize, minimizing the error rate (and therefore increasing accuracy) and maximizing fairness are often two irreconcilable goals.
$$
\begin{cases}
\text{optimal threshold~}&:~\mathbb{P}[\widehat{Y}\neq Y]=\displaystyle{\frac{6+1}{24}=29. 17\%}\text{ and }\displaystyle{\frac{\mathbb{P}[\widehat{Y}=1|S=\text{\textcolor{tikbleu}{\scriptsize\faCircle}}]}{\mathbb{P}[\widehat{Y}=1|S=\text{\textcolor{tikrouge}{\scriptsize\faStop}}]}=\frac{0}{24}=0\%}\\
\text{``fair'' threshold~}&:~\mathbb{P}[\widehat{Y}\neq Y]=\displaystyle{\frac{4+4}{24}=33. 33\%}\text{ and }\displaystyle{\frac{\mathbb{P}[\widehat{Y}=1|S=\text{\textcolor{tikbleu}{\scriptsize\faCircle}}]}{\mathbb{P}[\widehat{Y}=1|S=\text{\textcolor{tikrouge}{\scriptsize\faStop}}]}=\frac{2\cdot16}{12\cdot4}=33.33\%}\\
\end{cases}
$$
We will return later to the trade-off that will often exist between the fairness of the models, and their accuracy (or predictive power), and the associated efficiency frontier.

Another shortcoming of this approach is that the desired independence between the sensitive variable $s$ and the prediction $\widehat{y}$ does not take into account the fact that the outcome $y$ may be correlated with the sensitive variable $s$. In other words, if the groups induced by $s$ have different underlying distributions for $y$, ignoring these dependencies may lead to results that would be considered fair, but not for the groups themselves. Quite naturally, an extension of the independence property is the ``separation" criterion which requires independence between the prediction $\widehat{y}$ and the sensitive variable $S$, conditional on the value of the target variable $y$, i.e. $\widehat{Y}\indep S$ conditional on $Y$.

This approach can amount to choosing a different threshold, with a lower threshold for blue individuals $\text{\textcolor{tikbleu}{\scriptsize\faCircle}}$ than for red individuals $\text{\textcolor{tikrouge}{\scriptsize\faStop}}$, as in Figure \ref{fig:circle:square:2}. In this case, 
$$
\mathbb{P}[\widehat{Y}=1|S=\text{\textcolor{tikbleu}{\scriptsize\faCircle}}]=\mathbb{P}[m(\boldsymbol{X})>{\textcolor{tikbleu}{\tau}}|S=\text{\textcolor{tikbleu}{\scriptsize\faCircle}}]=50\%=\mathbb{P}[\widehat{Y}=1|S=\text{\textcolor{tikrouge}{\scriptsize\faStop}}]=\mathbb{P}[m(\boldsymbol{X})>{\textcolor{tikrouge}{\tau}}|S=\text{\textcolor{tikrouge}{\scriptsize\faStop}}].
$$

\begin{figure}[!ht]
    \centering
  \tikzset{every picture/.style={line width=0.75pt}} 

\begin{tikzpicture}[x=0.75pt,y=0.75pt,yscale=-1,xscale=1]

\draw    (27.37,40.1) -- (501.37,40.7) ;
\draw    (28.37,32.1) -- (28.37,49.1) ;
\draw    (501.37,33.1) -- (501.37,50.1) ;
\draw  [fill={rgb, 255:red, 74; green, 144; blue, 226 }  ,fill opacity=1 ] (61,40) .. controls (61,36.69) and (63.69,34) .. (67,34) .. controls (70.31,34) and (73,36.69) .. (73,40) .. controls (73,43.31) and (70.31,46) .. (67,46) .. controls (63.69,46) and (61,43.31) .. (61,40) -- cycle ;
\draw  [fill={rgb, 255:red, 74; green, 144; blue, 226 }  ,fill opacity=1 ] (88,40) .. controls (88,36.69) and (90.69,34) .. (94,34) .. controls (97.31,34) and (100,36.69) .. (100,40) .. controls (100,43.31) and (97.31,46) .. (94,46) .. controls (90.69,46) and (88,43.31) .. (88,40) -- cycle ;
\draw  [fill={rgb, 255:red, 74; green, 144; blue, 226 }  ,fill opacity=1 ] (106,40) .. controls (106,36.69) and (108.69,34) .. (112,34) .. controls (115.31,34) and (118,36.69) .. (118,40) .. controls (118,43.31) and (115.31,46) .. (112,46) .. controls (108.69,46) and (106,43.31) .. (106,40) -- cycle ;
\draw  [fill={rgb, 255:red, 74; green, 144; blue, 226 }  ,fill opacity=1 ] (125,40) .. controls (125,36.69) and (127.69,34) .. (131,34) .. controls (134.31,34) and (137,36.69) .. (137,40) .. controls (137,43.31) and (134.31,46) .. (131,46) .. controls (127.69,46) and (125,43.31) .. (125,40) -- cycle ;
\draw  [fill={rgb, 255:red, 74; green, 144; blue, 226 }  ,fill opacity=1 ] (143,40) .. controls (143,36.69) and (145.69,34) .. (149,34) .. controls (152.31,34) and (155,36.69) .. (155,40) .. controls (155,43.31) and (152.31,46) .. (149,46) .. controls (145.69,46) and (143,43.31) .. (143,40) -- cycle ;
\draw  [fill={rgb, 255:red, 74; green, 144; blue, 226 }  ,fill opacity=1 ] (162,40) .. controls (162,36.69) and (164.69,34) .. (168,34) .. controls (171.31,34) and (174,36.69) .. (174,40) .. controls (174,43.31) and (171.31,46) .. (168,46) .. controls (164.69,46) and (162,43.31) .. (162,40) -- cycle ;
\draw  [fill={rgb, 255:red, 74; green, 144; blue, 226 }  ,fill opacity=1 ] (251,40) .. controls (251,36.69) and (253.69,34) .. (257,34) .. controls (260.31,34) and (263,36.69) .. (263,40) .. controls (263,43.31) and (260.31,46) .. (257,46) .. controls (253.69,46) and (251,43.31) .. (251,40) -- cycle ;
\draw  [fill={rgb, 255:red, 74; green, 144; blue, 226 }  ,fill opacity=1 ] (269,40) .. controls (269,36.69) and (271.69,34) .. (275,34) .. controls (278.31,34) and (281,36.69) .. (281,40) .. controls (281,43.31) and (278.31,46) .. (275,46) .. controls (271.69,46) and (269,43.31) .. (269,40) -- cycle ;
\draw [color={rgb, 255:red, 65; green, 117; blue, 5 }  ,draw opacity=1 ]   (353.5,17) -- (354.5,87) ;
\draw  [fill={rgb, 255:red, 208; green, 2; blue, 27 }  ,fill opacity=1 ] (183,34) -- (194.5,34) -- (194.5,46) -- (183,46) -- cycle ;
\draw  [fill={rgb, 255:red, 208; green, 2; blue, 27 }  ,fill opacity=1 ] (200,34) -- (211.5,34) -- (211.5,46) -- (200,46) -- cycle ;
\draw  [fill={rgb, 255:red, 208; green, 2; blue, 27 }  ,fill opacity=1 ] (218,34) -- (229.5,34) -- (229.5,46) -- (218,46) -- cycle ;
\draw  [fill={rgb, 255:red, 208; green, 2; blue, 27 }  ,fill opacity=1 ] (234,34) -- (245.5,34) -- (245.5,46) -- (234,46) -- cycle ;
\draw  [fill={rgb, 255:red, 208; green, 2; blue, 27 }  ,fill opacity=1 ] (288,34) -- (299.5,34) -- (299.5,46) -- (288,46) -- cycle ;
\draw  [fill={rgb, 255:red, 208; green, 2; blue, 27 }  ,fill opacity=1 ] (305,34) -- (316.5,34) -- (316.5,46) -- (305,46) -- cycle ;
\draw  [fill={rgb, 255:red, 208; green, 2; blue, 27 }  ,fill opacity=1 ] (323,34) -- (334.5,34) -- (334.5,46) -- (323,46) -- cycle ;
\draw  [fill={rgb, 255:red, 208; green, 2; blue, 27 }  ,fill opacity=1 ] (339,34) -- (350.5,34) -- (350.5,46) -- (339,46) -- cycle ;
\draw  [fill={rgb, 255:red, 208; green, 2; blue, 27 }  ,fill opacity=1 ] (357,34) -- (368.5,34) -- (368.5,46) -- (357,46) -- cycle ;
\draw  [fill={rgb, 255:red, 208; green, 2; blue, 27 }  ,fill opacity=1 ] (372,34) -- (383.5,34) -- (383.5,46) -- (372,46) -- cycle ;
\draw  [fill={rgb, 255:red, 208; green, 2; blue, 27 }  ,fill opacity=1 ] (390.5,34) -- (402,34) -- (402,46) -- (390.5,46) -- cycle ;
\draw  [fill={rgb, 255:red, 208; green, 2; blue, 27 }  ,fill opacity=1 ] (405,34) -- (416.5,34) -- (416.5,46) -- (405,46) -- cycle ;
\draw  [fill={rgb, 255:red, 208; green, 2; blue, 27 }  ,fill opacity=1 ] (423,34) -- (434.5,34) -- (434.5,46) -- (423,46) -- cycle ;
\draw  [fill={rgb, 255:red, 208; green, 2; blue, 27 }  ,fill opacity=1 ] (439,34) -- (450.5,34) -- (450.5,46) -- (439,46) -- cycle ;
\draw  [fill={rgb, 255:red, 208; green, 2; blue, 27 }  ,fill opacity=1 ] (456,34) -- (467.5,34) -- (467.5,46) -- (456,46) -- cycle ;
\draw  [fill={rgb, 255:red, 208; green, 2; blue, 27 }  ,fill opacity=1 ] (474,34) -- (485.5,34) -- (485.5,46) -- (474,46) -- cycle ;
\draw [color={rgb, 255:red, 65; green, 117; blue, 5 }  ,draw opacity=1 ]   (140.5,18) -- (141.5,88) ;

\draw (107,55) node [anchor=north west][inner sep=0.75pt]  [xscale=0.9,yscale=0.9] [align=left] {1};
\draw (145,55) node [anchor=north west][inner sep=0.75pt]  [xscale=0.9,yscale=0.9] [align=left] {1};
\draw (163,55) node [anchor=north west][inner sep=0.75pt]  [xscale=0.9,yscale=0.9] [align=left] {1};
\draw (218,55) node [anchor=north west][inner sep=0.75pt]  [xscale=0.9,yscale=0.9] [align=left] {1};
\draw (253,55) node [anchor=north west][inner sep=0.75pt]  [xscale=0.9,yscale=0.9] [align=left] {1};
\draw (271,55) node [anchor=north west][inner sep=0.75pt]  [xscale=0.9,yscale=0.9] [align=left] {1};
\draw (339,56) node [anchor=north west][inner sep=0.75pt]  [xscale=0.9,yscale=0.9] [align=left] {1};
\draw (356,56) node [anchor=north west][inner sep=0.75pt]  [xscale=0.9,yscale=0.9] [align=left] {1};
\draw (372,56) node [anchor=north west][inner sep=0.75pt]  [xscale=0.9,yscale=0.9] [align=left] {1};
\draw (391,56) node [anchor=north west][inner sep=0.75pt]  [xscale=0.9,yscale=0.9] [align=left] {1};
\draw (423,56) node [anchor=north west][inner sep=0.75pt]  [xscale=0.9,yscale=0.9] [align=left] {1};
\draw (440,56) node [anchor=north west][inner sep=0.75pt]  [xscale=0.9,yscale=0.9] [align=left] {1};
\draw (458,56) node [anchor=north west][inner sep=0.75pt]  [xscale=0.9,yscale=0.9] [align=left] {1};
\draw (475,56) node [anchor=north west][inner sep=0.75pt]  [xscale=0.9,yscale=0.9] [align=left] {1};
\draw (63,55) node [anchor=north west][inner sep=0.75pt]  [xscale=0.9,yscale=0.9] [align=left] {0};
\draw (88,55) node [anchor=north west][inner sep=0.75pt]  [xscale=0.9,yscale=0.9] [align=left] {0};
\draw (126,55) node [anchor=north west][inner sep=0.75pt]  [xscale=0.9,yscale=0.9] [align=left] {0};
\draw (183,56) node [anchor=north west][inner sep=0.75pt]  [xscale=0.9,yscale=0.9] [align=left] {0};
\draw (200,56) node [anchor=north west][inner sep=0.75pt]  [xscale=0.9,yscale=0.9] [align=left] {0};
\draw (235,55) node [anchor=north west][inner sep=0.75pt]  [xscale=0.9,yscale=0.9] [align=left] {0};
\draw (406,56) node [anchor=north west][inner sep=0.75pt]  [xscale=0.9,yscale=0.9] [align=left] {0};
\draw (289,55) node [anchor=north west][inner sep=0.75pt]  [xscale=0.9,yscale=0.9] [align=left] {0};
\draw (305,55) node [anchor=north west][inner sep=0.75pt]  [xscale=0.9,yscale=0.9] [align=left] {0};
\draw (323,55) node [anchor=north west][inner sep=0.75pt]  [xscale=0.9,yscale=0.9] [align=left] {0};
\draw (24,11) node [anchor=north west][inner sep=0.75pt]  [xscale=0.9,yscale=0.9] [align=left] {0};
\draw (497,12) node [anchor=north west][inner sep=0.75pt]  [xscale=0.9,yscale=0.9] [align=left] {1};
\draw (326,90) node [anchor=north west][inner sep=0.75pt]  [color={rgb, 255:red, 65; green, 117; blue, 5 }  ,opacity=1 ,xscale=0.9,yscale=0.9] [align=left] {\begin{minipage}[lt]{39.03pt}\setlength\topsep{0pt}
\begin{center}
\textcolor[rgb]{0.82,0.01,0.11}{thresold}
\end{center}

\end{minipage}};
\draw (114,90) node [anchor=north west][inner sep=0.75pt]  [color={rgb, 255:red, 65; green, 117; blue, 5 }  ,opacity=1 ,xscale=0.9,yscale=0.9] [align=left] {\begin{minipage}[lt]{39.03pt}\setlength\topsep{0pt}
\begin{center}
\textcolor[rgb]{0.29,0.56,0.89}{thresold}
\end{center}

\end{minipage}};

\end{tikzpicture}
    \vspace{-.8cm}
    \caption{$s\in\{\text{\textcolor{tikbleu}{\scriptsize\faCircle}},\text{\textcolor{tikrouge}{\scriptsize\faStop}}\}$, $y\in\{0,1\}$, via \cite{kearns2019ethical}.}
    \label{fig:circle:square:2}
\end{figure}

In Figure \ref{fig:circle:square:6}, we can visualize the rate of false positives in each class, as a function of threshold, on the left, and the rate of true positives on the right. This last case is called ``equality of opportunity'' by \cite{hardt2016equality}.

\begin{figure}[!ht]
    \centering
     \includegraphics[width=\textwidth]{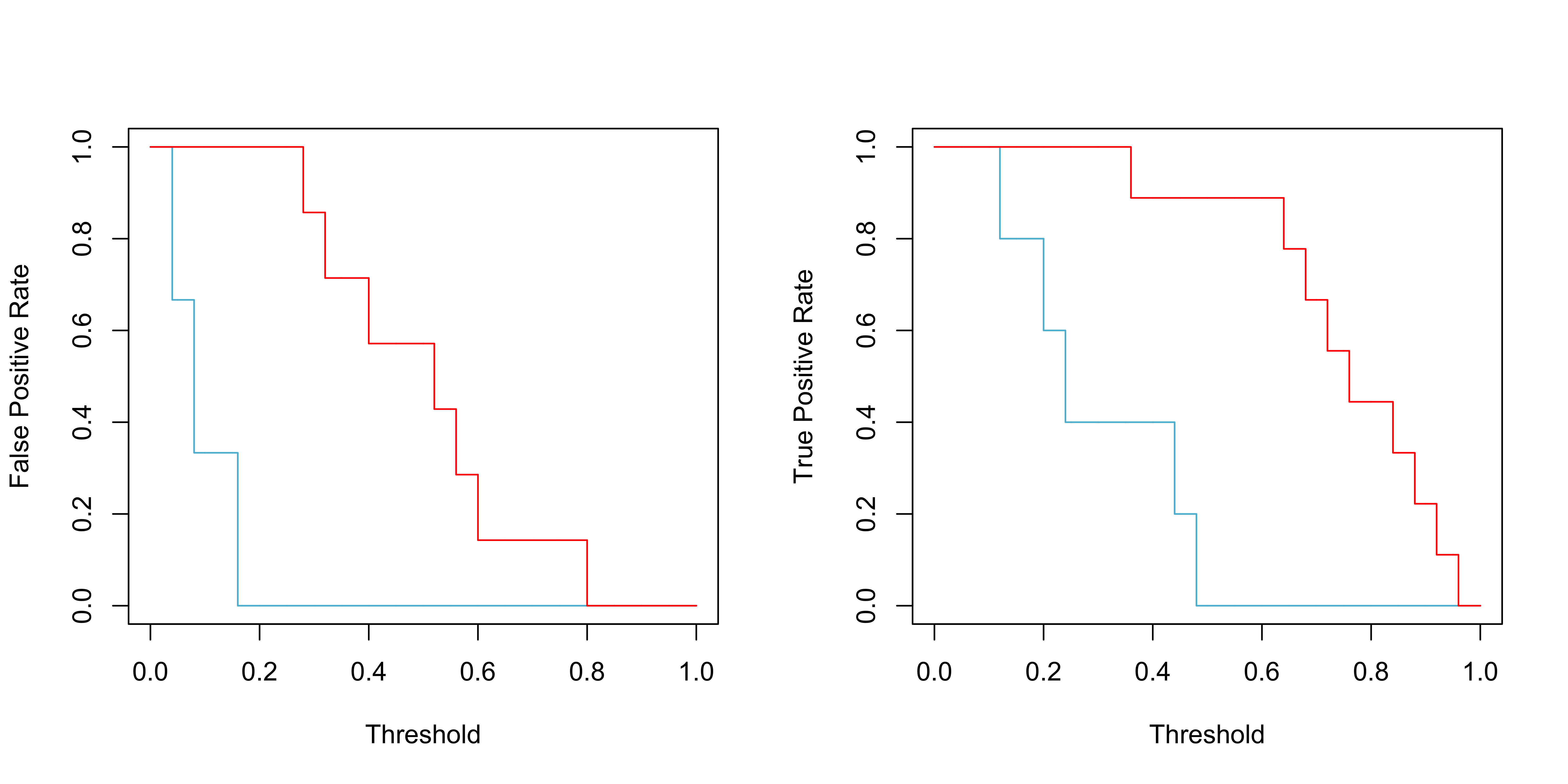}
    \caption{False positive rate, left, and true positive rate, right.}
    \label{fig:circle:square:6}
\end{figure}

Demographic parity (or statistical parity) suggests that a predictor is unbiased if the prediction $\widehat{y}$ is independent of the sensitive attribute $s$, 
\begin{equation*}
    \widehat{Y} \indep S
\end{equation*}
so that
$$
\begin{cases}
\mathbb{P}(\widehat{Y}=1|S=0)=\mathbb{P}(\widehat{Y}=1|S=1)=\mathbb{P}(\widehat{Y}=1),\\
\mathbb{P}(m(\boldsymbol{X})\leq \mu|S=0)=\mathbb{P}(\widehat{Y}\leq \gamma|S=1)=\mathbb{P}(m(\boldsymbol{X})\leq \mu),~\forall \mu\in[0,1],\
\end{cases}
$$
We will also speak here of ``group fairness". 
Note that in the first case, the condition is equivalent to having 
$$
\mathbb{E}(\widehat{Y}|S=0)=
\mathbb{E}(\widehat{Y}|S=1)=
\mathbb{E}(\widehat{Y})
$$
but not in the second, when $y$ is continuous. The latter will be true, but it will not be enough.
Here, the same proportion of each population is classified as positive. However, this may result in different false positive and true positive rates if the true outcome $y$ does indeed vary with the sensitive attribute $s$.

An alternative to the independence assumption $\widehat{Y} \indep S$ is to require that $\widehat{Y}$ and $S$ have zero mutual information,
$$
IM(\widehat{Y},S)=\mathcal{E}(\widehat{Y})+\mathcal{E}(S) - \mathcal{E}(\widehat{Y},S)=0,
$$
where $\mathcal{E}$ denotes the entropy, that is
$$
IM(\widehat{Y},S)=\displaystyle{ \sum_{\widehat{y},s}\mathbb{P}(\widehat{y},s)\log {\frac{\mathbb{P}(\widehat{y},s)}{\mathbb{P}(\widehat{y}})}},
$$
For example, if we want to maintain a threshold at 60\% for the advantaged population (curve \textcolor{colrwarouge}{red} on the left, $s=0$), we must slightly lower the threshold for the disadvantaged population (curve \textcolor{colrwableu}{blue} on the right, $s=1$), with here a threshold slightly lower than 50\%. In other words, with such a choice $\mathbb{P}(\widehat{Y}=1|S=1)=\mathbb{P}(\widehat{Y}=1|S=0)$.
The disadvantage of this method is that the true and false positive rates may be completely different in the two subpopulations. 

\subsection{Equalized odds (and other related concepts)}\label{subsec:equal:opp}

Equal opportunity and equality of opportunity are not so much a measure of fairness as a potential definition of fairness. Equal opportunity is achieved when the predicted target variable of a $\widehat{y}$ model and the label of a sensitive category $s$ are statistically independent of each other, conditional on the actual value of the target variable $y$. In a binary classification task, this can be simplified by requiring that true positive rates and false positive rates be equal between groups, where the groups are determined by the protected category. A slightly less demanding fairness criterion is equal opportunity, in which only the probability of the true positive is equalized across groups in a protected category.
Formally, we have the following definitions, where we require parity of false or true positives (Figures \ref{fig:ROC:4:0} and \ref{fig:ROC:4:1} respectively).

\begin{definition}[True positive equality, Equalized Odds, \cite{hardt2016equality}] We will speak of equality of opportunity, or parity of true positives, if
$$
\mathbb{P}[\widehat{Y}=1|S=0,Y=1] = 
\mathbb{P}[\widehat{Y}=1|S=1,Y=1]
$$
or equivalently
$$
\text{TPR}_0=\frac{\text{TP}_0}{\text{FN}_0 + \text{TP}_0}=
\frac{\text{TP}_1}{\text{FN}_1 + \text{TP}_1}=\text{TPR}_1.
$$
\end{definition}

\begin{definition}[False positive equality, \cite{hardt2016equality}] We will speak of equality of false positives if
$$
\mathbb{P}[\widehat{Y}=1|S=0,Y=0] = 
\mathbb{P}[\widehat{Y}=1|S=1,Y=0],
$$
or equivalently
$$
\text{FPR}_0=\frac{\text{FP}_0}{\text{TN}_0 + \text{FP}_0}=
\frac{\text{FP}_1}{\text{TN}_1 + \text{FP}_1}=\text{FPR}_1.
$$
\end{definition}

\begin{figure}[!ht]
    \centering
     \includegraphics[width=\textwidth]{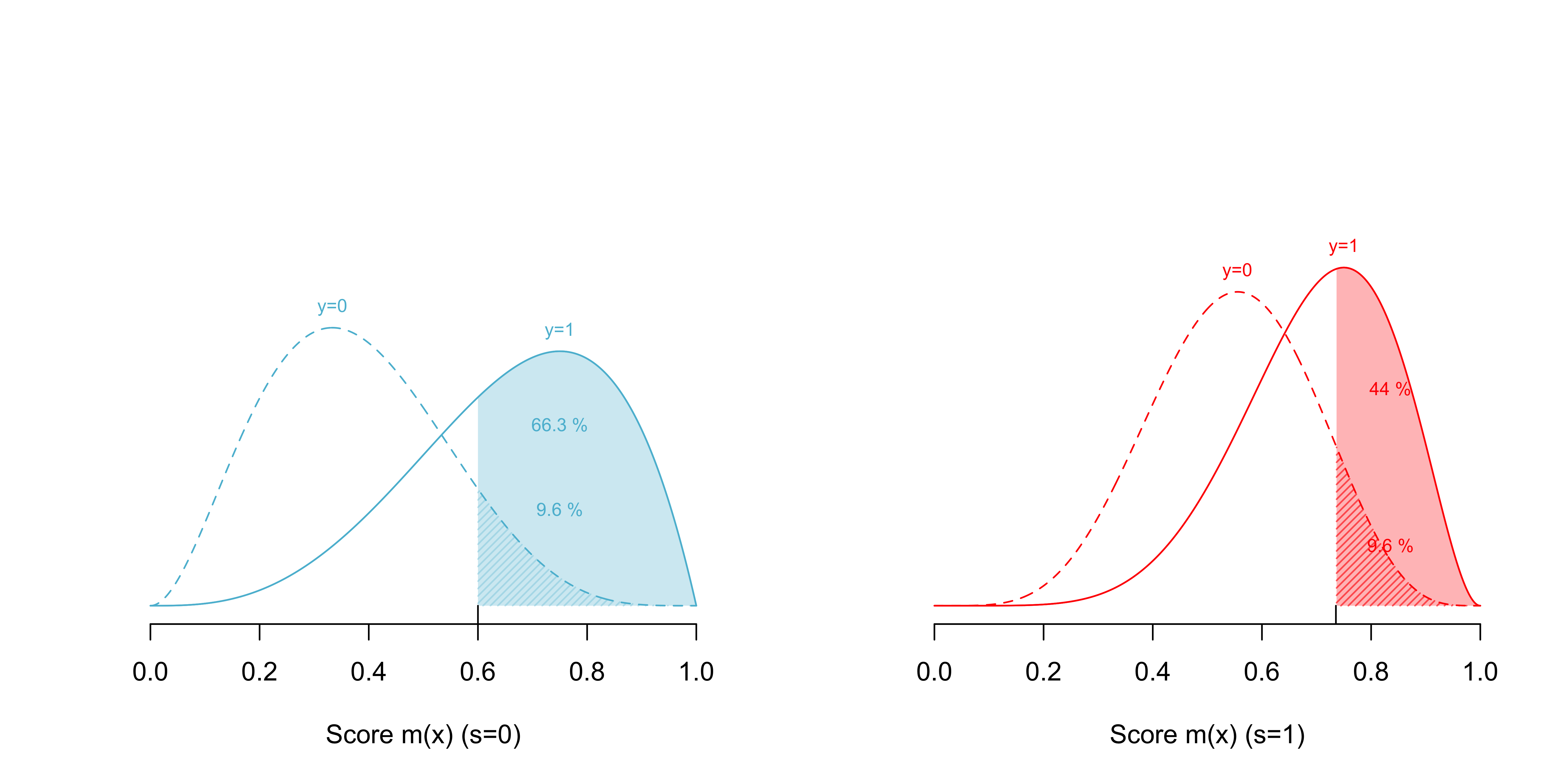}
    \caption{Distributions of $m(\boldsymbol{X})$ conditional on $y=1,s=0$ and $s$ conditional on $y=0,s=0$, left (assumed favored population), and distributions of $s$ conditional on $y=1, s=1$ and conditional on $y=0,s=1$, on the right (supposedly disadvantaged population), with $\mathbb{P}(\widehat{Y}=1|S=1, Y=0)=\mathbb{P}(\widehat{Y}=1|S=0,Y=0)$ (here of the order of 9.6\% false positives). }
    \label{fig:ROC:4:0}
\end{figure}

\begin{figure}[!ht]
    \centering
     \includegraphics[width=\textwidth]{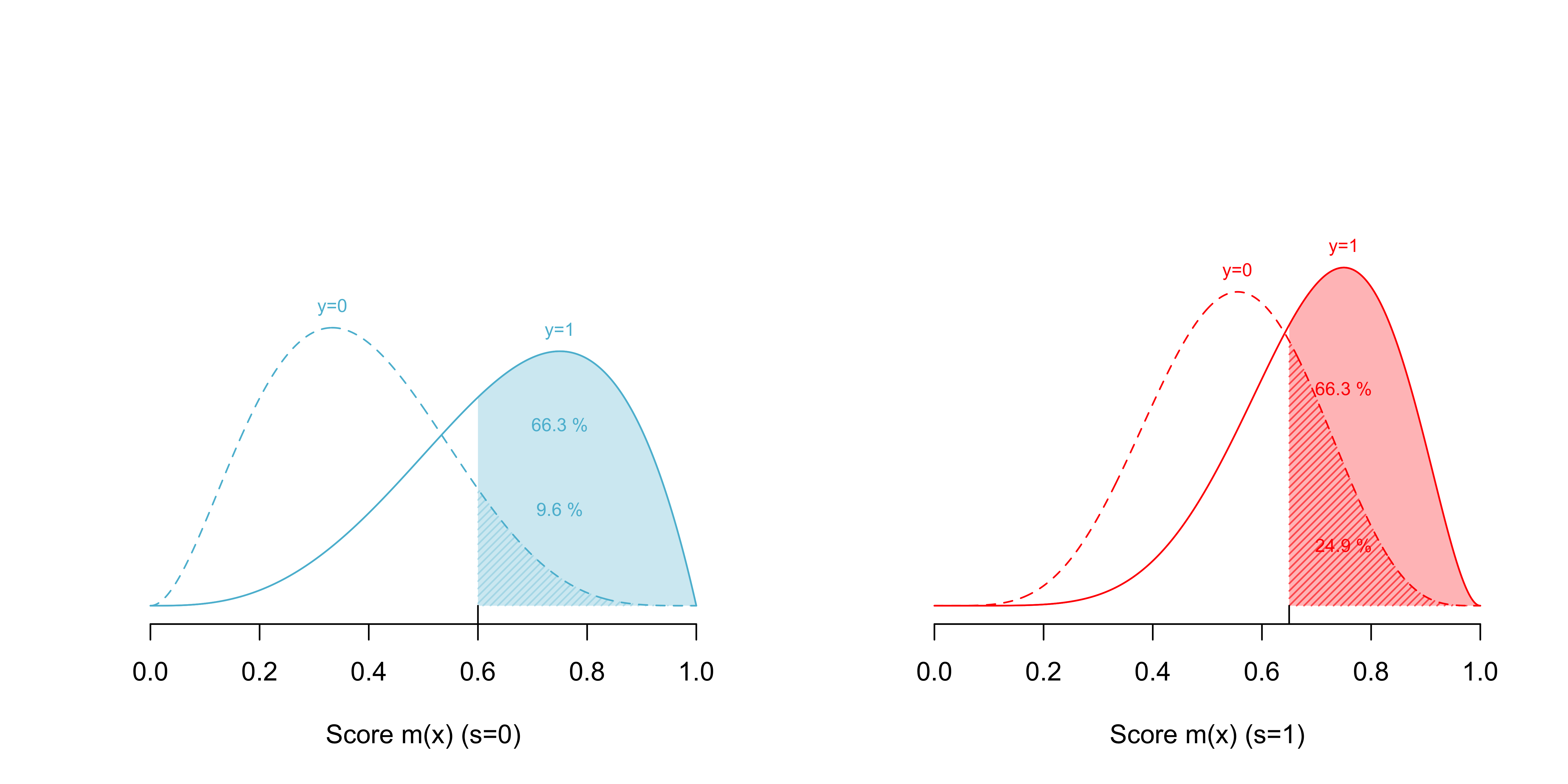}
    \caption{Distributions of $m(\boldsymbol{X})$ conditional on $y=1,s=0$ and $ y=0,s=0$, left (assumed favored population), and distributions of $m(\boldsymbol{X})$ conditional on $y=1, s=1$ and $y=0,s=1$, on the right (supposedly disadvantaged population), with $\mathbb{P}(\widehat{Y}=1|S=1,Y=1)=\mathbb{P}(\widehat{Y}=1|S=0,Y=1)$ (here of the order of 66.3\% true positives). }
    \label{fig:ROC:4:1}
\end{figure} 

\begin{definition}[Equalized of opportunity, \cite{hardt2016equality}]
The parity of false positives and true positives is called equality of opportunity,
$$
\begin{cases}
\mathbb{P}[\widehat{Y}=1|S=0,Y=1] = 
\mathbb{P}[\widehat{Y}=1|S=1,Y=1]\\
\mathbb{P}[\widehat{Y}=1|S=0,Y=0] = 
\mathbb{P}[\widehat{Y}=1|S=1,Y=0]
\end{cases}
$$
or
$$
\mathbb{P}[\widehat{Y}=1|S=0,Y=y] = 
\mathbb{P}[\widehat{Y}=1|S=1,Y=y], ~\forall y\in\{0,1\}
$$
in other words, $\widehat{Y}\indep S$ conditionally on $Y$.
\end{definition}

On can also use any measure based on confusion matrices, such as $\phi$, introduced by \cite{matthews1975comparison},

\begin{definition}[$\phi$-fairness, \cite{chicco2020advantages}]
We will have $\phi$-fairness if $\phi_1=\phi_0$, where $\phi_s$ denotes Matthews correlation coefficient for the $s$ group,
$$
\phi_s={\displaystyle ={\frac {\text {TP}_s \cdot \text {TN}_s -\text {FP}_s \cdot \text {FN}_s }{\sqrt {(\text {TP}_s +\text {FP}_s )(\text {TP}_s +\text {FN}_s )\cdot(\text {TN}_s +\text {FP}_s )(\text {TN}_s +\text {FN}_s )}}}} 
$$
\end{definition}

All those measures are based on some choice of thresholds, but it is also possible to consider a global measures of calibration, such as the area under the curve,

\begin{definition}[AUC fairness, \cite{borkan2019nuanced}]
We will have AUC fairness if $\text{AUC}_1=\text{AUC}_0$, where $\text{AUC}_s$ is the AUC for the $s$ group.
\end{definition}

We find a similar idea in \cite{beutel2019fairness}. 
The problem with the AUC is that we can have identical AUCs, but very different underlying ROC curves. So, it can be interesting to consider a notion of fairness based on the ROC curves.
As a reminder, we had defined the ROC curve as $t\mapsto\text{TPR}\circ \text{FPR}^{-1}(t)$.

\begin{definition}[Equality of ROC curves, \cite{vogel2021learning}]
Let $\text{FRP}_s(t)=\mathbb{P}[m(\boldsymbol{X})> t|Y=0,S=s]$ and $\text{TPR}_s(t)=\mathbb{P}[m(\boldsymbol{X})> t|Y=1, S=s]$. Set $\Delta_{TPR}(t)=\text{TPR}_1\circ\text{TPR}_0^{-1}(t)-t$ et $\Delta_{FRP}(t)=\text{FPR}_1\circ\text{FPR}_0^{-1}(t)-t$.
We will have an fairness of ROC curves if $\|\Delta_{TPR}\|_{\infty}=\|\Delta_{FPR}\|_{\infty}=0$.
\end{definition}

An (implicit) assumption made here is that class 1 (or $\text{\textcolor{tikbleu}{\scriptsize\faCircle}}$ in our illustration) in the sensitive attribute $S$ represents a socially sensitive group, i.e., a minority group that is discriminated against), such that disparate impact is defined by positive (i.e., desirable) outcomes.
Equal opportunity is satisfied if the prediction $\widehat{Y}$
is conditionally independent of the protected attribute $S$, given the actual value $Y$,
\begin{equation*}\label{eq:equalized:odds}
    \forall y:~\widehat{Y} \indep S~|~Y=y
\end{equation*}
or
$$
\mathbb{P}(\widehat{Y}=1|S=0,Y=y)=\mathbb{P}(\widehat{Y}=1|S=1,Y=y)=\mathbb{P}(\widehat{Y}=1|Y=y),~\forall y\in\{0,1\}
$$
which is equivalent to having 
$$
\mathbb{E}(\widehat{Y}|S=0,Y=y)=
\mathbb{E}(\widehat{Y}|S=1,Y=y)=
\mathbb{E}(\widehat{Y}|Y=y),~\forall y\in\{0,1\},
$$
The latter will be true, but it will not be enough.
This means that the true positive rate and the false positive rate will be the same for each population; each type of error is matched between each group. 

In our illustration, equality of opportunity is impossible to achieve. Indeed, this definition of fairness suggests that the false positive and true positive rates be the same for both populations. This may be reasonable, but in the illustrative example, it is impossible because the two ROC curves do not intersect. Note that if the curves did cross, this could impose threshold choices that would be unattractive in practice (with acceptance rates potentially much too low, or too high).

Equality of opportunity, defined by \cite{hardt2016equality}, has the same mathematical formulation as equality of opportunity, for a classifier, but it focuses on a particular label,
$$
    \exists y:~\widehat{Y} \indep S~|~Y=y
$$
Typically, we will focus on the $1$ label of the true $y$ value, to define equal opportunity, so that
$$
\mathbb{P}(\widehat{Y}=0|Y=1,S=s)=\mathbb{P}(\widehat{Y}=0|Y=1),~\forall S
s\in\{0,1\},
$$
which is the same as comparing the rates of negative rates.
In this case, we want the true positive rate $\mathbb{P}(\widehat{Y}=1|Y=1)$ to be the same for each population without taking into account the errors when $y=0$. In effect, this means that the same proportion of each population receives the ``good" result $y=1$. 

The deviation from equality of opportunity is measured by the difference in equality of opportunity:
$$
EOD=\mathbb{P}(\widehat{Y}=1|Y=1,S=1)- \mathbb{P}(\widehat{Y}=1|Y=1,S=0)
$$
As we have here a binary variable $y$, the condition on the probability of $\widehat{y}$ will be here equivalent to the condition on the expectation
$$
\mathbb{E}(\widehat{Y}|Y=1,S=s)=\mathbb{E}(\widehat{Y}|Y=1),~\forall s\in\{0,1\},
$$

In our illustration, equality of opprtunities is equivalent to finding equivalent levels of positive true rates on the ROC curves.

\begin{figure}[!ht]
    \centering
     \includegraphics[width=\textwidth]{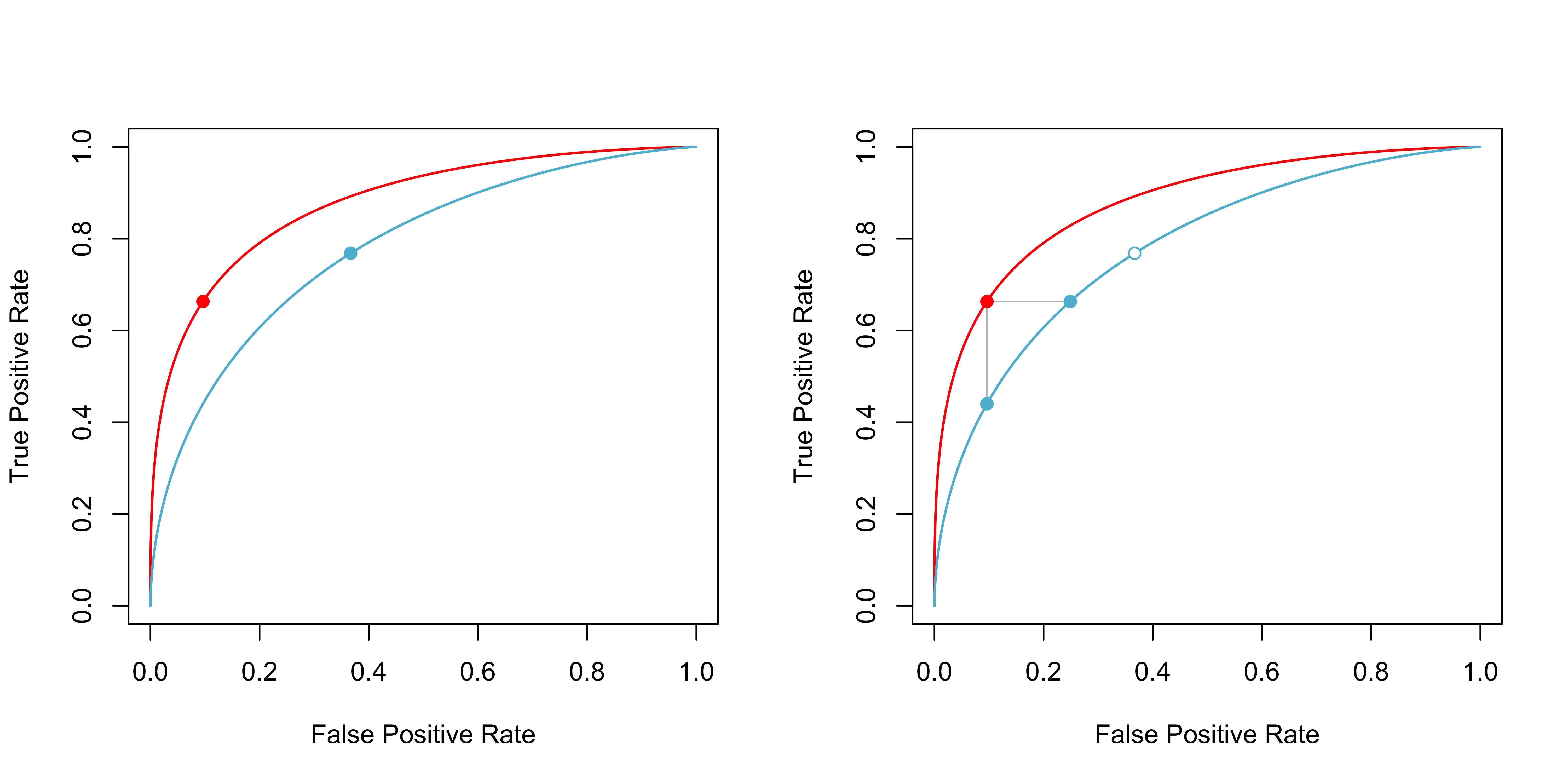}
    \caption{ROC curves for the two subpopulations, the assumed \textcolor{colrwarouge}{disfavoured} (or $s=1$) and the assumed \textcolor{colrwableu}{favoured} (or $s=0$). On the left, the two points correspond to a threshold of 60\%, identical for the two populations (strategy of blinding to the protected criterion). On the right, the case of equality of opportunity, where the threshold for the disadvantaged population is the threshold induced by $\textcolor{colrwableu}{\bullet}$, chosen so as to have the same rate of true positives as on the favoured population $\textcolor{colrwarouge}{\bullet}$.}
    \label{fig:ROC:6}
\end{figure}

As visible on the right side of Figure \ref{fig:ROC:6}, the thresholds are chosen so that the rate of true positives is the same for both populations. In other words, we must have the same proportion offered credit in each group (advantaged and disadvantaged). For example, here we keep the threshold of 60\% on the score for the advantaged population (corresponding to a true positive rate of about 63\%), and we must use a threshold of about 70\% on the score for the disadvantaged population (corresponding to a true positive rate of about 63\%).

Finally, instead of focusing solely on one quantity (false positive rate for instance), it is possible to consider some function of a pair of those quantities.

\begin{definition}[Equal treatment, \cite{berk2021fairness}]
We have equality of treatment, the rate of false positives and false negatives are identical in the protected groups,
$$
\frac{\mathbb{P}[\widehat{Y}=1|S=0,Y=0]}{\mathbb{P}[\widehat{Y}=0|S=0,Y=1]}=
\frac{\mathbb{P}[\widehat{Y}=1|S=1,Y=0]}{\mathbb{P}[\widehat{Y}=0|S=1,Y=1]}
$$
\end{definition}

\cite{berk2021fairness} uses the processing term in connection with causal inference which we will discuss next. If the classifier produces more false negatives than false positives for the supposedly privileged group, this means that more disadvantaged individuals receive a favorable outcome than the reverse. 

A slightly different version had been proposed by \cite{jung2020fair},

\begin{definition}[Equalizing Disincentives, \cite{jung2020fair}]
The difference between the true positive rate and the false positive rate must be the same in the protected groups,
$$
{\mathbb{P}[\widehat{Y}=1|S=0,Y=1]}-
{\mathbb{P}[\widehat{Y}=1|S=0,Y=0]} =
{\mathbb{P}[\widehat{Y}=1|S=1,Y=1]}-
{\mathbb{P}[\widehat{Y}=1|S=1,Y=0]}
$$
\end{definition}

\subsection{Conditional Demographic Parity}\label{sub:con:demo:parity}

All the discussion we have just had can be extended by also conditioning on some explanatory variables

\begin{definition}[Conditional demographic parity, \cite{corbettdavies2017algorithmic}]
We will have a conditional demographic parity if 
$$
\mathbb{P}[\widehat{Y}= y|\boldsymbol{X}_{\ell}=\boldsymbol{x},S=0] = 
\mathbb{P}[\widehat{Y}= y|\boldsymbol{X}_{\ell}=\boldsymbol{x},S=1],~\forall y\in\{0,1\} 
$$
where $\ell$ denotes a ``legitimate" subset of unprotected covariates.
\end{definition}

\subsection{Class balance and calibration}\label{sub:class:balance:2}

Instead of predicting the value of $\widehat{y}$ (conditional on $y$ and $s$), \cite{kleinberg2016inherent} had suggested predicting the average value of $m(\boldsymbol{x})$:

\begin{definition}[Class balance, \cite{kleinberg2016inherent}]
We will have class balance in the weak sense if
$$
\mathbb{E}[m(\boldsymbol{X})|Y=y,S=0] =
\mathbb{E}[m(\boldsymbol{X})|Y=y,S=1], ~\forall y\in\{0,1\} 
$$
or in the strong sense if
$$
\mathbb{P}[m(\boldsymbol{X})\leq \mu|Y=y,S=0] =
\mathbb{P}[m(\boldsymbol{X})\leq \mu|Y=y,S=1] , ~\forall \mu\in[0,1],~\forall y\in\{0,1\} .
$$
\end{definition}


A third commonly used criterion is sometimes called ``sufficiency", which requires independence between the target $Y$ and the sensitive variable $S$, conditional on a given score $m(\boldsymbol{X})$ (or forecast), $\widehat{Y}$, introduced by \cite{sokolova2006beyond}, and later taken up by \cite{kleinberg2016inherent} and \cite{zafar2017fairness}.
In most of the definitions we had seen, we were interested (only) in $\widehat{y}$, but it is also possible to use the score $m(\boldsymbol{X})$. Therefore, we aim at
\begin{equation*}
    Y \indep m(\boldsymbol{X})~|~S
\end{equation*}
so that
$$
\mathbb{P}(Y=1|m(\boldsymbol{X})=\mu,S=0)=\mathbb{P}(Y=1|m(\boldsymbol{X})=\mu,S=1)=\mathbb{P}(Y=1|m(\boldsymbol{X})=\mu),~\forall \mu\in[0,1].
$$

\begin{definition}[Calibration (or accuracy) parity, \cite{kleinberg2016inherent}, \cite{zafar2017fairness}] We have calibration parity if
$$
\mathbb{P}[Y=1|m(\boldsymbol{X})=\mu,S=0]=\mathbb{P}[Y=1|s(\boldsymbol{X})=\mu,S=1],~\forall \mu\in[0,1].
$$
\end{definition}

We can go further by asking for a little more, by asking not only for parity, but also for a good calibration

\begin{definition}[Good calibration, \cite{Kleinberg17}]
We have an fairness of good calibration if
$$
\mathbb{P}[Y=1|m(\boldsymbol{X})=\mu,S=0]=\mathbb{P}[Y=1|m(\boldsymbol{X})=\mu,S=1]=\mu,~\forall \mu\in[0,1].
$$
\end{definition}

This ``good calibration" property of the model $m$, also called ``well-calibration" in \cite{dawid1982well}, and ``autocalibration" in \cite{van2019calibration}, \cite{kruger2021generic} and \cite{denuit2021autocalibration} in the context of regression, i.e. $\mathbb{E}[Y|m(\boldsymbol{X})=\mu]=\mu$, is a standard property in econometrics, in generalized linear models, but not in most machine learning algorithms. 


A weaker version would be

\begin{definition}[Predictive parity (1), \cite{chouldechova2017fair}]
We have a predictive parity if
$$
\mathbb{P}[Y=1|\widehat{Y}=1,S=0]=\mathbb{P}[Y=1|\widehat{Y}=1,S=1].
$$
\end{definition}

Note that if $\widehat{y}$ is not a perfect classifier ($\mathbb{P}[\widehat{Y}\neq Y]>0$), and if the two groups are not balanced ($\mathbb{P}[S=0]\neq \mathbb{P}[S=1]$), then it is impossible to have predictive parity and equal opportunity at the same time. Note that
$$
\text{PPV}_s=\frac{\text{TPR}\cdot \mathbb{P}[S=s]}{\text{TPR}\cdot \mathbb{P}[S=s]+\text{FPR}\cdot (1-\mathbb{P}[S=s])},~\forall s\in\{0,1\},
$$
such that $\text{PPV}_0=\text{PPV}_1$ implies that either $\text{TPR}$ or $\text{FPR}$ is zero, and since
$$
\text{NPV}_s=\frac{(1-\text{FPR})\cdot (1-\mathbb{P}[S=s])}{(1-\text{TPR})\cdot \mathbb{P}[S=s]+(1-\text{FPR})\cdot (1-\mathbb{P}[S=s])},~\forall s\in\{0,1\},
$$
such that
$\text{NPV}_0\neq\text{NPV}_1$, and predictive parity cannot be achieved.

Continuing the formalism of \cite{chouldechova2017fair}, \cite{barocas-hardt-narayanan} proposed an extension to predictive parity

\begin{definition}[Predictive parity (2), \cite{barocas-hardt-narayanan}]
$$
\begin{cases}
\mathbb{P}[{Y}=1|S=0,\widehat{Y}=1] = 
\mathbb{P}[{Y}=1|S=1,\widehat{Y}=1]~&~\text{ positive prediction}\\
\mathbb{P}[{Y}=1|S=0,\widehat{Y}=0] = 
\mathbb{P}[{Y}=1|S=1,\widehat{Y}=0]~&~\text{ negative prediction}
\end{cases}
$$
or
$$
\mathbb{P}[{Y}=1|S=0,\widehat{Y}=\widehat{y}] = 
\mathbb{P}[{Y}=1|S=1,\widehat{Y}=\widehat{y}], ~\forall \widehat{y}\in\{0,1\}
$$
\end{definition}

Finally, let us note that \cite{Kleinberg17} introduced a notion of balance for positive / negative class.
$$
\begin{cases}
\mathbb{E}(m(\boldsymbol{X})|Y=1,S=1) =
\mathbb{E}(m(\boldsymbol{X})|Y=1,S=0),~\text{balance for positive class}\\
\mathbb{E}(m(\boldsymbol{X})|Y=0,S=1) =
\mathbb{E}(m(\boldsymbol{X})|Y=0,S=0),~\text{equilibrium for the negative class.}\\  
\end{cases}
$$




\subsection{Principle of non-reconstruction}\label{sub:non:reconstruct}

A last approach can be inspired by \cite{kim2017auditing}, for whom, another way to define if a classification is fair, or not, is to say that we cannot tell from the result if the subject was member of a protected group or not. In other words, if
an individual's score does not allow us to predict that individual's attributes better than guessing them without any information, we can say that the score was assigned fairly.

\begin{definition}[Non-reconstruction of the protected attribute, \cite{kim2017auditing}]
If we cannot tell from the result ($\boldsymbol{x}$, $m(\boldsymbol{x})$, $y$ and $\widehat{y}$) whether the subject was a member of a protected group or not, we will talk about fairness by non-reconstruction of the protected attribute
$$
\mathbb{P}[S=0|\boldsymbol{X},m(\boldsymbol{X}),\widehat{Y},Y] =
\mathbb{P}[S=1|\boldsymbol{X},m(\boldsymbol{X}),\widehat{Y},Y]  .
$$
\end{definition}

\subsection{Comparison of fairness criteria}\label{sug:comparison:group}

Demographic parity would result in 
$$
\mathbb{P}(\widehat{Y}=1|m(\boldsymbol{X})=\mu)=\mathbb{P}(\widehat{Y}=1),~\forall \mu,
$$
or $\text{TP}+\text{FP}$ must be identical on both groups ($s=0$ and $s=1$). On the confusion matrices of Figure \ref{fig:confusion:compare}, it is the case, because the positive rate is $50\%$ in both groups. The notion of equality of opportunity means
$$
\mathbb{P}(\widehat{Y}=1|S=0,Y=y)=\mathbb{P}(\widehat{Y}=1|S=1,Y=y)=\mathbb{P}(\widehat{Y}=1|Y=y),~\forall y\in\{0,1\},
$$
in other words, the false positive and false negative rates must be identical:
$$\frac{\text{FP}}{\text{TN}+\text{FP}}\text{ and }
\frac{\text{FN}}{\text{TP}+\text{FN}},
$$
must be identical on both groups.
This is the case on confusion matrices of Figure \ref{fig:confusion:compare}, because the false positive rate is 50\%, whether $s$ is $0$ or $1$ (with 30/60 versus 20/40), and the false negative rate is also 50\% in both groups. For predictive parity
$$
\mathbb{P}(Y=1|S=0,\widehat{Y}=y)=\mathbb{P}(Y=1|S=1,\widehat{Y}=y)=\mathbb{P}(Y=1|\widehat{Y}=y),~\forall y~
$$
in other words, the positive and negative predictive values must be identical:
$$\frac{\text{TP}}{\text{TP}+\text{FP}}\text{ and }
\frac{\text{TN}}{\text{TN}+\text{FN}}
$$
must be identical on both groups.
But here the positive predictive values are respectively 60\% and 40\%, depending on the value of $s$ (respectively 30/50 and 20/50). For the global accuracy, 
$$
\mathbb{P}(\widehat{Y}=Y|S=s)=\mathbb{P}(\widehat{Y}=Y),~\forall s
$$
or $\text{TP}+\text{TN}$ must be identical on both groups ($s=0$ and $s=1$). This is the case here because the rate of well classified observations is 50 
$$
\mathbb{P}(\widehat{Y}=Y|m(\boldsymbol{X})=\mu)=\mathbb{P}(\widehat{Y}=Y),~\forall \mu\in[0,1],
$$
or $\text{FP}/\text{FN}$ must be identical for both groups ($s=0$ and $s=1$). But here, the rates are respectively 3/2 and 2/3, which are not equal.

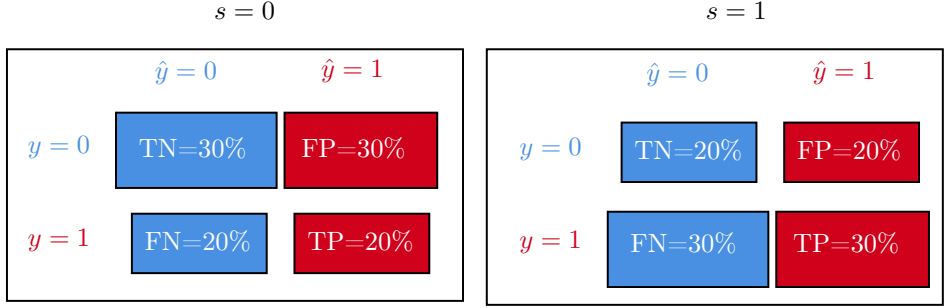
\begin{figure}
    \centering
    \tikzset{every picture/.style={line width=0.75pt}} 

\begin{tikzpicture}[x=0.75pt,y=0.75pt,yscale=-1,xscale=1]

\draw  [fill={rgb, 255:red, 74; green, 144; blue, 226 }  ,fill opacity=1 ] (76,63) -- (157,63) -- (157,101) -- (76,101) -- cycle ;
\draw  [fill={rgb, 255:red, 74; green, 144; blue, 226 }  ,fill opacity=1 ] (84,114) -- (152,114) -- (152,144) -- (84,144) -- cycle ;
\draw  [fill={rgb, 255:red, 208; green, 2; blue, 27 }  ,fill opacity=1 ] (161,63) -- (238,63) -- (238,101) -- (161,101) -- cycle ;
\draw  [fill={rgb, 255:red, 208; green, 2; blue, 27 }  ,fill opacity=1 ] (166,114) -- (234,114) -- (234,144) -- (166,144) -- cycle ;
\draw  [fill={rgb, 255:red, 74; green, 144; blue, 226 }  ,fill opacity=1 ] (331,68) -- (399,68) -- (399,98) -- (331,98) -- cycle ;
\draw  [fill={rgb, 255:red, 74; green, 144; blue, 226 }  ,fill opacity=1 ] (329,115) -- (397,115) -- (397,145) -- (329,145) -- cycle ;
\draw  [fill={rgb, 255:red, 208; green, 2; blue, 27 }  ,fill opacity=1 ] (413,68) -- (481,68) -- (481,98) -- (413,98) -- cycle ;
\draw  [fill={rgb, 255:red, 208; green, 2; blue, 27 }  ,fill opacity=1 ] (411,115) -- (479,115) -- (479,145) -- (411,145) -- cycle ;
\draw  [fill={rgb, 255:red, 74; green, 144; blue, 226 }  ,fill opacity=1 ] (324,113) -- (405,113) -- (405,151) -- (324,151) -- cycle ;
\draw  [fill={rgb, 255:red, 208; green, 2; blue, 27 }  ,fill opacity=1 ] (409,113) -- (486,113) -- (486,151) -- (409,151) -- cycle ;
\draw   (21,31) -- (251,31) -- (251,158) -- (21,158) -- cycle ;
\draw   (263,31) -- (493,31) -- (493,160) -- (263,160) -- cycle ;

\draw (178,34.4) node [anchor=north west][inner sep=0.75pt]  [color={rgb, 255:red, 208; green, 2; blue, 27 }  ,opacity=1 ]  {$\hat{y} =1$};
\draw (94,35.4) node [anchor=north west][inner sep=0.75pt]  [color={rgb, 255:red, 74; green, 144; blue, 226 }  ,opacity=1 ]  {$\hat{y} =0$};
\draw (30,73.4) node [anchor=north west][inner sep=0.75pt]  [color={rgb, 255:red, 74; green, 144; blue, 226 }  ,opacity=1 ]  {$y=0$};
\draw (30,120.4) node [anchor=north west][inner sep=0.75pt]  [color={rgb, 255:red, 208; green, 2; blue, 27 }  ,opacity=1 ]  {$y=1$};
\draw (86,74) node [anchor=north west][inner sep=0.75pt]  [color={rgb, 255:red, 255; green, 255; blue, 255 }  ,opacity=1 ] [align=left] {TN=30\%};
\draw (89,121) node [anchor=north west][inner sep=0.75pt]  [color={rgb, 255:red, 255; green, 255; blue, 255 }  ,opacity=1 ] [align=left] {FN=20\%};
\draw (168,74) node [anchor=north west][inner sep=0.75pt]  [color={rgb, 255:red, 255; green, 255; blue, 255 }  ,opacity=1 ] [align=left] {FP=30\%};
\draw (171,121) node [anchor=north west][inner sep=0.75pt]  [color={rgb, 255:red, 255; green, 255; blue, 255 }  ,opacity=1 ] [align=left] {TP=20\%};
\draw (426,35.4) node [anchor=north west][inner sep=0.75pt]  [color={rgb, 255:red, 208; green, 2; blue, 27 }  ,opacity=1 ]  {$\hat{y} =1$};
\draw (342,36.4) node [anchor=north west][inner sep=0.75pt]  [color={rgb, 255:red, 74; green, 144; blue, 226 }  ,opacity=1 ]  {$\hat{y} =0$};
\draw (278,74.4) node [anchor=north west][inner sep=0.75pt]  [color={rgb, 255:red, 74; green, 144; blue, 226 }  ,opacity=1 ]  {$y=0$};
\draw (278,121.4) node [anchor=north west][inner sep=0.75pt]  [color={rgb, 255:red, 208; green, 2; blue, 27 }  ,opacity=1 ]  {$y=1$};
\draw (336,75) node [anchor=north west][inner sep=0.75pt]  [color={rgb, 255:red, 255; green, 255; blue, 255 }  ,opacity=1 ] [align=left] {TN=20\%};
\draw (334,122) node [anchor=north west][inner sep=0.75pt]  [color={rgb, 255:red, 255; green, 255; blue, 255 }  ,opacity=1 ] [align=left] {FN=30\%};
\draw (418,75) node [anchor=north west][inner sep=0.75pt]  [color={rgb, 255:red, 255; green, 255; blue, 255 }  ,opacity=1 ] [align=left] {FP=20\%};
\draw (416,122) node [anchor=north west][inner sep=0.75pt]  [color={rgb, 255:red, 255; green, 255; blue, 255 }  ,opacity=1 ] [align=left] {TP=30\%};
\draw (124,5.4) node [anchor=north west][inner sep=0.75pt]    {$s=0$};
\draw (372,5.4) node [anchor=north west][inner sep=0.75pt]    {$s=1$};

\end{tikzpicture}
    \vspace{-.8cm}\caption{fairness of a classifier from the confusion matrices, on the two subpopulations, $s=0$ on the left and $s=1$ on the right.}
    \label{fig:confusion:compare}
\end{figure}

We will now consider the different possible ways of setting these thresholds that result in different senses of fairness. We emphasize that we are not advocating any particular criteria, but simply exploring the ramifications of different choices.
For demographic parity, the threshold could be chosen so that the same proportion of each group is classified as $\widehat{y}=1$.
For equal opportunity, thresholds are chosen so that the true positive rate is the same for both populations (Figure 4). Of those who repay the loan, the same proportion are offered credit in each group. For the two ROC curves, this means that the thresholds are chosen so that the vertical position on each curve is the same regardless of the horizontal position (Figure 2c). However, this means that different proportions of the blue and yellow groups receive loans (Figure 4b).


The different notions of fairness can be summarized in Table \ref{tab:def:group}.
\begin{table}[!ht]
     \begin{tabular}{|lllc|}\hline
         {\em statistical parity} & \cite{dwork2012fairness} & 
         $\mathbb{P}[\widehat{Y}=1|S=s]=\text{cst},~\forall s$
         & independence\\
          {\em conditional statistical parity} & \cite{corbettdavies2017algorithmic} & 
         $\mathbb{P}[\widehat{Y}=1|S=s,X=x]=\text{cst}_x,~\forall s,y$
         & $\widehat{Y}\indep S$ \\ \hline
         {\em equalized odds} & \cite{hardt2016equality} & 
         $\mathbb{P}[\widehat{Y}=1|S=s,Y=y]=\text{cst}_y,~\forall s,y$
         & separation\\
        {\em equalized opportunity} & \cite{hardt2016equality} & 
         $\mathbb{P}[\widehat{Y}=1|S=s,Y=1]=\text{cst},~\forall s$
         & \\
         {\em predictive equality} & \cite{corbettdavies2017algorithmic} & 
         $\mathbb{P}[\widehat{Y}=1|S=s,Y=0]=\text{cst},~\forall s$
         & $\widehat{Y}\indep S~|~Y$ \\
         {\em balance (positive)} & \cite{Kleinberg17} & 
         $\mathbb{E}[m(\boldsymbol{X})|S=s,Y=1]=\text{cst},~\forall s$
         & $m(\boldsymbol{X})\indep S~|~Y$ \\
        {\em balance (negative)} & \cite{Kleinberg17} & 
         $\mathbb{E}[m(\boldsymbol{X})|S=s,Y=0]=\text{cst},~\forall s$
         & \\ \hline\hline
          {\em conditional accuracy equality} & \cite{berk2017convex} & 
         $\mathbb{P}[Y=y|S=s,\widehat{Y}=y]=\text{cst}_y,~\forall s,y$
         & sufficiency\\
         {\em predictive parity} & \cite{chouldechova2017fair} & 
         $\mathbb{P}[Y=1|S=s,\widehat{Y}=1]=\text{cst},~\forall s$ & \\
         {\em calibration} & \cite{chouldechova2017fair} & 
         $\mathbb{P}[Y=1|m(\boldsymbol{X})=\mu,S=s]=\text{cst}_\mu,~\forall \mu,s$
         & $Y\indep S~|~\widehat{Y}$\\
        {\em well-calibration} & \cite{chouldechova2017fair} & 
         $\mathbb{P}[Y=1|m(\boldsymbol{X})=\mu,S=s]=\mu,~\forall \mu,s$
         & \\ \hline\hline
        {\em accuracy equality} & \cite{berk2017convex} &  
         $\mathbb{P}[\widehat{Y}=Y|S=s]=\text{cst},~\forall s$
         & \\
        {\em treatment equality} & \cite{berk2017convex} & 
         $\displaystyle{\frac{\text{FNR}_s}{\text{FPR}_s}=\text{cst}_s},~\forall s$
         & \\ \hline
    \end{tabular}
    \caption{Group Fairness Definitions.}
    \label{tab:def:group}
\end{table}

\subsection{Relaxation and confidence intervals}\label{sub:conf:int}

From a statistical perspective, achieving fairness is impossible since it requires equality between probabilities, or strict independence. Thus, we will discuss practical use of those concepts.

\subsubsection{Exogenous threshold and relaxation}

We had seen that the demographic fairness is translated by the equality
$$
\frac{\mathbb{P}[\widehat{Y}=1|S=0]}{
\mathbb{P}[\widehat{Y}=1|S=1]}=1=\frac{\mathbb{P}[\widehat{Y}=1|S=1]}{
\mathbb{P}[\widehat{Y}=1|S=0]}
$$
If this approach is intellectually interesting, the statistical reality is that having a perfect equality between two (predictive) probabilities is often impossible.

\begin{definition}[Disparate impact, \cite{feldman2015certifying}]
A decision function $\widehat{Y}$ has a disparate impact, for a given threshold $d$, if, 
$$
\min\lbrace
\frac{\mathbb{P}[\widehat{Y}=1|S=0]}{
\mathbb{P}[\widehat{Y}=1|S=1]},
\frac{\mathbb{P}[\widehat{Y}=1|S=1]}{
\mathbb{P}[\widehat{Y}=1|S=0]}\rbrace<d~\text{(usually 80\%)}.
$$
\end{definition}

This so-called ``{\em four-fifths rule}", coupled with the $d=80\%$ threshold, was originally defined by the State of California Fair Employment Practice Commission (FEPC) Technical Advisory Committee on Testing, which issued the California State Guidelines on Employee Selection Procedures in October 1972, as recalled in
\cite{feldman2015certifying}, \cite{mercat2016discrimination} or \cite{biddle2017adverse}. This standard was later adopted in the 1978 Uniform Guidelines on Employee Selection Procedures, used by the Equal Employment Opportunity Commission (EEOC), the U.S. Department of Labor, and the U.S. Department of Justice. An important point here is that this form of discrimination occurred even when the employer did not intend to discriminate, but by looking at employment statistics (on gender or racial grounds), it was possible to observe (and correct) discriminatory bias.

For example, on the data in Figure \ref{fig:circle:square:3},
$$
\frac{\mathbb{P}[\widehat{Y}=1|S=\text{\textcolor{tikbleu}{\scriptsize\faCircle}}]}{\mathbb{P}[\widehat{Y}=1|S=\text{\textcolor{tikrouge}{\scriptsize\faStop}}]} = \frac{1}{3} \ll 80\%.
$$
Another approach, suggested to relax the equality $\mathbb{P}(\widehat{Y}=1|S=0)=\mathbb{P}(\widehat{Y}=1|S=1)$, 
consists in introducing a notion of $\varepsilon$-fairness
$$
|\mathbb{P}(\widehat{Y}=1|S=0)-\mathbb{P}(\widehat{Y}=1|S=1)|<\varepsilon. 
$$
The left deviation is sometimes called ``statistical parity difference" ($SPD$). \cite{zliobaite2015relation} suggests normalizing the statistical parity difference, 
$$
NSPD=\frac{SPD}{D_{\max}}\text{ where }D_{\max}=\min\lbrace\frac{\mathbb{P}(\widehat{Y}=1)}{\mathbb{P}(S=1)},\frac{\mathbb{P}(\widehat{Y}=0)}{\mathbb{P}(S=0)}\rbrace
$$
so that $NSPD=1$ for maximum discrimination.

\subsubsection{Endogenous threshold and confidence intervals}

\cite{besse2018confidence} proposes another approach, based on confidence intervals for fairness criteria. For example, for the disparate impact, we have seen that we should calculate
$$
T=\frac{\mathbb{P}[\widehat{Y}=1|S=0]}{\mathbb{P}[\widehat{Y}=1|S=1]}
$$
whose empirical version is
$$
\widehat{t}_n = \frac{\sum_{i}\widehat{y}_i\boldsymbol{1}(s_i=0)}{\sum_{i}\widehat{y}_i\boldsymbol{1}(s_i=1)}\cdot
\frac{\sum_{i}\boldsymbol{1}(s_i=1)}{\sum_{i}\boldsymbol{1}(s_i=0)}$$
which can be used to construct a confidence interval for $T$ (\cite{besse2018confidence} proposes an asymptotic test, but resampling methods are possible).

\subsection{Implementation and comparison}\label{sub:implementation}

On the toy example (with 24 observations) of Figure \ref{fig:circle:square:1}, repeated in table \ref{tab:ex:book:1}, we obtain the values in table \ref{tab:ex:book:2}. Note that the `diff' column gives the absolute difference between the two probabilities (expressed as percentages) and the `(\%)' column gives the relative difference between the two probabilities, expressed as a percentage.  \ref{tab:ex:book:2}.

\begin{table}[!ht]
    \centering
 \begin{tabular}{|l|cccccccccc|cccccccccccccc|}\hline
   $s$ & $\text{\textcolor{tikbleu}{\scriptsize\faCircle}}$  & $\text{\textcolor{tikbleu}{\scriptsize\faCircle}}$  & $\text{\textcolor{tikbleu}{\scriptsize\faCircle}}$  & $\text{\textcolor{tikbleu}{\scriptsize\faCircle}}$  & $\text{\textcolor{tikbleu}{\scriptsize\faCircle}}$  & $\text{\textcolor{tikbleu}{\scriptsize\faCircle}}$ & $\text{\textcolor{tikrouge}{\scriptsize\faStop}}$ & $\text{\textcolor{tikrouge}{\scriptsize\faStop}}$ & $\text{\textcolor{tikrouge}{\scriptsize\faStop}}$ & $\text{\textcolor{tikrouge}{\scriptsize\faStop}}$ 
     & $\text{\textcolor{tikbleu}{\scriptsize\faCircle}}$  & $\text{\textcolor{tikbleu}{\scriptsize\faCircle}}$ & $\text{\textcolor{tikrouge}{\scriptsize\faStop}}$& $\text{\textcolor{tikrouge}{\scriptsize\faStop}}$& $\text{\textcolor{tikrouge}{\scriptsize\faStop}}$& $\text{\textcolor{tikrouge}{\scriptsize\faStop}}$& $\text{\textcolor{tikrouge}{\scriptsize\faStop}}$& $\text{\textcolor{tikrouge}{\scriptsize\faStop}}$& $\text{\textcolor{tikrouge}{\scriptsize\faStop}}$& $\text{\textcolor{tikrouge}{\scriptsize\faStop}}$& $\text{\textcolor{tikrouge}{\scriptsize\faStop}}$& $\text{\textcolor{tikrouge}{\scriptsize\faStop}}$& $\text{\textcolor{tikrouge}{\scriptsize\faStop}}$& $\text{\textcolor{tikrouge}{\scriptsize\faStop}}$ \\ 
     $y$ & 0 & 0 & 1& 0 & 1& 1& 0 & 0 & 1& 0 
     & 1& 1& 0 & 0 & 0 & 1& 1& 1& 1& 0& 1& 1& 1& 1\\
     $\widehat{y}$ & 0 & 0 & 0 & 0 & 0 & 0 & 0 & 0 & 0 & 0 & 1 & 1 & 1 & 1 & 1 & 1 & 1 & 1 & 1 & 1 & 1 & 1& 1 & 1\\\hline
    \end{tabular}
    \caption{Data from Figure \ref{fig:circle:square:1}, ordered according to their $S$-score (not shown here) with an identical threshold for both groups ($s\in\{\text{\textcolor{tikbleu}{\scriptsize\faCircle}},\text{\textcolor{tikrouge}{\scriptsize\faStop}}\}$), and with $\widehat{y}=m_{\tau}(\boldsymbol{x})$, for some threshold $\tau$.}
    \label{tab:ex:book:1}
\end{table}

\begin{table}[]
    \centering
    \begin{tabular}{|ll|rr|rr|}\hline
   Name & Probabilistic formula & $\text{\textcolor{tikbleu}{\scriptsize\faCircle}}$~~~~ & $\text{\textcolor{tikrouge}{\scriptsize\faStop}}$~~~~ & diff & (\%)~~~ \\ \hline
 statistical parity &$\mathbb{P}[\widehat{Y}=1|S=\circ]$& 25.0\% & 75.0\% & 50.0 & +200.0\% \\
 equalized opportunity&$\mathbb{P}[\widehat{Y}=1|S=\circ,Y=1]$& 40.0\% & 88.9\% & 48.9 & +122.2\% \\
 predictive equality &$\mathbb{P}[\widehat{Y}=1|S=\circ,Y=0]$& 0.0\% & 57.1\% & 57.1 & -\\
 conditional accuracy &$\mathbb{P}[Y=0|S=\circ,\widehat{Y}=0]$& 50.0\% & 75.0\% & 25.0 & +50.0\% \\
 predictive parity &$\mathbb{P}[Y=1|S=\circ,\widehat{Y}=1]$& 100.0\% & 66.7\% & -33.3 & -33.3\% \\
 accuracy equality &$\mathbb{P}[\widehat{Y}=Y|S=\circ]$& 62.5\% & 68.8\% & 6.2 & +10.0\% \\
 treatment equality &$\text{FN}_{\circ}/\text{FP}_{\circ}$& - ~~ & 25.0\% & -~ & -~~ \\\hline
    \end{tabular}
    \caption{Data from Figure \ref{fig:circle:square:1} and Table \ref{tab:ex:book:1}, with the different concepts of fairness, the values of the measures for the two groups, $s\in\{\text{\textcolor{tikbleu}{\scriptsize\faCircle}},\text{\textcolor{tikrouge}{\scriptsize\faStop}}\}$, the absolute difference, and the relative difference (as a percentage).}
    \label{tab:ex:book:2}
\end{table}

On the continuous example of Figure \ref{fig:ROC:2}, we get the values in the table \ref{tab:ex:book:3}, with the same threshold of 60\% for both groups, as in Figure \ref{fig:ex:book:3}, or with two different thresholds, with $55\%$ and $65\%$ respectively for the bullet $\text{\textcolor{tikbleu}{\scriptsize\faCircle}}$ (or $s=0$) and square $\text{\textcolor{tikrouge}{\scriptsize\faStop}}$ (or $s=1$) groups, as in Figure \ref{fig:ex:book:4}. In Figures \ref{fig:ex:book:3} and \ref{fig:ex:book:4}, at the top are the conditional densities of $m(\boldsymbol{X})| Y=1,S=\text{\textcolor{tikrouge}{\scriptsize\faStop}}$ (solid line) and $m(\boldsymbol{X})| Y=0,S=\text{\textcolor{tikrouge}{\scriptsize\faStop}}$ (dashed line), on the left (assumed \textcolor{tikrouge}{disadvantaged} population), and the conditional densities of $m(\boldsymbol{X})| Y=1,S=\text{\textcolor{tikbleu}{\scriptsize\faCircle}}$ and $m(\boldsymbol{X})| Y=0,S=\text{\textcolor{tikbleu}{\scriptsize\faCircle}}$, on the right (population assumed to be \textcolor{tikbleu}{favored}) At the bottom are the survival functions, $t\mapsto\mathbb{P}(m(\boldsymbol{X})>t| Y=y,S=\circ)$, with, in Figure \ref{fig:ex:book:3},
$$
\widehat{y}=\begin{cases}
\boldsymbol{1}_{[60\%,100\%]}(s)\text{ if }s=\text{\textcolor{tikbleu}{\scriptsize\faCircle}}\text{ or }0\\
\boldsymbol{1}_{[60\%,100\%]}(s)\text{ if }s=\text{\textcolor{tikrouge}{\scriptsize\faStop}}\text{ or }1
\end{cases}
$$
and on Figure \ref{fig:ex:book:4},
$$
\widehat{y}=\begin{cases}
\boldsymbol{1}_{[55\%,100\%]}(s)\text{ if }s=\text{\textcolor{tikbleu}{\scriptsize\faCircle}}\text{ or }0\\
\boldsymbol{1}_{[65\%,100\%]}(s)\text{ if }s=\text{\textcolor{tikrouge}{\scriptsize\faStop}}\text{ or }1\\
\end{cases}
$$
The probabilities indicated on the survival functions are the probabilities of ``true" positives or negatives, according to, for example A VERIFIER !!!!!!!
$$
\mathbb{P}(\widehat{Y}=0|Y=0,S=\text{\textcolor{tikrouge}{\scriptsize\faStop}})=90.4\%\text{ et }
\mathbb{P}(\widehat{Y}=1|Y=1,S=\text{\textcolor{tikrouge}{\scriptsize\faStop}})=66.3\%
$$
while the rates of negative or positive ``falses" are respectively
$$
\mathbb{P}(\widehat{Y}=1|Y=0,S=\text{\textcolor{tikrouge}{\scriptsize\faStop}})=9.6\%\text{ et }
\mathbb{P}(\widehat{Y}=0|Y=1,S=\text{\textcolor{tikrouge}{\scriptsize\faStop}})=33.7\%.
$$

\begin{figure}
    \centering
    \includegraphics[width=\textwidth]{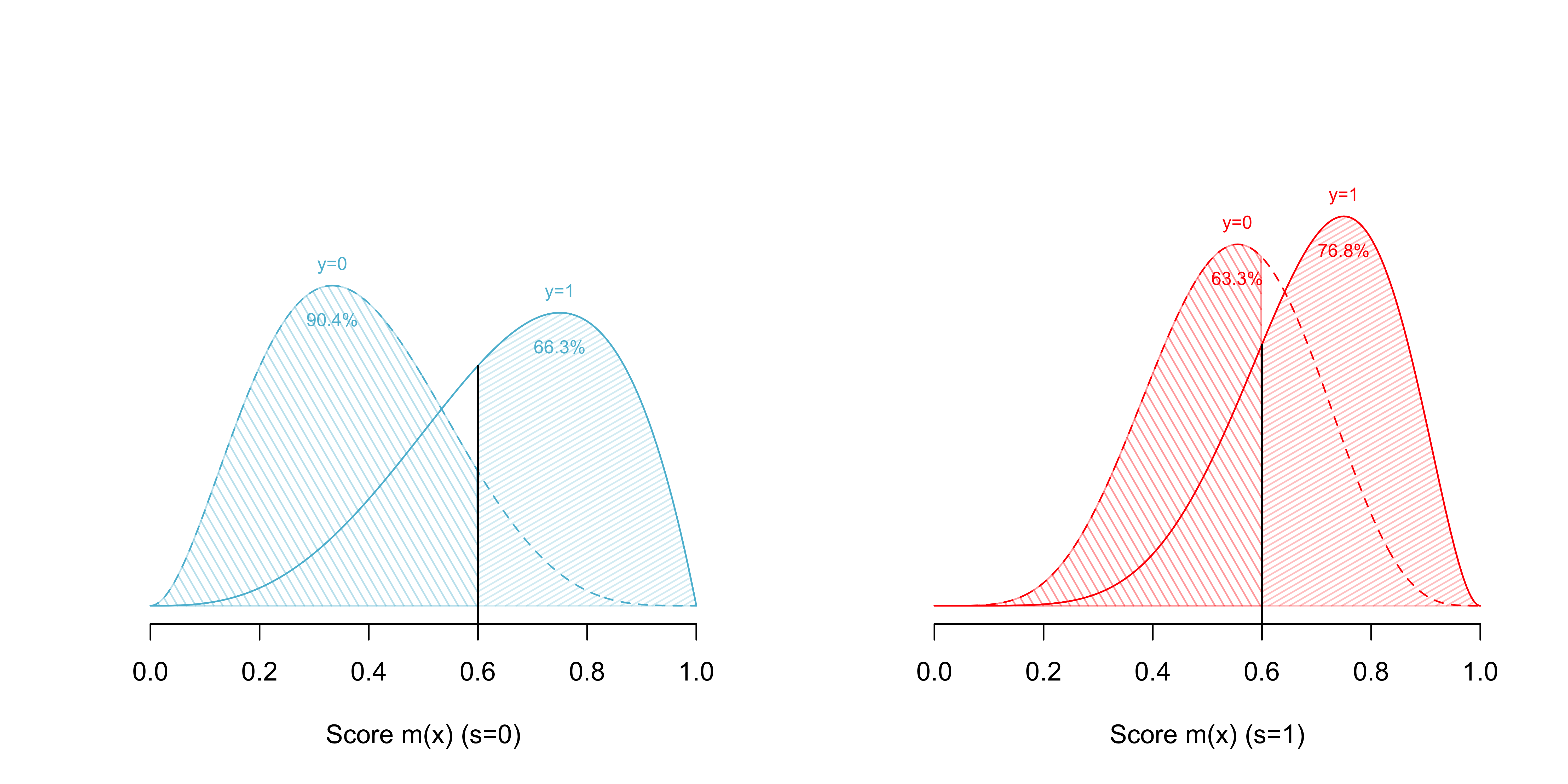}
    
    \includegraphics[width=\textwidth]{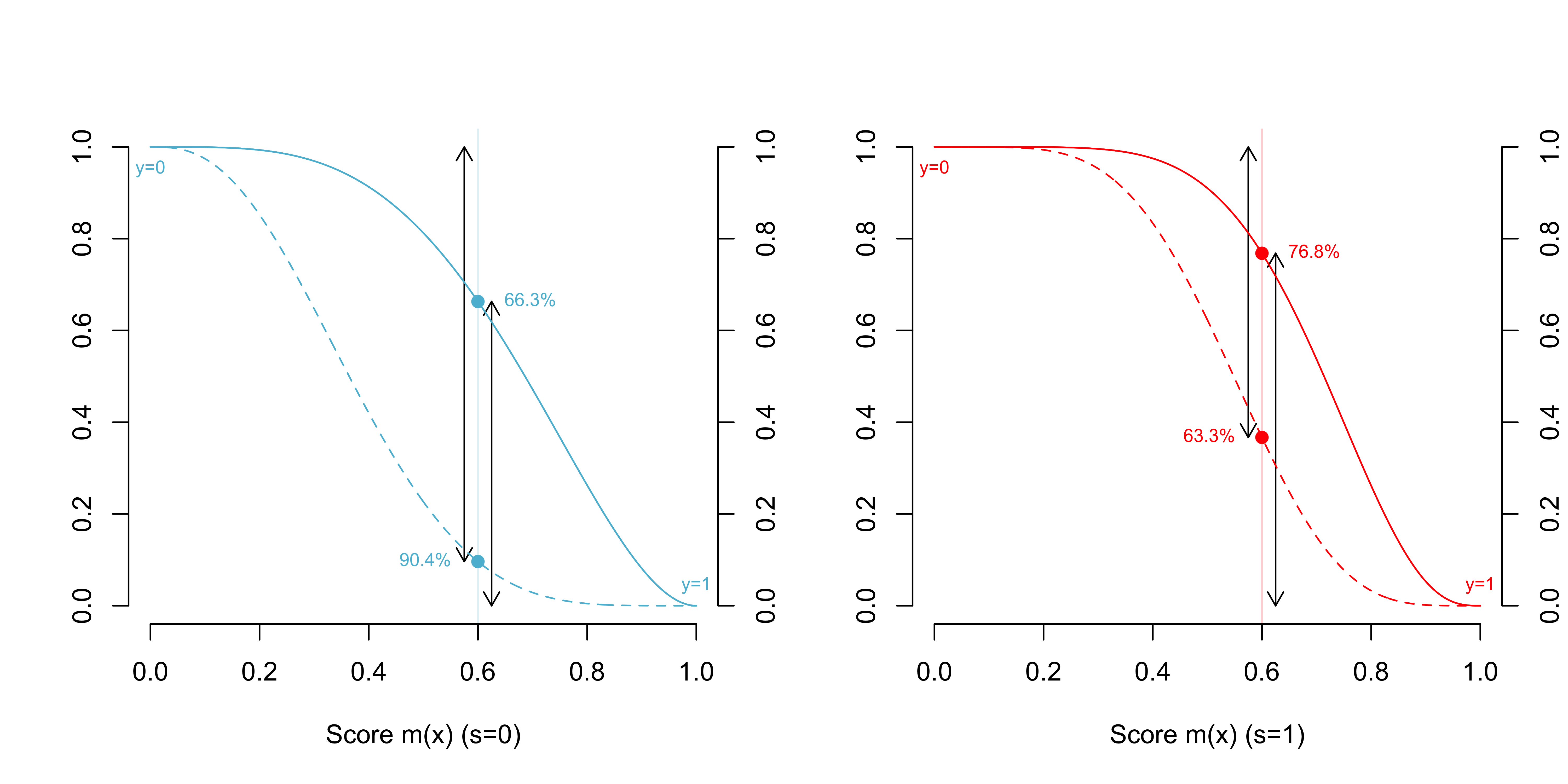}
    \caption{Continuous example, with the same single ($60\%$) in both groups, $s\in\{\text{\textcolor{tikbleu}{\scriptsize\faCircle}},\text{\textcolor{tikrouge}{\scriptsize\faStop}}\}$ or $p\in\{0,1\}$, with the density of $m(\boldsymbol{X})$ on top, and its survival function on the bottom, and $\widehat{y}=
\boldsymbol{1}_{[60\%,100\%]}(m(\boldsymbol{x}))$. }
    \label{fig:ex:book:3}
\end{figure}

\begin{table}[!ht]
    \centering
    \begin{tabular}{|ll|rr|rr|}\hline
   Name & Probabilistic formula & $\text{\textcolor{tikbleu}{\scriptsize\faCircle}}$~~~~ & $\text{\textcolor{tikrouge}{\scriptsize\faStop}}$~~~~ & diff & (\%)~~~ \\ \hline
 statistical parity &$\mathbb{P}[\widehat{Y}=1|S=\circ]$& 38\% & 56.8\% & 18.8 & 49.5 \% \\
 equalized opportunity&$\mathbb{P}[\widehat{Y}=1|S=\circ,Y=1]$& 66.3\% & 76.8\% & 10.5 & 15.9\% \\
 predictive equality &$\mathbb{P}[\widehat{Y}=1|S=\circ,Y=0]$& 9.6\% & 36.7\% & 27.1 & 281.2\% \\
 conditional accuracy &$\mathbb{P}[Y=0|S=\circ,\widehat{Y}=0]$& 72.8\% & 73.2\% & 0.4 & 0.5\% \\
 predictive parity &$\mathbb{P}[Y=1|S=\circ,\widehat{Y}=1]$& 87.3\% & 67.7\% & -19.6 & -22.5\% \\
 accuracy equality &$\mathbb{P}[\widehat{Y}=Y|S=\circ]$& 62\% & 56.8\% & -5.3 & -8.5\% \\
 treatment equality &$\text{FN}_{\circ}/\text{FP}_{\circ}$& 350.1 & 63.2 & -286.9 & -82\% \\\hline
    \end{tabular}
    \caption{Different concepts of fairness based on Figure \ref{fig:ex:book:3}, with the values of the measures for both groups, $s\in\{\text{\textcolor{tikbleu}{\scriptsize\faCircle}},\text{\textcolor{tikrouge}{\scriptsize\faStop}}\}$ or $s\in\{0.1\}$, the absolute difference and the relative difference (in percent), with the same cutoff (60\%) in both groups to obtain $\widehat{y}$ from the $m$ score. } 
    \label{tab:ex:book:3}
\end{table}

\begin{figure}
    \centering
    \includegraphics[width=\textwidth]{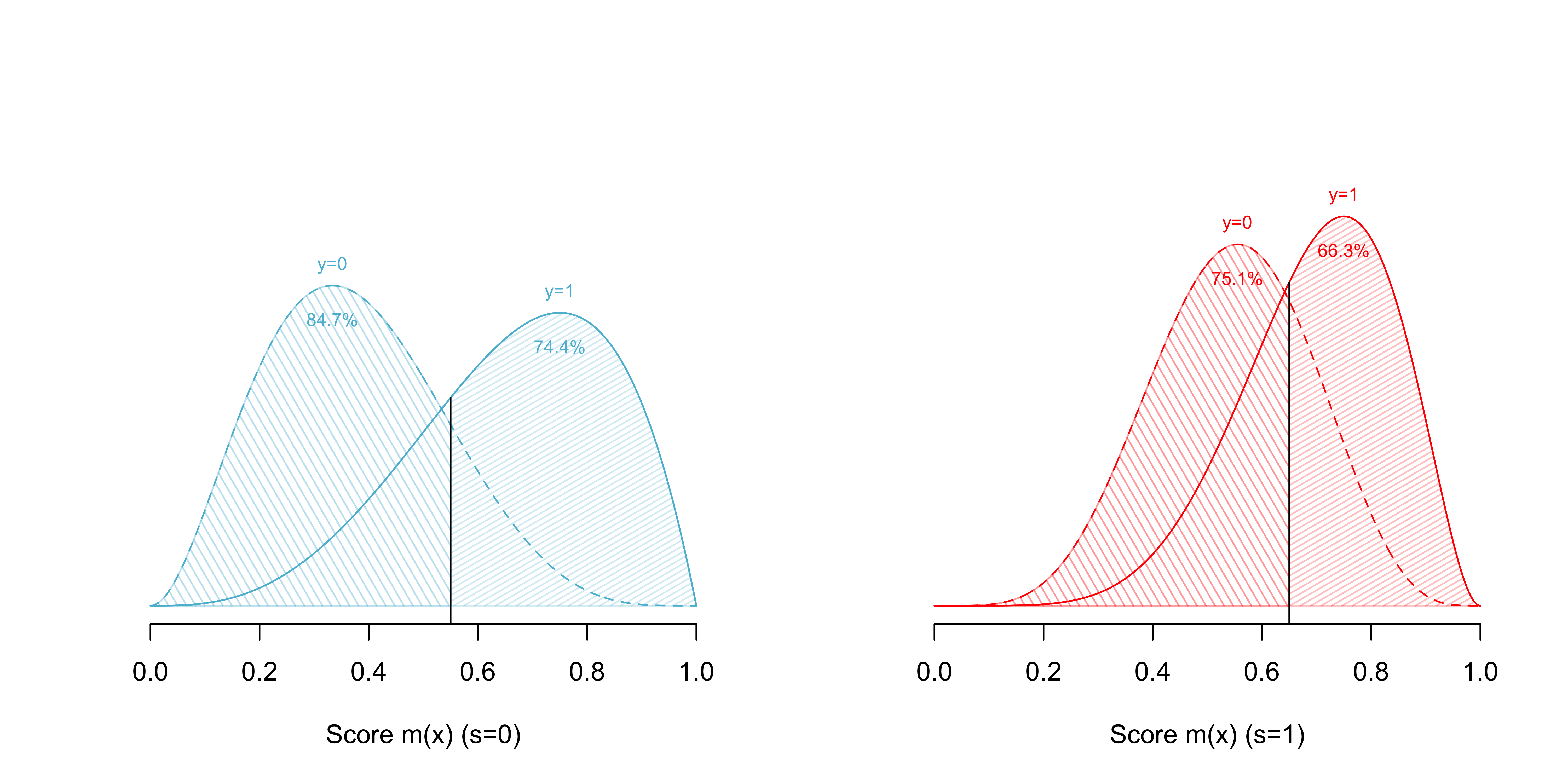}
    \includegraphics[width=\textwidth]{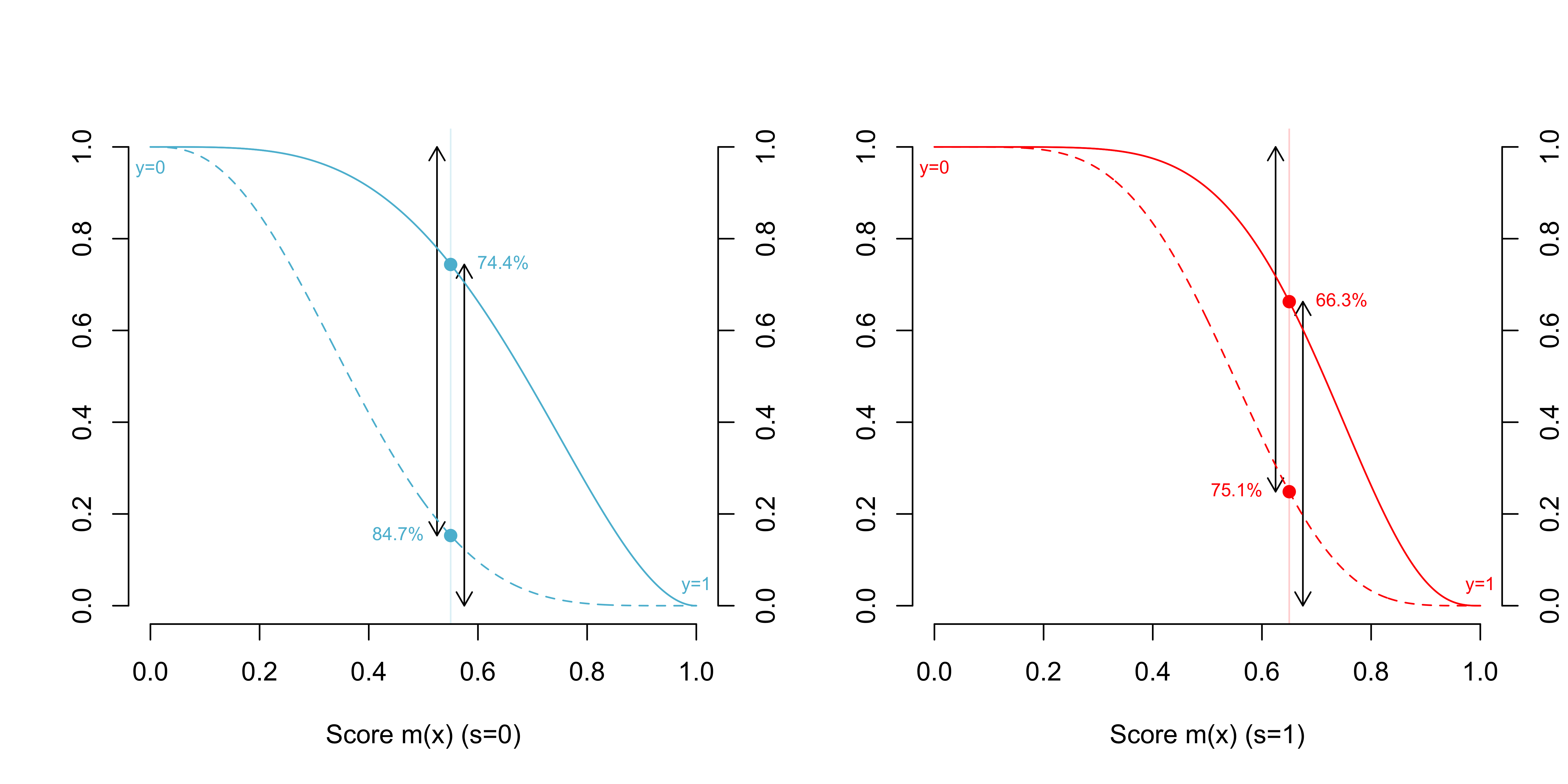}
    \caption{Continuous example, with different thresholds ($55\%$ and $65\%$) in the two groups, $s\in\{\text{\textcolor{tikbleu}{\scriptsize\faCircle}},\text{\textcolor{tikrouge}{\scriptsize\faStop}}\}$ or $s\in\{0,1\}$, with the density of $m(\boldsymbol{X})$ at the top, and its survival function at the bottom, and $\widehat{y}=\boldsymbol{1}_{[55\%,100\%]}(m(\boldsymbol{x}))$ if $s=\text{\textcolor{tikbleu}{\scriptsize\faCircle}}$ and $\widehat{y}=\boldsymbol{1}_{[65\%,100\%]}(m(\boldsymbol{x}))$ if $s=\text{\textcolor{tikrouge}{\scriptsize\faStop}}$.}
    \label{fig:ex:book:4}
\end{figure}

\begin{table}[!ht]
    \centering
    \begin{tabular}{|ll|rr|rr|}\hline
   Name & Probabilistic formula & $\text{\textcolor{tikbleu}{\scriptsize\faCircle}}$ / 0~~~~ & $\text{\textcolor{tikrouge}{\scriptsize\faStop}}$ / 1~~~~ & diff & (\%)~~~ \\\hline
 statistical parity   &$\mathbb{P}[\widehat{Y}=1|S=\circ]$&  25.0\% & 75.0\% &  50.0 & +200.0\% \\
 equalized opportunity&$\mathbb{P}[\widehat{Y}=1|S=\circ,Y=1]$&  40.0\% & 88.9\% &  48.9 & +122.2\% \\
 predictive equality  &$\mathbb{P}[\widehat{Y}=1|S=\circ,Y=0]$&   0.0\% & 57.1\% &  57.1 &   -\\
 conditional accuracy &$\mathbb{P}[Y=0|S=\circ,\widehat{Y}=0]$&  50.0\% & 75.0\% &  25.0 &  +50.0\% \\
 predictive parity    &$\mathbb{P}[Y=1|S=\circ,\widehat{Y}=1]$& 100.0\% & 66.7\% & -33.3 & -33.3\% \\
 accuracy equality    &$\mathbb{P}[\widehat{Y}=Y|S=\circ]$&  62.5\% & 68.8\% &   6.2 &  +10.0\% \\
 treatment equality   &$\text{FN}_{\circ}/\text{FP}_{\circ}$&       - ~~  & 25.0\% &  -~ &   -~~ \\\hline
    \end{tabular}
    \caption{Different concepts of fairness based on Figure \ref{fig:ex:book:4}, with the values of the measures for the two groups, $s\in\{\text{\textcolor{tikbleu}{\scriptsize\faCircle}},\text{\textcolor{tikrouge}{\scriptsize\faStop}}\}$, the absolute difference, and the relative difference (in percent), with a different threshold (55\% if $s=\text{\textcolor{tikbleu}{\scriptsize\faCircle}}$ and 65\% if $s=\text{\textcolor{tikrouge}{\scriptsize\faStop}}$) in the two groups to obtain $\widehat{y}$ from the score $m(\boldsymbol{X})$. } 
    \label{tab:ex:book:4}
\end{table}

\section{Individual-level fairness}\label{sec:eq:indiv}

In the previous section, we were interested in a notion of ``group" fairness, with subgroups constituted by the values of $y$, $\widehat{y}$ and $s$, since the three variables are categorical. 
Individual fairness is a relatively different concept. The criteria discuessed in the previous section were all group-based, whereas individual fairness, as the name suggests, is based on the individual observations. It was first proposed in \cite{dwork2012fairness}. The notion of individual fairness emphasizes that similar individuals (on unprotected attributes) should be treated similarly.




\subsection{Lipschitz property}\label{sub:lipschitz}

The natural idea, found in \cite{duivesteijn2008nearest}, is that two ``close" individuals (in the sense of unprotected characteristics $\boldsymbol{x}$) must have the same forecast. Formally, let us consider two distance, one on $\{0,1\}\times\{0,1\}$ noted $d_y$, and one on $\mathcal{X}$ noted $d_x$, such that we will have individual fairness on a database of size $n$ if we have the following property (called Lipschitz property)

\begin{definition}[Lipschitz property, \cite{duivesteijn2008nearest}, \cite{luong2011k}]
A decision function $\widehat{Y}$ satistfies the Lipschitz property if
$$
d_y(\widehat{y}_i,\widehat{y}_j) \leq d_x(\boldsymbol{x}_i,\boldsymbol{x}_j),~\forall i,j=1,\cdots,n.
$$
\end{definition}

\cite{duivesteijn2008nearest} talked about monotonic classification. It is difficult to determine which metric to use to measure the similarity of two individuals (i.e. between $\boldsymbol{x}_i$ and $\boldsymbol{x}_j$), as explained by \cite{kim2018fairness}. The most usual is to use a Mahalanobis type distance, to take into account the different scales between the variables.

\subsection{Counterfactual Fairness}\label{sub:counterfactual}

\subsubsection{Causal inference}

Consider, as in \cite{rubin1974estimating} or
\cite{hernan2010causal}, the following framework: let $t$ denote some binary treatment ($t\in\{0,1\}$, with respectively, the control and the treatment). Let $\boldsymbol{x}$  be some covariates,  $y$ the observed outcome, with $y_{T\leftarrow 1}^\star$ and $y_{T\leftarrow 0}^\star$ the potential outcomes (also denoted $y(1)$ and $y(0)$ in \cite{imbens2015causal} or \cite{imai2018quantitative}, or $y^1$ and $y^0$ in \cite{morgan2015counterfactuals} or \cite{cunningham2021causal}, even $y_{t=1}$ and $y_{t=0}$ in \cite{pearl2018book}), realized either under treatment condition ($t=1$) or under control condition ($t=0$). Note that the observed outcome is $y=y_{T\leftarrow t}^\star$, or $y=t\cdot y_{T\leftarrow 1}^\star+(1-t)\cdot y_{T\leftarrow 0}^\star$. An illustration is reported in Table~\ref{Tab:potential:outcome}.

\begin{table}[h!]
    \begin{tabular}{lcccccccc}\hline\hline
    & Treatment & \multicolumn{3}{c}{Outcome} & Age & Gender & Height & Weight \\
    \cmidrule(lr){3-5}
    & $t_i$ & $y_i$ & $y_{i,T\leftarrow 1}^\star$ & $y_{i,T\leftarrow o}^\star$ & $x_{1,i}$ &  $x_{2,i}$ &  $x_{3,i}$ &  $x_{4,i}$ \\ \midrule
    1 & 1 & 121 & 121 & {\bf ?} & 37 & F & 160 & 56 \\
    2 & 0 & 109 & {\bf ?} & 109 & 28 & F & 156 & 54 \\
    3 & 1 & 162 & 162 & {\bf ?} & 53 & M & 190 & 87 \\ \hline\hline
    \end{tabular}
    \caption{Potential outcome framework of causal inference, with one binary treatment $t_i$, the observed outcome variable $y_i$ and the two potential outcomes $y_{i,T\leftarrow 1}^\star$ and $y_{i,T\leftarrow 0}^\star$, as well as some covariates $\boldsymbol{x}_i$. One of the two potential outcomes is observed, and the other is missing, indicated by the question mark in the table.}\label{Tab:potential:outcome}
\end{table}

We will use the term ``treatment'' (and letter $t$) even if interventions are not possible, so it is no {\em per se} a ``treatment''. In this article, we try to answer a hypothetical question, like most questions asked at the third level of the ``ladder of causality". For instance, in a context of quantifying discrimination, the ``treatment" will denote the sensitive attribute, such as the race of an individual, \textit{e.g.}, ``{\em what would have been the outcome if that person had been Afro-American?}" Since our approach proposes an improvement on the metrics used in causal inference literature, we will use similar notations. 
There will be a significant impact of treatment $t$ on $y$ if $y^\star_{T\leftarrow0}\neq y^\star_{T\leftarrow1}$. More specifically, the causal effect for individual $i$ is ${\tau_i=y^\star_{i,T\leftarrow1} - y^\star_{i,T\leftarrow0}}$, as discussed in \cite{charpentier2023transport}.

A model satisfies the ``counterfactual fairness'' property if ``{\em had the protected attributes (e.g., race) of the individual been different, other things being equal, the decision would have remained the same}''. Also, a classifier will be counterfactually fair if for all individuals the outcome is equal to the outcome of its counterfactual individual (i.e., the same individual with one protected attribute reversed). 
$$
\mathbb{P}[Y^\star_{S\leftarrow 0}=1 |\boldsymbol{X}=\boldsymbol{x}]=
\mathbb{P}[Y^\star_{S\leftarrow 1}=1 |\boldsymbol{X}=\boldsymbol{x}],
$$
where $Y^\star_{S\leftarrow z}$ is the prediction of the classifier if $s$ takes a specific value (corresponding to some sort of ``intervention").

\begin{definition}[Counterfactual fairness, \cite{Kusner17}]
If the prediction in the real world is the same as the prediction in the counterfactual world where the individual would have belonged to a different demographic group, we have counterfactual fairness, i.e.
$$
\mathbb{P}[Y^\star_{S\leftarrow s}=y|\boldsymbol{X}=\boldsymbol{x}]=
\mathbb{P}[Y^\star_{S\leftarrow s'}=y|\boldsymbol{X}=\boldsymbol{x}],~\forall s,s',\boldsymbol{x},y.
$$
\end{definition}





\begin{table}[!ht]
    \begin{tabular}{|lll|}\hline
         {\em fairness through awareness} & \cite{dwork2012fairness} & 
         $D(\widehat{y}_i,\widehat{y}_j)\leq d(\boldsymbol{x}_i,\boldsymbol{x}_j),~\forall i,j$ \\ \hline\hline
        {\em counterfactual fairness} & \cite{Kusner17} & $\mathbb{P}[Y^\star_{S\leftarrow s}=y|\boldsymbol{X}=\boldsymbol{x}]=\text{cst}_y,~\forall s$ \\
{\em no proxy discrimination} & \cite{kilbertus2017avoiding} & $\mathbb{P}[\widehat{Y}=y|do(S=s)]=\text{cst}_y,~\forall s$\\\hline
    \end{tabular}
    \caption{Definitions of individual fairness, with the $do$ operator of \cite{pearl1988probabilistic}.}
    \label{tab:def:indiv}
\end{table}






\section{Correcting discrimination}\label{sec:correction}




Once a discrimination of a model $m$ with respect to some sensitive attribute $s$ is observed, one might try to correct the discrimination

\subsection{Pre-processing approaches}\label{sub:pre:processing}

A straightforward approach to removing bias from datasets would be to remove the protected attribute and other data elements that are suspected of containing related information. Unfortunately, such removal is rarely sufficient. In the literature, there are four approaches to removing bias by manipulating the data set. Respectively, these approaches modify the labels $y$
the observed data $\boldsymbol{x}$, the data/label pairs $\boldsymbol{x},y\}$ and the weighting of these pairs.

\subsubsection{Manipulation of labels}

\citeauthor{kamiran2009classifying} \citeyear{kamiran2009classifying}  and \citeyear{kamiran2012data} proposed to modify some of the training labels, which they call data manipulation. They compute a classifier on the original dataset and find examples close to the decision surface. They then swap the labels so that a positive outcome for the disadvantaged group is more likely and retrain. This is a heuristic approach that empirically improves fairness at the expense of accuracy.

\subsubsection{Manipulation of observed data}

\cite{feldman2015certifying} proposed to manipulate the individual dimensions of the $\boldsymbol{x}$ data in a way that depends on the sensitive attribute $s$. The idea is to align the cumulative distributions $F_0[\boldsymbol{x}]$ and $F_1[\boldsymbol{x}]$ for the feature $\boldsymbol{x}$ when the sensitive attribute $s$ is $0$ and $1$, respectively, to a median cumulative distribution $F_m[\boldsymbol{x}]$. This method is similar to the normalization of test scores across different schools, and is called ``disparate impact suppression''. This approach has the disadvantage of treating each input variable $\boldsymbol{x}$ separately, and ignores their (possible) interactions.

\subsubsection{Manipulation of labels and data}

\cite{calmon2017optimized} learns a $\psi$-transformation that transforms $\{\boldsymbol{x},y\}$ data pairs into new $\{\boldsymbol{x}',y'\}$ data values in a way that explicitly depends on the sensitive attribute $s$. \cite{calmon2017optimized} formulates this problem as an optimization problem in which the change in data utility must be minimized, subject to bounds on the harm and distortion of the original values. 
Unlike disparate impact removal, this method takes into account interactions between all dimensions of the data. However, the randomized transformation is formulated as a probability table, which is only suitable for data sets with a small number of discrete input and output variables. 

\subsubsection{Pairwise weighting of data}

\cite{kamiran2012data} proposes to reweight the $\{\boldsymbol{x},y\}$ observations in the training dataset so that cases where the sensitive attribute $s$ predicts that the disadvantaged group will have a positive outcome are more heavily weighted. They then train a classifier that uses these weights in its cost function. They also propose to resample the training data according to these weights and use a standard classifier.

\subsection{Reprocessing or in-processing algorithms}\label{sub:in:processing}

In the previous section, we introduced the latent prejudice measure based on the mutual information between the data $\boldsymbol{x}$ and the sensitive attribute $s$. Similarly, we can measure the dependency between the labels $y$ and the sensitive attribute $s$
$$
IP=\sum_{y,s}\mathbb{P}(y,s)\log\frac{\mathbb{P}(y,s)}{\mathbb{P}(y)\mathbb{P}(s)}
$$
which \cite{Kamishima2011Fairness} calls the ``{\em indirect prejudice}". Intuitively, if there is no way to predict the labels from the sensitive attribute and vice versa, then there is no possibility of bias. 

One approach to eliminating bias during training is to explicitly remove this dependency using adversarial learning. Other approaches involve penalizing mutual information using regularization, fitting the model under the constraint that it is unbiased. We will briefly discuss each of these approaches.

\subsubsection{Adversarial debiasing}

Adversarial debiasing (introduced in \cite{beutel2017data} or \cite{zhang2018mitigating}) reduces evidence of sensitive attributes in predictions by simultaneously trying to fool a second classifier that is trying to guess the sensitive attribute $s$. \cite{beutel2017data} forces both classifiers to use a shared representation and therefore minimizing the performance of the adversary classifier means removing all information about the sensitive attribute from that representation.
\cite{beutel2017data} proposes a representation for classification that was also used to predict the sensitive attribute. The system was trained adversarially, encouraging good system performance but punishing correct classification of the sensitive attribute. In this way, a representation that does not contain information about the sensitive attribute is learned.


\subsubsection{Suppression of bias by regularization}

\cite{Kamishima2011Fairness} proposed adding an additional regularization condition to the output of the logistic regression classifier that attempts to minimize the mutual information between the sensitive attribute and the $\widehat{y}$ prediction. 

In econometrics and machine learning, we try to maximize accuracy by solving
$$
{\underset{\boldsymbol{\theta}\in\Theta}{\text{argmin}}\left\lbrace \mathcal{L}\big(m_{\boldsymbol{\theta}}(\boldsymbol{x}),{y}\big)
\right\rbrace},\text{ where } \mathcal{L}\big(m_{\boldsymbol{\theta}}(\boldsymbol{x}),{y}\big)=\sum_{i=1}^n \ell\big(m_{\boldsymbol{\theta}}(\boldsymbol{x}_i),{y_i}\big)
$$
for some loss function $\ell$ (that might be related to minus the log-likelihood). Following \cite{hastie2015statistical} it is possible achieve parsimony be introducing some penalty in the objective function~: given a penalty $\mathcal{P}$
$$
\underset{\boldsymbol{\theta}\in\Theta}{\text{argmin}}\left\lbrace  \mathcal{L}\big(m_{\boldsymbol{\theta}}(\boldsymbol{x}),{y}\big)+\lambda \mathcal{P}(m_{\boldsymbol{\theta}})
\right\rbrace,
$$
for instance $\mathcal{P}(m_{\boldsymbol{\theta}})=\text{dim}(\boldsymbol{\theta})$, the (true) dimension of $\boldsymbol{\theta}$, when removing null values.
Inspired by \cite{Goodfellow2018} (but also \cite{bechavod2017penalizing} or \cite{cho2020fair}), {to avoid un-fairness}, it is natural penalize according to a discrimination measure. We have seen (see e.g. Table \ref{tab:def:group}) that most group fairness principles are related to independence, or conditional independence, and therefore, can be quantified using correlation related measures, as discussed in section \ref{sec:indep}. For {demographic parity}, characterized by independence between $\widehat{y}$ and the sensitive attribute $s$. For instance, we can solve
$$
\underset{\boldsymbol{\theta}\in\Theta}{\text{argmin}}\left\lbrace  \mathcal{L}\big(m_{\boldsymbol{\theta}}(\boldsymbol{x}),{y}\big)+\lambda \text{cor}(m_{\boldsymbol{\theta}}(\boldsymbol{y}),s)
\right\rbrace,
$$
i.e. 
$$
\underset{\boldsymbol{\theta}\in\Theta}{\text{argmin}}\left\lbrace  \mathcal{L}\big(m_{\boldsymbol{\theta}}(\boldsymbol{x}),{y}\big)+\lambda \text{cor}^\star(m_{\boldsymbol{\theta}}(\boldsymbol{y}),s)
\right\rbrace,
$$
that can be written (in a very general context, where $s$ can be non binary)
$$
\underset{\boldsymbol{\theta}\in\Theta}{\text{max}}\left\lbrace  \underset{g\in\mathcal{S}_y,h\in\mathcal{S}_s}{\text{argmin}}\left\lbrace  \mathcal{L}\big(m_{\boldsymbol{\theta}}(\boldsymbol{x}),{y}\big)+\lambda \text{cor}(g(m_{\boldsymbol{\theta}}(\boldsymbol{y})),h(s))
\right\rbrace\right\rbrace,
$$
that can be solved using either neural networks or some function basis for $\mathcal{S}_y$ and $\mathcal{S}_s$, as in \cite{grari2022}.
For {equalized odds}), characterized by independence between $\widehat{y}$ and the sensitive attribute $s$, conditional on $y$, i.e. when $y\in\{0,1\}$
$$
\underset{\boldsymbol{\theta}\in\Theta}{\text{argmin}}\left\lbrace  \mathcal{L}\big(m_{\boldsymbol{\theta}}(\boldsymbol{x}),{y}\big)+\lambda_0 \text{cor}(m_{\boldsymbol{\theta}}(\boldsymbol{y}),s|y=0)+
\lambda_1 \text{cor}(m_{\boldsymbol{\theta}}(\boldsymbol{y}),s|y=1)
\right\rbrace,
$$
that could extended to maximal correlation.

\subsection{Post-processing}\label{sub:post:processing}

Several techniques used to perform post-processing of the output scores of the classifier to make decisions fairer were introduced in
\cite{corbettdavies2017algorithmic}, \cite{dwork2018decoupled}, \cite{menon2018cost}, \cite{lohia2019bias} and \cite{awasthi2020equalized}. For instance, 
\cite{hardt2016equality} considered a technique for flipping some decisions of a classifier to enhance equalized odds, or equalized opportunity, while \cite{menon2018cost} suggested selecting district thresholds $\tau$ for each group (as discussed earlier), in a manner that maximizes accuracy and minimizes demographic parity.

\section*{Acknowledgments}
Arthur Charpentier acknowledges the financial support of the AXA Research Fund through the joint research initiative {\em use and value of unusual data in actuarial science}, as well as NSERC grant 2019-07077.

\bibliographystyle{apalike}
\bibliography{biblio.bib}

\end{document}